\documentclass[11pt,a4paper]{article}

\usepackage[backend=bibtex, style=authoryear, maxbibnames=9]{biblatex}
\usepackage{pdfpages}
\usepackage{color}
\usepackage{diagbox}
\usepackage{amsthm, amsmath, amsfonts,graphicx,url}
\usepackage{amssymb}
\usepackage{enumitem}
\usepackage{bm, bbm}
\usepackage{blkarray}
\usepackage{multirow}
\usepackage[margin=1in]{geometry}
\usepackage{changepage}
\usepackage{authblk}
\usepackage[font=small,labelfont=bf]{caption}
\usepackage[font=footnotesize,labelfont=bf]{subcaption}
\numberwithin{equation}{section}
\usepackage{hyperref}
\hypersetup{
	colorlinks,
	citecolor=blue,
	linkcolor=cornflowerblue,
	filecolor=yaleblue,      
	urlcolor=cornflowerblue
}
\usepackage{cleveref}
\usepackage{footnotebackref}
\crefformat{footnote}{#2\footnotemark[#1]#3}
\usepackage{easyReview}
\allowdisplaybreaks

\DeclareMathOperator*{\GG}{GG}

\DeclareMathOperator*{\BIC}{BIC}
\DeclareMathOperator*{\IMSE}{IMSE}
\DeclareMathOperator*{\const}{const}

\DeclareMathOperator*{\unif}{Unif}
\DeclareMathOperator*{\Ga}{Gamma}
\DeclareMathOperator*{\Beta}{Beta}
\DeclareMathOperator*{\iid}{iid}

\DeclareMathOperator*{\opt}{opt}
\DeclareMathOperator*{\mle}{mle}
\DeclareMathOperator*{\lasso}{lasso}

\DeclareMathOperator*{\rank}{rank}

\definecolor{cornflowerblue}{rgb}{0.39, 0.58, 0.93}
\definecolor{yaleblue}{rgb}{0.06, 0.3, 0.57}

\title{\textbf{MM Algorithms for Statistical Estimation in Quantile Regression}}
\author{Yifan CHENG \thanks{y.cheng@u.nus.edu}  }
\author{Anthony KUK Yung Cheung}
\affil{\footnotesize Department of Statistics and Data Science, National University of Singapore} 
\date{}
\begin{document}
	\maketitle	
	\vspace{-5mm}
	\begin{abstract}
		Quantile regression \parencite{Koenker1978} is a robust and practically useful way to efficiently model quantile varying correlation and predict varied response quantiles of interest. This article constructs and tests MM algorithms, which are simple to code and have been suggested superior to some other prominent quantile regression methods in nonregularized problems \parencite{Pietrosanu2017}, in an array of linear quantile regression settings. Simulation studies comparing MM to existing tested methods and applications to various real data sets have corroborated our algorithms' effectiveness. 
	\end{abstract}
	\tableofcontents
	
	\section{Introduction}\label{sect1}	
	\par Predicting varied quantiles of the response variable is desirable in diverse domains (e.g., \cite{Buchinsky1995}, \cite{Taylor1998}; \cite{Yu2003}; \cite{Koenker2004}), as quantiles closer to the tails are of more interest than $q=0.5$ in some applications. Various quantile estimation approaches (e.g., \cite{Cole1988}; \cite{Efron1991}; \cite{He1997}) have been proposed to model non-constant correlation patterns between the dependent and independent variables across different outcome quantiles, among which \cite{Koenker1978}'s method remains foremost. Quantile regression introduced by \textcite{Koenker1978} not only excels in efficiently portraying a comprehensive picture of the regressor-regressand's dependency structure but also stays more robust against outliers than ordinary regression of the mean. 
	Denote $n$ and $p$ as the sample size and covariate dimension (including the intercept), respectively. Suppose we have observed a sample ($\bm{x}_i,y_i),\, i=1,2,\dots, n$. For any given quantile level $q\in (0,1)$ and any specified regression function $f(\bm{X}, \bm{\theta})$,  
    let $f_i(\bm{\theta}):\mathbb{R}^p \rightarrow \mathbb{R} =f(\bm{x}_i,\bm{\theta}) \text{ and } r_i(\bm{\theta})=y_i-f_i(\bm{\theta})$, $i=1,2,\dots,n$. \textcite{Koenker1978} defined the regression quantile at $q$ as the $p$-dimensional vector $\bm{\hat{\theta}}$ that minimizes the asymmetric $L_1$ loss function
	\begin{equation}\label{sample quantile loss function}
		L(\bm{\theta})=\sum_{i=1}^n \rho_q (r_i(\bm{\theta})),
	\end{equation}
	where 
	\begin{equation}\label{quantile loss rho} 
		\rho_q (r_i) = |r_i|\left[q\mathbbm{1}_{\{r_i\geq0\}}+(1-q)\mathbbm{1}_{\{r_i<0\}}\right]=qr_i-r_i\mathbbm{1}_{\{r_i<0\}},\quad i=1,2,\dots,n.
	\end{equation}   
	Compared to traditional least-squares estimators, the minimizers of \Cref{sample quantile loss function} are much more robust to an array of non-normal error distributions while performing similarly well for Gaussian linear models \parencite{Koenker1978}.
	\par The prevailing most well-adopted quantile regression package in R is probably still \texttt{quantreg} developed by \textcite{Koenker2010}, which estimates empirical regression coefficients for specified quantile orders via a default simplex algorithm \parencite{Koenker1987} and an interior point method \parencite{Portnoy1997} triggering the Frisch-Newton algorithm for huge data sets. \textcite{Pietrosanu2017} advanced in questing robust accelerated quantile regression methods \parencite{Portnoy1997} by conducting simulation studies on four prominent approaches -- Alternating Direction Method of Multipliers (ADMM) algorithms, Coordinate Descent (CD) algorithms, MM algorithms, and algorithms adopted by \texttt{quantreg}. They found out that MM produced comparably accurate parameter estimates markedly faster than the other three methods in non-regularized quantile and composite quantile regression. 
	\par Although termed ``algorithm", MM, which stands for Majorize-Minimize for minimization problems and Minorize-Maximize for maximization problems, is in fact a strategy for creating an easy-to-minimize/maximize surrogate function that majorizes/minorizes the objective function. Given an objective function $L(\bm{\theta}):\mathbb{R}^p \rightarrow \mathbb{R}$ that is difficult to be minimized directly and a current iteration's parameter estimate $\bm{\theta}^{(t)}$ ($t\in \mathbb{N}$), a surrogate function $Q\left(\bm{\theta | \theta}^{(t)}\right):\mathbb{R}^p \rightarrow \mathbb{R}$ is called a majorizer of $L(\bm{\theta})$ at $\bm{\theta}^{(t)}$ if it satisfies 
	\begin{equation}\label{tangent ppty1}
		Q\big(\bm{\theta | \theta}^{(t)}\big)\geq L\left(\bm{\theta}\right) \text{ for all } \bm{\theta}
	\end{equation}
	and in particular
	\begin{equation}\label{tangent ppty2}
		Q\big(\bm{\theta}^{(t)} | \bm{\theta}^{(t)}\big) = L\big(\bm{\theta}^{(t)}\big).
	\end{equation}
	If the value $\bm{\theta}^{(t+1)}$ at which $Q\big(\bm{\theta | \theta}^{(t)}\big)$ is minimized can be conveniently calculated, then $Q\big(\bm{\theta | \theta}^{(t)}\big)$ yields a successful MM algorithm which drives the objective function $L(\bm{\theta})$ further downhill each iteration thanks to the descent property
	\begin{equation}\label{descent ppty}
		L\big(\bm{\theta}^{(t+1)}\big)\leq Q\big(\bm{\theta}^{(t+1)}|\bm{\theta}^{(t)}\big)\leq Q\big(\bm{\theta}^{(t)}|\bm{\theta}^{(t)}\big)=L\big(\bm{\theta}^{(t)}\big).
	\end{equation}
	To construct suitable majorizing/minorizing surrogate functions and thus significantly simplify difficult optimization problems, MM typically exploits convexity/concavity and miscellaneous inequalities \parencite{Hunter2004} such as Jensen's Inequality (which brings about the well-known EM algorithm, a special case of MM), Supporting Hyperplanes, Quadratic Upper Bound \parencite{Bohning1988}, the Arithmetic-Geometric Mean Inequality, and the Cauchy-Schwartz Inequality. The MM algorithm enjoys implementation simplicity and remarkable stability guaranteed by \Cref{descent ppty} while possessing complementary computation speed merits \parencite{Hunter2004} compared to the widely-adopted Newton-Raphson algorithm, and has hence gained application attention in diversified fields (e.g., \cite{Sabatti2002}; \cite{Hunter2002}; \cite{Hunter2004a}; \cite{May2005}).

	\section{MM Algorithms in Linear Quantile Regression}\label{sect2}
	\subsection{An MM Algorithm in Quantile Regression}\label{sect2a}
	\par For any given quantile level $q\in(0,1)$, the loss function $L(\bm{\theta})$ in \Cref{sample quantile loss function} is hard to directly minimize since it may accommodate multiple minima and its components \eqref{quantile loss rho} are non-differentiable when $r_i=0$ \parencite{Hunter2000}. An MM algorithm may be constructed to bypass the non-differentiability problem by setting (\cite{Hunter2000}; \cite{Hunter2004})
	\begin{equation}\label{typical qr MM}
		Q\left(\bm{\theta}|\bm{\theta}^{(t)}\right)=\sum_{i=1}^n \zeta_q\left(y_i-f_i(\bm{\theta})|y_i-f_i\big(\bm{\theta}^{(t)}\big)\right)=\sum_{i=1}^n \zeta_q\left(r_i(\bm{\theta})|r_i\big(\bm{\theta}^{(t)}\big)\right),
	\end{equation}
	where
	\begin{equation}\label{zeta}
		\zeta_q\left(r_i(\bm{\theta})|r_i\big(\bm{\theta}^{(t)}\big)\right)=\frac{1}{4}\left\lbrace \frac{r_i^2(\bm{\theta})}{\big|r_i\big(\bm{\theta}^{(t)}\big)\big|}+(4q-2)r_i(\bm{\theta})+\big|r_i\big(\bm{\theta}^{(t)}\big)\big|\right\rbrace. 
	\end{equation}
	It can be shown that the surrogate function in \Cref{typical qr MM} is a majorizer of $L(\bm{\theta})$ at $\bm{\theta}^{(t)}$ (We leave the proof to \textbf{Proposition A.1.} in \Cref{appenA}), yet this MM method suffers from the vexing drawback that $Q\big(\bm{\theta}|\bm{\theta}^{(t)}\big)$ is not defined when one $r_i\big(\bm{\theta}^{(t)}\big)=0$, $i\in\{1,2,\dots,n\}$. This is a realistic concern that should be settled, as some $f_i\big(\bm{\theta}^{(t)}\big)$ may well converge to $y_i$ as  $t\rightarrow \infty.$ For example, in the simplest case when all $f_i\big(\bm{\theta}^{(t)}\big)$, $i=1,2,...,n$ denote the sample mean $\mu^{(t)}$, $\mu^{(t)} \text{ should }\rightarrow \text{ one } y_i \text{ as } t\rightarrow \infty$ if $nq \not\in \mathbb{N}$ \parencite{Hunter2000}.  
	\par \textcite{Hunter2000} addressed this issue by introducing a close-approximate function $L_{\epsilon}(\bm{\theta})$, where $\epsilon>0$ is a small real number that denotes a perturbation, of the actual loss function \Cref{sample quantile loss function}
	\begin{equation}\label{close approximate sample qr loss function}
		L_{\epsilon}(\bm{\theta})=\sum_{i=1}^n \rho_q^{\epsilon}(r_i(\bm{\theta}))=\sum_{i=1}^n \left\lbrace \rho_q(r_i(\bm{\theta}))-\frac{\epsilon}{2}\ln(\epsilon+|r_i(\bm{\theta})|) \right\rbrace 
	\end{equation}
	and inventing a corresponding majorizer
	\begin{equation}\label{Qepsilon}
		Q_{\epsilon}\left(\bm{\theta}|\bm{\theta}^{(t)}\right)=\sum_{i=1}^n \zeta_q^{\epsilon}\left(r_i(\bm{\theta})|r_i\big(\bm{\theta}^{(t)}\big)\right)=\frac{1}{4}\sum_{i=1}^n\left\lbrace \frac{r_i^2(\bm{\theta})}{\epsilon+\big|r_i\big(\bm{\theta}^{(t)}\big)\big|}+(4q-2)r_i(\bm{\theta})+{\const}_i \right\rbrace, 
	\end{equation}
	where ${\const}_i,\:i=1,2,\ldots,n$ are chosen to ensure that
	\begin{equation}\label{def zeta equation} \zeta_q^{\epsilon}\left(r_i\big(\bm{\theta}^{(t)}\big)|r_i\big(\bm{\theta}^{(t)}\big)\right)=\frac{1}{4}\left\lbrace \frac{r_i^2\big(\bm{\theta}^{(t)}\big)}{\epsilon+\big|r_i\big(\bm{\theta}^{(t)}\big)\big|}+(4q-2)r_i\big(\bm{\theta}^{(t)}\big)+{\const}_i\right\rbrace =\rho_q^{\epsilon}\left(r_i\big(\bm{\theta}^{(t)}\big)\right).
	\end{equation}
	By definition, $Q_{\epsilon}\big(\bm{\theta}|\bm{\theta}^{(t)}\big)$ satisfies \Cref{tangent ppty2} as $\zeta_q^{\epsilon}\big(r_i\big(\bm{\theta}^{(t)}\big)|r_i\big(\bm{\theta}^{(t)}\big)\big)=\rho_q^{\epsilon}\big(r_i\big(\bm{\theta}^{(t)}\big)\big)$ for all $ i\in\{1,2,\dots,n\}$. Given current residual value $r^{(t)}=r\big(\bm{\theta}^{(t)}\big)$ at iteration t, we let 
	\begin{equation}
		h(r)=\zeta_q^{\epsilon}\big(r|r^{(t)}\big)-\rho_q^{\epsilon}(r)=\frac{1}{4}\left\lbrace \frac{r^2}{\epsilon+\big|r^{(t)}\big|}-2r+{\const}+2\epsilon\cdot \ln(\epsilon+|r|) \right\rbrace +r\mathbbm{1}_{\{r<0\}}.\nonumber
	\end{equation}
	Then
	\begin{align}\label{h'(r)1}
		&\text{For } r\geq0,\; h'(r)=\frac{1}{2}\left( \frac{r}{\epsilon+|r^{(t)}|}-1+\frac{\epsilon}{\epsilon+r} \right)=\frac{1}{2}\cdot \frac{r\big(r-\big|r^{(t)}\big|\big)}{\big(\epsilon+\big|r^{(t)}\big|\big)(\epsilon+r)};\\
		\label{h'(r)2}&\text{For } r<0, \; h'(r)=\frac{1}{2}\left( \frac{r}{\epsilon+\big|r^{(t)}\big|}+1-\frac{\epsilon}{\epsilon-r} \right)=-\frac{1}{2}\cdot \frac{r\big(r+\big|r^{(t)}\big|\big)}{\big(\epsilon+\big|r^{(t)}\big|\big)(\epsilon-r)}.
	\end{align} 
	Hence $h'(r)=0\iff r=0 \text{ or } r=\pm\: \big|r^{(t)}\big| \iff r=0 \text{ or } r=\pm\: r^{(t)}$ since the sets $\left\lbrace \big|r^{(t)}\big|,-\big|r^{(t)}\big|\right\rbrace $ and $\left\lbrace r^{(t)},-r^{(t)}\right\rbrace $ are the same. We can see from \Cref{h'(r)1,h'(r)2} that $h'(r)\leq0 \text{ for } r\in \big(-\infty,-\big|r^{(t)}\big|\big]\cup\big[0,\big|r^{(t)}\big|\big] \text{ and } \geq0 \text{ for } r\in \big[-\big|r^{(t)}\big|,0\big]\cup\big[\big|r^{(t)}\big|,\infty\big),$ i.e., $h(r)$ is non-increasing for $r\in \big(-\infty,-\big|r^{(t)}\big|\big]\cup\big[0,\big|r^{(t)}\big|\big] \text{ and non-decreasing for } r\in \big[-\big|r^{(t)}\big|,0\big]\cup\big[\big|r^{(t)}\big|,\infty\big).$
	Observe that $h(r)$ is an even function and thus by \Cref{def zeta equation},
	\begin{equation*}
		h\big(-r^{(t)}\big)=h\big(r^{(t)}\big)=0\iff h\big(-\big|r^{(t)}\big|\big)=h\big(\big|r^{(t)}\big|\big)=0\Rightarrow h(r)\geq 0\text{ for all }r. \qquad 
	\end{equation*}
	Hence,
	$$
	\zeta_q^{\epsilon}\big(r|r^{(t)}\big)\geq\rho_q^{\epsilon}(r)\text{ for all }r \Rightarrow Q_{\epsilon}\big(\bm{\theta}|\bm{\theta}^{(t)}\big)\geq L_{\epsilon}(\bm{\theta})\text{ for all }\bm{\theta} .
	$$
	By the above reasoning, $Q_{\epsilon}\big(\bm{\theta}|\bm{\theta}^{(t)}\big)$ satisfies \Cref{tangent ppty1} as well and thus indeed majorizes $L_{\epsilon}(\bm{\theta})$ at $\bm{\theta}^{(t)}$. 
	
	\subsection{Estimating Different Quantile Coefficients Separately} \label{sect2b}
	\par For linear quantile regression, $f_i(\bm{\theta})=\bm{x}_i^T\bm{\theta} \text{ for } i=1,2,\dots,n$ and $\bm{r}(\bm{\theta})_{n \times 1}=\left(r_1(\bm{\theta}),\ldots, r_n(\bm{\theta})\right)^T=\bm{y}_{n\times 1}-\bm{X}_{n\times p}\bm{\theta}_{p \times 1}$ for any given quantile level $q\in(0,1)$. $Q_{\epsilon}(\bm{\theta}|\bm{\theta}^{(t)})$ is quadratic in $\bm{\theta}$ and the MM algorithm can proceed in simplicity by explicitly solving for $\bm{\theta}^{(t+1)}$ that minimizes $Q_{\epsilon}\big(\bm{\theta}|\bm{\theta}^{(t)}\big)$ and updating the current parameter estimate $\bm{\theta}^{(t)}$ to $\bm{\theta}^{(t+1)}$. 
	\par Given $q\in(0,1)$ and current parameter estimate $\bm{\theta}^{(t)}$, let 
	\begin{equation}\label{Wdiag}
		\bm{W}^{(t)} = 
		\begin{pmatrix}
			\frac{1}{\epsilon+\big|r_1\big(\bm{\theta}^{(t)}\big)\big|}& & & \\
			& \frac{1}{\epsilon+\big|r_2\big(\bm{\theta}^{(t)}\big)\big|} &  & \\
			& & \ddots & \\
			& & & \frac{1}{\epsilon+\big|r_n\big(\bm{\theta}^{(t)}\big)\big|}
		\end{pmatrix},\;
		\bm{c}_{n\times 1}=
		\begin{pmatrix}
			4q-2\\
			4q-2\\
			\vdots \\
			4q-2
		\end{pmatrix}.
	\end{equation}
	Then \Cref{Qepsilon} becomes 
	\begin{align}\label{matrix form res}
		Q_{\epsilon}\big(\bm{\theta}|\bm{\theta}^{(t)}\big)=&\frac{1}{4}\left( \bm{r}(\bm{\theta})^T\bm{W}^{(t)}\bm{r}(\bm{\theta})+\bm{c}^T\bm{r}(\bm{\theta})\right)+\const\nonumber \\
		= & \frac{1}{4}\left( \bm{(y-X\theta)}^T\bm{W}^{(t)}\bm{(y-X\theta)}+\bm{c}^T\bm{(y-X\theta)}\right)+\const.
	\end{align}
	Thus minimizing $Q_{\epsilon}\big(\bm{\theta}|\bm{\theta}^{(t)}\big)$ is equivalent to minimizing 
	\begin{equation}\label{g theta}
		g\big(\bm{\theta}|\bm{\theta}^{(t)}\big)=\bm{(y-X\theta)}^T\bm{W}^{(t)}\bm{(y-X\theta)}+\bm{c}^T\bm{(y-X\theta)},
	\end{equation}
	which can be done directly as $g\big(\bm{\theta}|\bm{\theta}^{(t)}\big)$ is quadratic in $\bm{\theta}$. Setting
	$$
	\frac{\partial}{\partial\bm{\theta}}g(\bm{\theta}|\bm{\theta}^{(t)}) \bigg|_{\bm{\theta}=\bm{\theta}^{(t+1)}}=\left(2\bm{X}^T\bm{W}^{(t)}\bm{X\theta}-2\bm{X}^T\bm{W}^{(t)}\bm{y}-\bm{X}^T\bm{c}\right)_{\bm{\theta}=\bm{\theta}^{(t+1)}}=\bm{0}\, \text{ yields} 
	$$
	\begin{equation}\label{theta t+1}
		\bm{\theta}^{(t+1)}=(\bm{X}^T\bm{W}^{(t)}\bm{X})^{-1}\left(\bm{X}^T\bm{W}^{(t)}\bm{y}+\frac{1}{2}\bm{X}^T\bm{c}\right).
	\end{equation}
    Suppose that $\bm{X}_{n\times p}$ ($n>p$) is of full column rank. Then for all $p\times 1$ vector $\bm{v}\neq \bm{0}_{p\times 1}$, $\bm{Xv}\neq \bm{0}_{n\times 1}$ and thus $\bm{v}^T\bm{X}^T\bm{W}^{(t)}\bm{Xv}=\bm{(Xv)}^T\bm{W}^{(t)}\bm{(Xv)}>0$, as $\bm{W}^{(t)}$ is a diagonal matrix whose diagonal entries are all strictly positive. This shows that $\bm{X}^T\bm{W}^{(t)}\bm{X}$ is positive definite and thus invertible. Hence, $\bm{\theta}^{(t+1)}$ given in \Cref{theta t+1} is indeed well-defined and minimizes \Cref{g theta,matrix form res}, as the Hessian matrix 
    $\frac{\partial^2}{\partial\bm{\theta}\partial\bm{\theta}^T}g\big(\bm{\theta}|\bm{\theta}^{(t)}\big)=2\bm{X}^T\bm{W}^{(t)}\bm{X}$ 
    is positive definite for all $\bm{\theta}$ and in particular $\bm{\theta}^{(t+1)}$.

	\subsection{Estimating Different Quantile Coefficients Simultaneously}\label{sect2c}
	\par It is of tremendous interest as well as utmost importance to efficiently discern informative signals from and discard unwanted noises in the empirical sample quantile functions. To date, most relevant efforts either resort to varied smoothing techniques on the empirical quantile functions (e.g.,\cite{Cheng1995}; \cite{Cheng1997}) or adjust nonparametric fitting techniques to quantile regression (e.g.,\cite{Yu2016}), which more or less suffer from interpretation simplicity and prediction efficiency. \textcite{Frumento2016} addressed these concerns by bringing forward a parsimonious parametric approach that directly models the linear regression coefficients as smooth functions of $q$, which succeeds in effectively pooling information across quantile levels. The method was later built into R via the \texttt{qrcm} package \parencite{Package2020}.
	\par Rather than estimating the regression coefficients directly but only one $q$ a time as in the previous subsection, \textcite{Frumento2016} introduced a basis of $h\in \mathbb{N}$ known functions of the quantile order -- $\bm{b}(q)=(b_1(q),b_2(q),\dots,b_h(q))^T$, where $h$ is usually less than 10 and $\bm{b}(q)$ is required to induce a monotonically non-decreasing $Q(q|\bm{x,\theta})$ for some $\bm{\theta}$, wrote the regression coefficient as $\bm{\beta}_{q}=\bm{A}_{p\times h}\bm{b}(q)$ for all $q$, and estimated instead the $p\times h$ parameter matrix $\bm{A} = \big(\theta_{jl}\big)_{j=1,\ldots,p,\: l = 1,\ldots, h}$. 
	The residual vector can now be expressed as a function of $q$ -- $\bm{r}(q)=(r_1(q),r_2(q),\dots,r_n(q))^T=\bm{y}-\bm{XAb}(q)$. This way, entries of $\bm{\beta}_q$ are represented as linear combinations of basis functions of $q$ and are no longer estimated separately.
	\par The MM algorithm described in \Cref{sect2b} for linear quantile regression can be conveniently generalized to fit the model in this section by rewriting the parameter matrix $\bm{A}_{p\times h}$ as 
	\begin{equation}\label{para vector form}
		\bm{\theta}_{hp\times 1}=(\theta_{11},\theta_{21},\dots,\theta_{p1},\dots\dots,\theta_{1h},\theta_{2h},\dots,\theta_{ph})^T
	\end{equation}
	to form the parameter vector to be estimated and replacing the $\bm{X}_{n\times p}$ in \Cref{sect2b} by new design matrices $\bm{D}(q)_{n\times hp}$ for any $q\in(0,1)$:
	\begin{align}\label{D(q)}
		\bm{D}(q)_{n\times hp}&=\bm{b}(q)^T_{1\times h}\otimes \bm{X}_{n\times p}\\
		&=\begin{pmatrix}
			x_{11}b_1(q) & x_{12}b_1(q) & \dots & x_{1p}b_1(q)&
			\dots \dots&x_{11}b_h(q)&x_{12}b_h(q)&\dots&x_{1p}b_h(q)\\
			x_{21}b_1(q)&x_{22}b_1(q)&\dots&x_{2p}b_1(q)&
			\dots \dots&x_{21}b_h(q)&x_{22}b_h(q)&\dots&x_{2p}b_h(q)\\
			\vdots & \vdots & \ddots & \vdots &\ddots & \vdots&\vdots & \ddots & \vdots\\
			\vdots & \vdots & \ddots & \vdots &\ddots & \vdots&\vdots & \ddots & \vdots\\
			x_{n1}b_1(q)&x_{n2}b_1(q)&\dots&x_{np}b_1(q)&
			\dots \dots&x_{n1}b_h(q)&x_{n2}b_h(q)&\dots&x_{np}b_h(q)
		\end{pmatrix}.\nonumber
	\end{align}
	Then $\bm{r}(q,\bm{\theta})=(r_1(q,\bm{\theta}),r_2(q,\bm{\theta}),\dots,r_n(q,\bm{\theta}))^T=\bm{y}-\bm{XAb}(q)=\bm{y}-\bm{D}(q)\bm{\theta}$. Given $k\in \mathbb{N}$ quantile orders $q_1,q_2,\dots,q_k$ that we wish to simultaneously model, where $k$ is typically set large (e.g., 100, 1000) for better performances, the surrogate function with perturbation becomes 
	\begin{align}\label{Qepsilonk}
		Q_{\epsilon}\left(\bm{\theta}|\bm{\theta}^{(t)}\right)=&\sum_{a=1}^{k}\sum_{i=1}^n \zeta_{q_a}^{\epsilon}\left(r_i(q_a,\bm{\theta})|r_i\big(q_a,\bm{\theta}^{(t)}\big)\right)\nonumber \\
		=&\frac{1}{4}\sum_{a=1}^{k}\sum_{i=1}^n\left\lbrace \frac{r_i^2(q_a,\bm{\theta})}{\epsilon+\big|r_i\big(q_a,\bm{\theta}^{(t)}\big)\big|}+(4q_a-2)r_i(q_a,\bm{\theta})+{\const}_{ia} \right\rbrace. 
	\end{align}
	\par Given current parameter estimate $\bm{\theta}^{(t)},\text{ for all }a\in \{1,2,\dots,k\}$, let 
	\begin{equation*}
		(\bm{W}_a)_{n\times n}^{(t)} = 
		\begin{pmatrix}
			\frac{1}{\epsilon+\big|r_1\big(q_a,\bm{\theta}^{(t)}\big)\big|}& & & \\
			& \frac{1}{\epsilon+\big|r_2\big(q_a,\bm{\theta}^{(t)}\big)\big|} &  & \\
			& & \ddots & \\
			& & & \frac{1}{\epsilon+\big|r_n\big(q_a,\bm{\theta}^{(t)}\big)\big|}
		\end{pmatrix},\;
		(\bm{c}_a)_{n\times 1}=
		\begin{pmatrix}
			4q_a-2\\
			4q_a-2\\
			\vdots \\
			4q_a-2
		\end{pmatrix}.
	\end{equation*}
	Then \Cref{Qepsilonk} can be written as
	\begin{align}\label{matrix form res k}
		Q_{\epsilon}\big(\bm{\theta}|\bm{\theta}^{(t)}\big)=&\frac{1}{4}\sum_{a=1}^{k}\left\lbrace  \bm{r}(q_a,\bm{\theta})^T\bm{W}_a^{(t)}\bm{r}(q_a,\bm{\theta})+\bm{c}_a^T\bm{r}(q_a,\bm{\theta})+{\const}_a\right\rbrace \nonumber \\
		= & \frac{1}{4}\sum_{a=1}^{k}\left\lbrace  (\bm{y}-\bm{D}(q_a)\bm{\theta})^T\bm{W}_a^{(t)}(\bm{y}-\bm{D}(q_a)\bm{\theta})+\bm{c}_a^T(\bm{y}-\bm{D}(q_a)\bm{\theta})+{\const}_a\right\rbrace,
	\end{align}
	which is again quadratic in $\bm{\theta}$ and thus can be solved directly for $\bm{\theta}^{(t+1)}$. Let 
	\begin{equation}\label{g k}
		g\big(\bm{\theta}|\bm{\theta}^{(t)}\big)=\sum_{a=1}^{k}\left\lbrace  \bm{(y-D}(q_a)\bm{\theta)}^T\bm{W}_a^{(t)}\bm{(y-D}(q_a)\bm{\theta)}+\bm{c}_a^T\bm{(y-D}(q_a)\bm{\theta)}\right\rbrace .
	\end{equation}
	Then minimizing $g\big(\bm{\theta}|\bm{\theta}^{(t)}\big)$ is equivalent to minimizing $ Q_{\epsilon}\big(\bm{\theta}|\bm{\theta}^{(t)}\big)$.
	\begin{align}\label{g' k}
		&\frac{\partial}{\partial\bm{\theta}}g\big(\bm{\theta}|\bm{\theta}^{(t)}\big)\bigg|_{\bm{\theta}=\bm{\theta}^{(t+1)}}=\left(\sum_{a=1}^k \left\lbrace 2\bm{D}^T(q_a)\bm{W}_a^{(t)}\bm{D}(q_a)\bm{\theta}-2\bm{D}^T(q_a)\bm{W}_a^{(t)}\bm{y}-\bm{D}^T(q_a)\bm{c}_a \right\rbrace\right)_{\bm{\theta}=\bm{\theta}^{(t+1)}} =\bm{0}\nonumber\\
		&\qquad\Rightarrow \bm{\theta}^{(t+1)} = \left[  \sum_{a=1}^k \bm{D}^T(q_a)\bm{W}_a^{(t)}\bm{D}(q_a)\right]^{-1} \left[ \sum_{a=1}^k \left(\bm{D}^T(q_a)\bm{W}_a^{(t)}\bm{y}+\frac{1}{2}\bm{D}^T(q_a)\bm{c}_a \right) \right], 
	\end{align}
	which is well-defined for a full rank $\bm{X}_{n\times p}$ ($n>p$), a small $h\in\mathbb{N}$, and a big $k\in\mathbb{N}$. To see this, we note that $\sum_{a=1}^k \bm{D}^T(q_a)_{hp\times n}(\bm{W}_a)^{(t)}_{n\times n}\bm{D}(q_a)_{n\times hp} = \bm{D}^T_{hp\times nk}\bm{W}_{nk\times nk}^{(t)}\bm{D}_{nk\times hp}$, where
	\begin{align}\label{positive definite k}
		& \bm{D}_{nk\times hp} = 
		\begin{pmatrix}
			\bm{D}(q_1)_{n\times hp} \\
			\bm{D}(q_2)_{n\times hp} \\
			\vdots \\
			\bm{D}(q_k)_{n\times hp}
		\end{pmatrix} = 
		\begin{pmatrix}
			\bm{b}(q_1)^T_{1\times h}\otimes \bm{X}_{n\times p} \\
			\bm{b}(q_2)^T_{1\times h}\otimes \bm{X}_{n\times p} \\
			\vdots \\
			\bm{b}(q_k)^T_{1\times h}\otimes \bm{X}_{n\times p}
		\end{pmatrix} = 
		\begin{pmatrix}
			\bm{b}(q_1)^T \\
			\bm{b}(q_2)^T \\
			\vdots \\
			\bm{b}(q_k)^T
		\end{pmatrix}_{k\times h}\otimes \bm{X}_{n\times p} 
	\end{align}
	    and 
	\begin{align}
         \bm{W}_{nk\times nk}^{(t)}=
	    \begin{pmatrix}
	     (\bm{W}_1)^{(t)}_{n\times n} & & &\\
	     &(\bm{W}_2)^{(t)}_{n\times n} & &\\
	     & & \ddots &\\
	     & & & (\bm{W}_k)^{(t)}_{n\times n}
         \end{pmatrix}.
    \end{align}
    \begin{align*}
    	\rank(\bm{D}) = \rank\left(\begin{pmatrix}
    		\bm{b}(q_1)^T \\
    		\bm{b}(q_2)^T \\
    		\vdots \\
    		\bm{b}(q_k)^T
    	\end{pmatrix}\right) \times \rank(\bm{X}) = h\times p = hp\text{ implies that } \bm{D}_{nk\times hp}\text{ is full-rank}.    
    \end{align*}
	Since $\bm{W}_{nk\times nk}^{(t)}$ is a diagonal matrix whose diagonal entries are all strictly positive and $\bm{D}_{nk\times hp}$ ($nk>hp$) is of full column rank, $\bm{D}^T\bm{W}^{(t)}\bm{D}$ is positive definite and thus invertible. 
	$\bm{\theta}^{(t+1)}$ given in \Cref{g' k} indeed minimizes \Cref{g k,Qepsilonk}, as $\frac{\partial^2}{\partial\bm{\theta}\partial\bm{\theta}^T}g\big(\bm{\theta}|\bm{\theta}^{(t)}\big)=2\sum_{a=1}^k \bm{D}^T(q_a)\bm{W}_a^{(t)}\bm{D}(q_a) = 2\bm{D}^T\bm{W}^{(t)}\bm{D}$ is positive definite for all $\bm{\theta}$ including $\bm{\theta}^{(t+1)}$.
	\par With no additional assumptions, the parametric logistic basis 
	\begin{equation}\label{logistic basis}
		\bm{b}(q)=(1,\ln(q), \ln(1-q))^T
	\end{equation}
	has been suggested to perform well generally by \textcite{Frumento2016} and was thus adopted by us. Since there are situations, though not frequent, when linear combinations of \Cref{logistic basis} drastically deviate from the actual coefficient functions (see \Cref{sect3} when the errors follow a shifted rescaled $\Beta(0.5,0.5)$ distribution, for instance), we also resorted to flexible nonparametric bases. Natural splines (cubic splines that restrict both tails to be linear, abbreviated as ``ns"), which encompass both flexibility retained from ordinary cubic splines and stability achieved by imposing the tail linearity restriction \parencite{Perperoglou2019} and have thus been extensively utilized in regression analysis, were considered. We followed the truncated power basis form presented by \textcite{Hess1994} and defined
	\begin{align}\label{natural spline basis}
		& \bm{b}(q)_{h\times 1}=\big(1,q,S_1(q),S_2(q),\dots,S_{h-2}(q)\big)^T,\text{ where for all }l\in \{1,2,\dots,h-2\},\nonumber \\
		& S_l(q)=(q-q_l)_+^3 - \frac{(q-q_{h-1})_+^3(q_h-q_l)}{q_h-q_{h-1}} + \frac{(q-q_h)_+^3(q_{h-1}-q_l)}{q_h-q_{h-1}}
	\end{align}
	for ns bases with $h>2\:(h\in \mathbb{N})$ knots $q_1,q_2,\dots,q_h\in (0,1)$. 
	\par We can conveniently extend the linear quantile regression MM algorithm constructed in this subsection to parametrically model quantile regression problems with manifest complex nonlinear patterns by first flexibly transforming the original explanatory variables via for instance natural spline transformations \parencite{Perperoglou2019} again. More details are in \Cref{appenC}.

	\section{Some Numerical Experiments and Simulation Studies}\label{sect3}
	\par We implemented the MM algorithm in \Cref{sect2b} on the ``pollution" data set \parencite{McDonald1973} in R's \texttt{SMPracticals} package and compared our results with the ones produced by the \texttt{rq} function in the \texttt{quantreg} package \parencite{Koenker2010}. The data set consists of 60 observations and conveys information about a mortality measure ($y$) with 15 explanatory variables, out of which we considered $x_1$ (precipitation, or ``prec"), $x_9$ (percentage of non-white, or ``nonw"), $x_{10}$ (percentage of white collar, or ``wwdrk"), and $x_{14}$ ($SO_2$ potential, or ``so"), as these four variables have been demonstrated by \textcite{Peng2014}  to possess either fairly constant significant ($x_1, x_9$) or manifest varying ($x_{10},x_{14}$) effects across different quantile orders. Reassuringly, we arrived at extremely close regression coefficient estimates for various quantile order choices. A comparison of the two methods' estimated coefficients for quantile orders $q=0.1,\,0.3,\,0.5,\,0.7,\,0.9\,$ is presented in \Cref{qmm rq coef}. This lends us considerable confidence in the effectiveness of this aforementioned MM algorithm (which also turned out to be fast).
	\begin{table}[ph!]
		\centering
		\scalebox{0.75}{
			\begin{tabular}{|c|c| c c c c c |}
				\hline
				\multicolumn{2}{|c|}{\textbf{Quantile Level}} & $\bm{0.1}$ & $\bm{0.3}$ & $\bm{0.5}$ & $\bm{0.7}$ & $\bm{0.9}$ \\
				\hline
				\multirow{2}{*}{\textbf{Intercept}} & {MM} & 6.93236001846033 & 6.81129448032327 & 6.82937502691576 & 6.78261292514527 & 6.75103326513239 \\
				& \texttt{rq} & 6.93236000893714 & 6.81129447876662 & 6.82937503030063 & 6.78261292566128 & 6.75103326659995 \\
				\hline
				\multirow{2}{*}{\textbf{Prec}} & {MM} & 0.00176228378387 & 0.00176159829763 & 0.00219734102618 & 0.00256007248951 & 0.00379190890437 \\
				& \texttt{rq} & 0.00176228385391 & 0.00176159830142 & 0.00219734120251 & 0.00256007251365 & 0.00379190894470 \\
				\hline
				\multirow{2}{*}{\textbf{Nonw}} & {MM} & 0.00307198793255 & 0.00378855193952 & 0.00338337914962 & 0.00317285210519 & 0.00310426173527 \\
				& \texttt{rq} & 0.00307198786522 & 0.00378855195678 & 0.00338337911208 & 0.00317285194220 & 0.00310426167488 \\
				\hline
				\multirow{2}{*}{\textbf{Wwdrk}} & {MM} & -0.00564678398232 & -0.00247626438379 & -0.00285944783005 & -0.00170900792443 & -0.00108160523893 \\
				& \texttt{rq} & -0.00564678381819 & -0.00247626435265 & -0.00285944806563 & -0.00170900789901 & -0.00108160528760 \\
				\hline
				\multirow{2}{*}{\textbf{So}} & {MM} & 0.00051426030202 & 0.00040409834617 & 0.00037884373159 & 0.00035640682915 & 0.00019472676333 \\
				& \texttt{rq} & 0.00051426030253 & 0.00040409834526 & 0.00037884377530 & 0.00035640682973 & 0.00019472676437 \\
				\hline
		\end{tabular} }
		\caption{Comparison of selected quantile orders' coefficient estimates from \texttt{rq} and MM (with number of iterations $N=100000$ and perturbation $\epsilon= 10^{-10}$) for linear quantile regression on $y,x_1,x_9,x_{10},x_{14}$ of the ``pollution" data set.}
		\label{qmm rq coef}
	\end{table}
	
	\begin{figure}[ph!]
		\centering
		\includegraphics[width=0.95\textwidth]{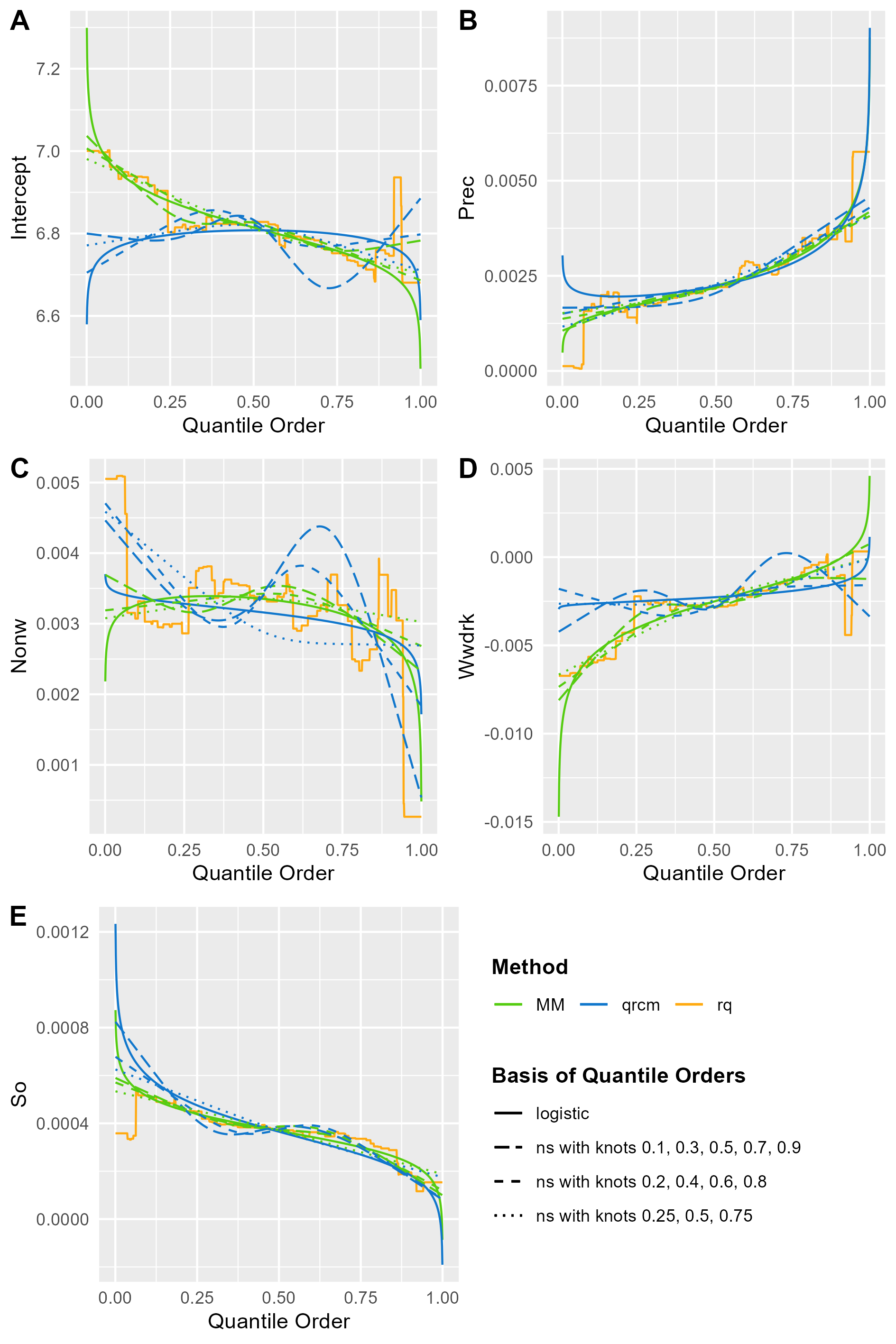}  		
		\caption{Estimated regression coefficients $\hat{\bm{\beta}}(q)$ versus quantile levels $q$ plot for our models with the logistic basis and 3 natural spline bases. The orange lines are \texttt{rq}'s empirical regression coefficient estimates at 1000 quantile levels. The blue and green lines denote estimated $\hat{\bm{\beta}}(q)$ produced by \texttt{qrcm} and our MM algorithm described in \Cref{sect2c}, respectively.}
		\label{beta plots pollution}
	\end{figure}
    \par
\begin{figure}[h!]
		\centering
		\includegraphics[width=0.95\textwidth]{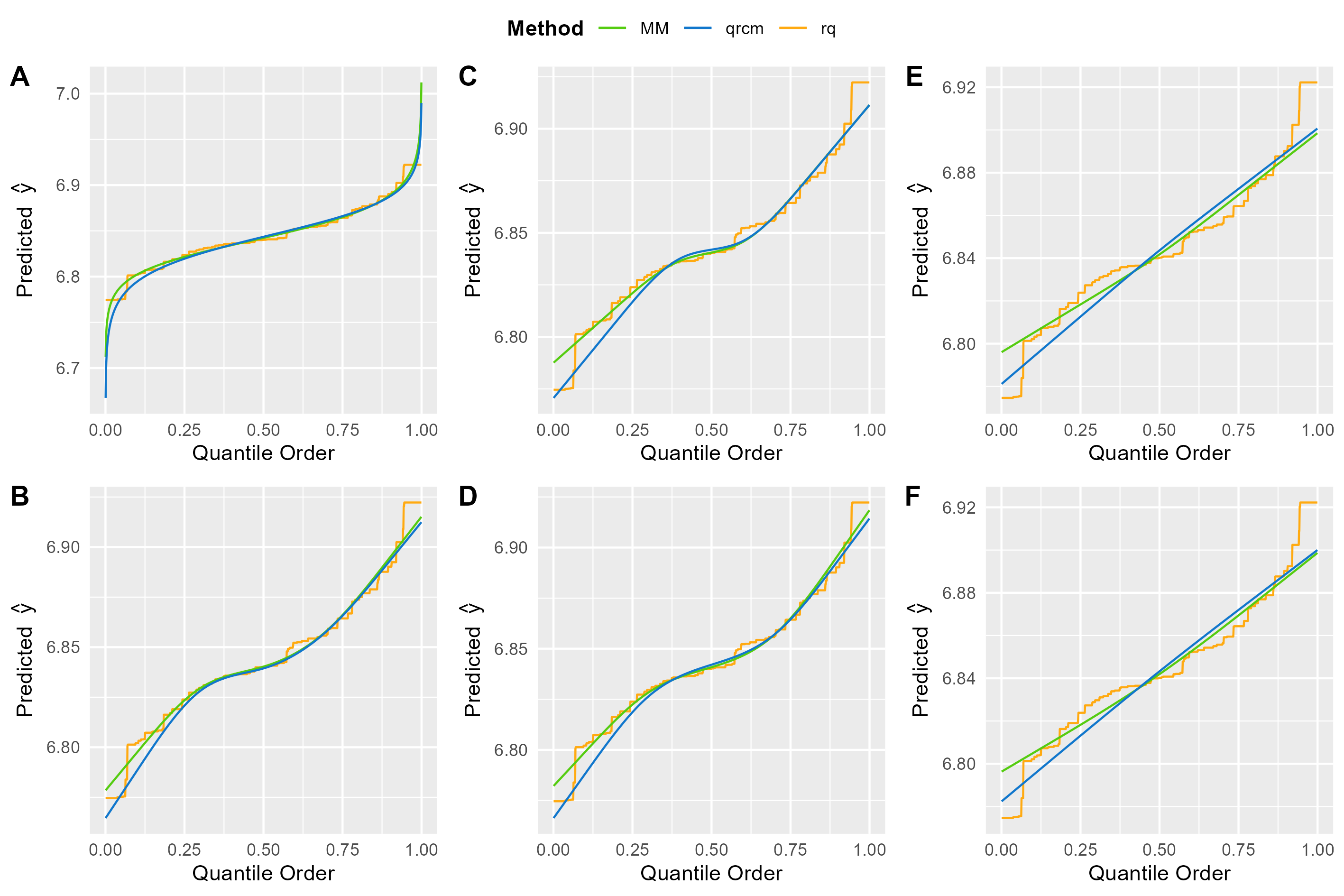}  		
		\caption{Quantile plots ($\hat Q_q(Y|\bm{x})=\hat{y}(q|\bm{x})$ versus $q$, $q\in (0,1)$) for six regression models' predicted $\hat{y}$, given the independent variables $\bm{x}_{p\times 1}=\sum_{i=1}^n \bm{x}_i/n\:(p=4+1=5,\:n=60)$. We have adopted the logistic basis (\textbf{A}) and natural spline bases with knots 0.1, 0.3, 0.5, 0.7, 0.9 (\textbf{B}), 0.2, 0.4, 0.6, 0.8 (\textbf{C}), 0.1, 0.3, 0.7, 0.9 (\textbf{D}), 0.25, 0.5, 0.75 (\textbf{E}), and 0.1, 0.5, 0.9 (\textbf{F}).}
		\label{median quantile plots linQuantRegSimulQs}
	\end{figure}

 
    We are equipped to extend our MM algorithm to model linear regression coefficients for different quantile orders concurrently rather than separately. We implemented the MM algorithm in \Cref{sect2c} on the ``pollution" data set with variables $y,x_1,x_9,x_{10},x_{14}$ again and carried out the corresponding \texttt{quantreg}, \texttt{qrcm} \parencite{Package2020} routines. The results validated our MM algorithm's appropriateness once more. 
    \Cref{beta plots pollution} depicts the estimated linear regression coefficients $\hat{\bm{\beta}}(q)_{p\times 1}$ as functions of the quantile order $q\in (0,1)$, from which we can see that MM's estimates capture efficiently \texttt{rq} empirical estimates' trends and might be slightly better than their \texttt{qrcm} counterparts for all models' five coefficients.  
    \Cref{median quantile plots linQuantRegSimulQs} presents estimated response values $\hat{y}(q|\bm{x})=\hat Q_q(Y|\bm{x})$ at $1000$ quantile levels $q\in(0,1)$, taking $\bm{x}=(x_1,x_9,x_{10},x_{14})$ as the observed sample mean vector. For all models here, MM effectively smooths the empirical functions produced by \texttt{rq} and agrees well to \texttt{qrcm}. 

	\par
\begin{figure}[ph!]
		\centering
		\includegraphics[width=0.85\textwidth]{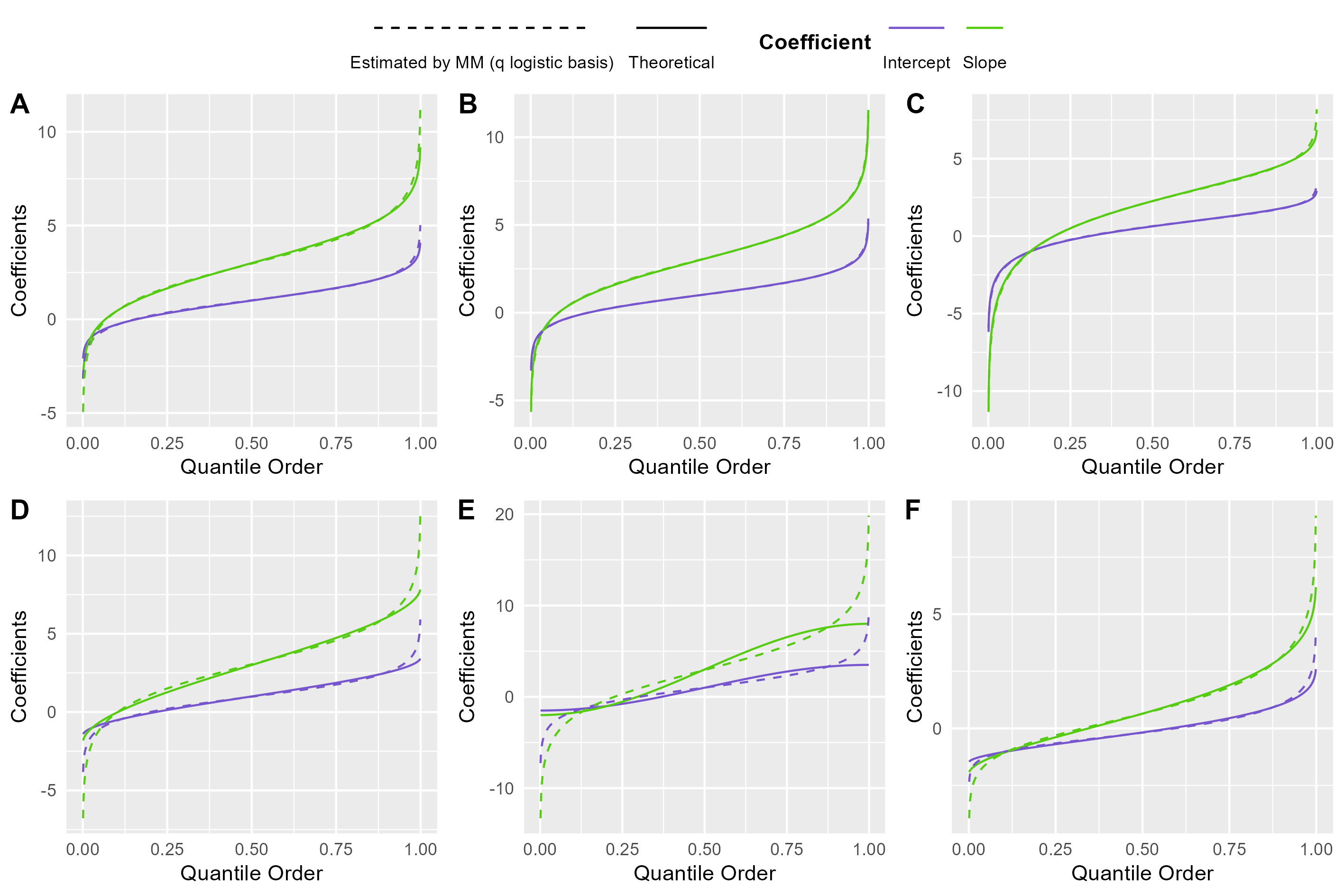}
		\caption{Mean estimated quantile coefficient functions (dashed lines) using the logistic basis for the six error distributions -- standard normal (\textbf{A}), $t(10)$ (\textbf{B}), standard log exponential (\textbf{C}), and short-tailed beta distributions $\Beta(2,2)$ (\textbf{D}), $\Beta(0.5,0.5)$ (\textbf{E}), $\Beta(2,5)$ (\textbf{F}). The solid lines are the theoretical quantile coefficient functions.}
		\label{quantile est coef function plots}
	\end{figure}

 We now carry out simulation studies using the heterogeneous linear regression model
	\begin{equation}\label{hetero linear reg model}
		Y_i = 1+3X_i+(1+2X_i)e_i,\; i=1,2,\dots,n,
	\end{equation}
	where the regressors $X_i\overset{\iid}{\sim}\unif(0,1)$ are one-dimensional uniform random variables independent of the errors $e_{1:n}$, which are IID following six distributions:
	\begin{enumerate}
		\item $e_i \sim N(0,1)$;\label{1}
		\item $e_i \sim t(10)$;\label{2}
		\item $e_i = 5(B_i-\frac{1}{2})$, where $B_i\sim \Beta(\alpha,\beta)$ with $(\alpha,\beta)\in\{(2,2), \left(\frac{1}{2},\frac{1}{2}\right), (2,5)\}$;\label{3}
		\item $e_i=\ln(\epsilon_i),\text{ where }\epsilon_i \sim \exp(1)$.\label{4}
	\end{enumerate}
	With the exact relationships between $Y_i$ and $X_i$, $e_i$ (\Cref{hetero linear reg model}) and theoretical distributions of $X_i$ and $e_i$ for all $i\in\{1,\ldots,n\}$, we can derive the theoretical conditional quantile function 
	\begin{align}\label{theoretical quantile function}
		Q_q(Y|x)=F_y^{-1}(q|x)&=1+3x+(1+2x)Q_q(e)=[1+Q_q(e)]+[3+2Q_q(e)]x,\:q\in(0,1).
	\end{align}
	In the special case where $e_i\overset{\iid}{\sim}\text{standard logistic}, Q_q(e)=\ln(q)-\ln(1-q) \text{  and hence  } Q_q(Y|x)=[1+\ln(q)-\ln(1-q)]+[3+2(\ln(q)-\ln(1-q))]x$, suggesting that the theoretical coefficients are exact linear combinations of logistic basis functions. Since the normal distribution, largely similar to t distributions, is very close to the logistic distribution, we expect the logistic basis to work well for normal and $t(10)$ errors. This indeed turned out to be the case (\Cref{quantile est coef function plots}).
	
	\par
\begin{figure}[ph!]
    	\centering
    	\includegraphics[width=0.866\textwidth]{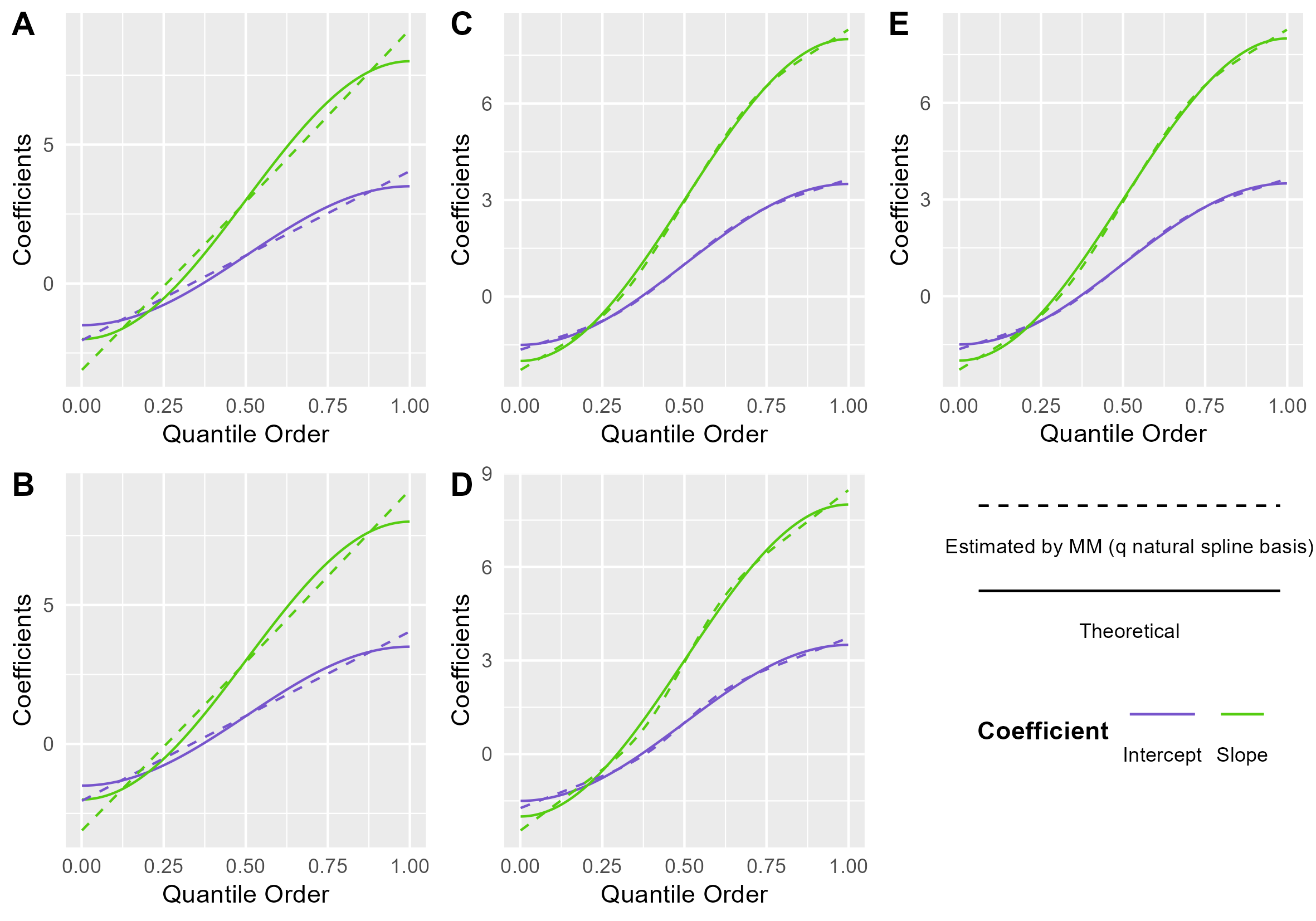}
    	\caption{Mean estimated quantile coefficient functions (dashed lines) for the short-tailed $\Beta(0.5,0.5)$ error distribution using ns basis with knots $0.1,0.5,0.9$ (\textbf{A}), $0.2,0.5,0.8$ (\textbf{B}), $0.1,0.3,0.7,0.9$ (\textbf{C}), $0.2,0.4,0.6,0.8$ (\textbf{D}), and $0.1,0.3,0.5,0.7,0.9$ (\textbf{E}). The solid lines are the theoretical quantile coefficient functions.}
    	\label{beta ns quantile est coef function plots}
    \end{figure}

 We calculated our deviation statistic -- the Integrated Mean Squared Error (IMSE)
	\begin{equation}\label{imse}
		\IMSE = \frac{1}{N}\sum_{l=1}^{N}\sum_{i=1}^{n}\left( Q_q(Y|x_i)-\hat{Q}_q(Y|x_i)\right)^2
	\end{equation}
	using $N$ random samples, each of size $n$, at quantile levels $q=0.1,\:0.2,\:\dots,\:0.9$ for a variety of linear quantile regression approaches. These methods include \texttt{rq} from \texttt{quantreg}, \texttt{qMM} (the MM algorithm in \Cref{sect2b}), \texttt{iqr} from \texttt{qrcm} and \texttt{qrcmMM} (the MM algorithm in \Cref{sect2c}) with the logistic basis and natural spline bases. 
	As shown in \Cref{normal t10 logexp imse,Beta imse}, \texttt{qMM} and \texttt{qrcmMM} IMSEs correspond closely to their \texttt{rq} and \texttt{qrcm} counterparts at all tested quantile levels for all six aforementioned $e_i$ distributions. Hence, we now focus on analyzing MM's results. 
    As expected, \texttt{qrcm} with the logistic basis significantly outperforms \texttt{qMM} as well as \texttt{qrcm} with nonparametric bases for normal, $t(10)$, and log exponential errors (\Cref{normal t10 logexp imse}). For the short-tailed $\Beta(0.5,0.5)$ error distribution, however, \texttt{qrcm} with the logistic basis loses out (\Cref{Beta imse}).
	As \Cref{quantile est coef function plots}\textbf{E} depicts, \texttt{qrcm} with the logistic basis is indeed inadequate for short-tailed $\Beta(0.5,0.5)$ errors. This happens likely because linear combinations of functions in the logistic basis concave up whereas the theoretical quantile coefficient functions for short-tailed $\Beta(0.5,0.5)$ errors concave down. Under such scenarios when the parametric logistic basis severely misspecifies the actual correlation patterns, flexible nonparametric ns bases with more than three knots closely estimate theoretical quantile coefficient functions (\Cref{beta ns quantile est coef function plots}) and improve upon separate quantile estimates (\Cref{Beta imse}), gaining us assurance in the effectiveness of modeling quantile coefficients as linear combinations of basis functions of $q$.
	\begin{table}[ph!]
		\centering
		\scalebox{0.7}{
			\begin{tabular}{|r|ccccccccc|}
				\hline
				\multicolumn{10}{|c|}{\bm{$e\sim N(0,1)$}}\\
				\hline
				\textbf{Quantile Level}& $\bm{0.1}$ & $\bm{0.2}$ & $\bm{0.3}$ & $\bm{0.4}$ & $\bm{0.5}$ & $\bm{0.6}$ & $\bm{0.7}$ & $\bm{0.8}$ & $\bm{0.9}$ \\ 
				\hline
				rq & 22.24821 & 16.10671 & 13.86712 & 13.04180 & 12.11576 & 13.24554 & 14.69919 & 16.47695 & 25.13818 \\ 
                    qMM & 22.19156 & 16.12816 & 13.86136 & 13.02504 & 12.11490 & 13.23653 & 14.69057 & 16.50089 & 25.13012 \\ 
				\hline
				qrcmlogistic & 17.16718 & 12.58532 & 11.40789 & 10.25537 & 9.81351 & 10.37832 & 11.74901 & 13.51759 & 20.52232 \\ 
                    MMlogistic & 17.16379 & 12.57550 & 11.40380 & 10.25596 & 9.81587 & 10.37933 & 11.74600 & 13.51015 & 20.52400 \\ 
				\hline
				qrcmspline 0.1\_0.5\_0.9 & 22.75182 & 27.87691 & 29.07420 & 16.82461 & 10.27289 & 16.55472 & 29.19284 & 29.24372 & 25.52883 \\ 
                    MMspline 0.1\_0.5\_0.9 & 22.89600 & 27.69450 & 28.93591 & 16.78205 & 10.27383 & 16.51647 & 29.05745 & 29.05795 & 25.67515 \\ 
				\hline
				qrcmspline 0.2\_0.5\_0.8 & 22.68471 & 27.93398 & 29.00131 & 16.87486 & 10.47318 & 16.60154 & 29.16428 & 29.36575 & 25.46089 \\ 
                    MMspline 0.2\_0.5\_0.8 & 22.83329 & 27.76421 & 28.87588 & 16.84127 & 10.47816 & 16.56359 & 29.02730 & 29.17648 & 25.59391 \\ 
				\hline
				qrcmspline 0.1\_0.3\_0.7\_0.9 & 26.69058 & 17.55810 & 12.59560 & 13.63645 & 10.67844 & 13.57804 & 12.85956 & 19.19861 & 30.41405 \\ 
                    MMspline 0.1\_0.3\_0.7\_0.9 & 26.47565 & 17.52878 & 12.58046 & 13.60387 & 10.67924 & 13.54767 & 12.84432 & 19.15434 & 30.17102 \\ 
				\hline
				qrcmspline 0.2\_0.4\_0.6\_0.8 & 22.13834 & 24.47002 & 11.10685 & 16.67830 & 10.92009 & 16.66907 & 11.49093 & 26.26063 & 25.48560 \\ 
                    MMspline 0.2\_0.4\_0.6\_0.8 & 21.98425 & 24.36851 & 11.11428 & 16.61919 & 10.92130 & 16.61226 & 11.49685 & 26.15133 & 25.32456 \\ 
				\hline
				qrcmspline 0.1\_0.3\_0.5\_0.7\_0.9 & 26.89851 & 18.15940 & 13.33193 & 13.67687 & 11.71228 & 13.57708 & 13.93839 & 19.83295 & 30.60826 \\ 
                    MMspline 0.1\_0.3\_0.5\_0.7\_0.9 & 26.67302 & 18.11682 & 13.31426 & 13.64351 & 11.71051 & 13.54258 & 13.92634 & 19.80081 & 30.35751 \\ 
				\hline
				\hline
				\multicolumn{10}{|c|}{\bm{$e\sim t(10)$}}\\
				\hline
				\textbf{Quantile Level}& $\bm{0.1}$ & $\bm{0.2}$ & $\bm{0.3}$ & $\bm{0.4}$ & $\bm{0.5}$ & $\bm{0.6}$ & $\bm{0.7}$ & $\bm{0.8}$ & $\bm{0.9}$ \\ 
				\hline
                    rq & 27.41375 & 17.30877 & 14.11940 & 13.45854 & 13.99761 & 13.78847 & 14.47089 & 17.90523 & 29.84599 \\ 
                    qMM & 27.35780 & 17.27099 & 14.11300 & 13.45640 & 13.98506 & 13.79151 & 14.44906 & 17.86305 & 29.86262 \\ 
                    \hline
                    qrcmlogistic & 21.69898 & 12.85838 & 10.89730 & 10.32029 & 10.22706 & 10.48377 & 11.33215 & 13.90138 & 24.58947 \\ 
                    MMlogistic & 21.70182 & 12.85770 & 10.89739 & 10.32139 & 10.22891 & 10.48606 & 11.33441 & 13.90230 & 24.58262 \\ 
                    \hline
                    qrcmspline 0.1\_0.5\_0.9 & 32.02033 & 35.37554 & 37.02666 & 19.49030 & 10.78879 & 20.87955 & 39.60333 & 38.42644 & 33.12124 \\ 
                    MMspline 0.1\_0.5\_0.9 & 32.25187 & 35.11533 & 36.83492 & 19.43591 & 10.79346 & 20.82594 & 39.40870 & 38.15584 & 33.34444 \\ 
                    \hline
                    qrcmspline 0.2\_0.5\_0.8 & 31.88831 & 35.47912 & 36.98001 & 19.54250 & 11.00530 & 20.94230 & 39.57463 & 38.53937 & 32.99737 \\ 
                    MMspline 0.2\_0.5\_0.8 & 32.10719 & 35.23188 & 36.80104 & 19.49557 & 11.01313 & 20.88979 & 39.38065 & 38.26954 & 33.21691 \\ 
                    \hline
                    qrcmspline 0.1\_0.3\_0.7\_0.9 & 35.92861 & 22.44336 & 13.94594 & 15.83932 & 11.41705 & 15.29400 & 13.96810 & 25.28172 & 40.99669 \\ 
                    MMspline 0.1\_0.3\_0.7\_0.9 & 35.58717 & 22.38098 & 13.91573 & 15.78334 & 11.41364 & 15.24511 & 13.94468 & 25.19740 & 40.58436 \\ 
                    \hline
                    qrcmspline 0.2\_0.4\_0.6\_0.8 & 28.06846 & 32.60728 & 11.86716 & 20.25510 & 11.79325 & 19.35562 & 12.44285 & 36.19840 & 32.12780 \\ 
                    MMspline 0.2\_0.4\_0.6\_0.8 & 27.85511 & 32.44387 & 11.87290 & 20.17121 & 11.79619 & 19.28283 & 12.45080 & 36.00611 & 31.87039 \\ 
                    \hline
                    qrcmspline 0.1\_0.3\_0.5\_0.7\_0.9 & 36.66713 & 22.56951 & 13.80907 & 16.17954 & 13.57117 & 15.47282 & 14.27477 & 25.05770 & 42.33289 \\ 
                    MMspline 0.1\_0.3\_0.5\_0.7\_0.9 & 36.30695 & 22.50168 & 13.78319 & 16.12620 & 13.56453 & 15.42476 & 14.25684 & 24.98746 & 41.93550 \\
                    \hline
				\hline
				\multicolumn{10}{|c|}{\bm{$e=\ln(\epsilon),\textbf{ where }\epsilon \sim \exp(1)$}}\\
				\hline
				\textbf{Quantile Level}& $\bm{0.1}$ & $\bm{0.2}$ & $\bm{0.3}$ & $\bm{0.4}$ & $\bm{0.5}$ & $\bm{0.6}$ & $\bm{0.7}$ & $\bm{0.8}$ & $\bm{0.9}$ \\ 
				\hline
                    rq & 84.77815 & 40.80326 & 27.10702 & 23.22761 & 19.01945 & 15.81423 & 14.60091 & 11.19701 & 13.37049 \\ 
                    qMM & 84.74011 & 40.79091 & 27.10065 & 23.22461 & 19.09301 & 15.79974 & 14.60748 & 11.19764 & 13.36569 \\ 
                    \hline
                    qrcmlogistic & 59.54423 & 30.07673 & 21.03025 & 17.01122 & 14.66868 & 12.99251 & 11.64239 & 10.45978 & 10.22578 \\ 
                    MMlogistic & 59.54887 & 30.06456 & 21.01875 & 17.00374 & 14.66513 & 12.99142 & 11.64145 & 10.45636 & 10.22149 \\
                    \hline
                    qrcmspline 0.1\_0.5\_0.9 & 81.12328 & 97.79237 & 82.03028 & 32.10191 & 15.35016 & 23.57249 & 28.63888 & 18.57495 & 15.27305 \\ 
                    MMspline 0.1\_0.5\_0.9 & 81.60985 & 97.10757 & 81.63525 & 32.02509 & 15.35337 & 23.51031 & 28.50896 & 18.46277 & 15.37037 \\ 
                    \hline
                    qrcmspline 0.2\_0.5\_0.8 & 82.69136 & 102.96539 & 86.03598 & 31.52170 & 15.94103 & 24.19164 & 26.33142 & 16.69839 & 14.57296 \\ 
                    MMspline 0.2\_0.5\_0.8 & 83.22019 & 102.19804 & 85.59968 & 31.44488 & 15.93875 & 24.11837 & 26.20658 & 16.60077 & 14.66135 \\ 
                    \hline
                    qrcmspline 0.1\_0.3\_0.7\_0.9 & 93.80122 & 78.26567 & 24.63783 & 26.60745 & 18.29061 & 15.05701 & 15.77836 & 10.00315 & 16.30409 \\ 
                    MMspline 0.1\_0.3\_0.7\_0.9 & 92.99538 & 77.91880 & 24.64041 & 26.50589 & 18.25890 & 15.03987 & 15.73332 & 10.00798 & 16.18954 \\ 
                    \hline
                    qrcmspline 0.2\_0.4\_0.6\_0.8 & 75.32165 & 107.35437 & 28.26206 & 32.33288 & 19.44216 & 18.35022 & 13.35708 & 12.08152 & 14.20136 \\ 
                    MMspline 0.2\_0.4\_0.6\_0.8 & 74.84933 & 106.73074 & 28.28204 & 32.19883 & 19.41683 & 18.30862 & 13.33282 & 12.05897 & 14.11047 \\ 
                    \hline
                    qrcmspline 0.1\_0.3\_0.5\_0.7\_0.9 & 113.13272 & 63.12735 & 28.48088 & 26.08760 & 18.65186 & 17.10657 & 12.48270 & 12.83113 & 13.74328 \\ 
                    MMspline 0.1\_0.3\_0.5\_0.7\_0.9 & 111.86556 & 62.95839 & 28.40124 & 25.99428 & 18.66299 & 17.06000 & 12.46006 & 12.80519 & 13.69187 \\
				\hline
			\end{tabular}
		}
		\caption{IMSE (as in \Cref{imse}) calculated with the number of samples $N=350$ and sample size $n=500$ when the errors follow a standard normal distribution (as in \Cref{1}), a heavy-tailed $t$ distribution with degrees of freedom 10 (as in \Cref{2}), or an asymmetric log exponential distribution (as in \Cref{4}). The methods \texttt{qrcmlogistic} and \texttt{MMlogistic} denote respectively \texttt{qrcm} and \texttt{qrcmMM} with the logistic basis \eqref{logistic basis}; the methods \texttt{qrcmspline} and \texttt{MMspline} denote \texttt{qrcm} and \texttt{qrcmMM} with natural spline bases \eqref{natural spline basis} containing the specified knots. We used test quantiles $q=0.001,0.002,\dots,0.999$ to fit all MM models concerned.}
		\label{normal t10 logexp imse}
	\end{table} 
	\begin{table}[ph!]
		\hspace{-1.05cm}
		\scalebox{0.7}{
			\begin{tabular}{|r|ccccccccc|}
				\hline 
				\multicolumn{10}{|c|}{\scalebox{1}{\bm{$e=5(B-0.5), \textbf{ where }B\sim \Beta(2,2)$}}}\\[2mm]
				\hline
				\textbf{Quantile Level}& $\bm{0.1}$ & $\bm{0.2}$ & $\bm{0.3}$ & $\bm{0.4}$ & $\bm{0.5}$ & $\bm{0.6}$ & $\bm{0.7}$ & $\bm{0.8}$ & $\bm{0.9}$ \\ 
				\hline
				rq & 21.46882 & 22.83887 & 22.14847 & 22.14426 & 21.71720 & 24.23045 & 22.78511 & 22.80701 & 21.72210 \\ 
                    qMM & 21.46618 & 22.82854 & 22.12376 & 22.13606 & 21.65883 & 24.23237 & 22.78041 & 22.81987 & 21.71124 \\ 
                    \hline
                    qrcmlogistic & 18.64965 & 34.49844 & 35.39456 & 24.56411 & 18.44312 & 22.48222 & 32.11153 & 31.99067 & 19.88999 \\ 
                    MMlogistic & 18.65464 & 34.46547 & 35.37453 & 24.55462 & 18.43648 & 22.47234 & 32.09613 & 31.97401 & 19.89186 \\ 
                    \hline
                    qrcmspline 0.1\_0.5\_0.9 & 19.95956 & 21.91054 & 23.53816 & 20.13629 & 18.91923 & 21.89850 & 26.13399 & 24.03974 & 20.83311 \\ 
                    MMspline 0.1\_0.5\_0.9 & 20.02664 & 21.83199 & 23.47603 & 20.11550 & 18.91446 & 21.87101 & 26.06103 & 23.95281 & 20.91032 \\ 
                    \hline
                    qrcmspline 0.2\_0.5\_0.8 & 20.00612 & 21.83760 & 23.33305 & 20.29628 & 19.43214 & 22.07042 & 25.90572 & 23.94302 & 20.88889 \\ 
                    MMspline 0.2\_0.5\_0.8 & 20.08070 & 21.76321 & 23.27201 & 20.27576 & 19.42954 & 22.04920 & 25.84415 & 23.87001 & 20.97013 \\ 
                    \hline
                    qrcmspline 0.1\_0.3\_0.7\_0.9 & 18.45833 & 16.74557 & 20.16599 & 21.48571 & 19.42050 & 20.97026 & 20.07219 & 17.96624 & 20.17941 \\ 
                    MMspline 0.1\_0.3\_0.7\_0.9 & 18.43188 & 16.75404 & 20.15165 & 21.46243 & 19.41759 & 20.96058 & 20.07270 & 17.97697 & 20.15668 \\ 
                    \hline
                    qrcmspline 0.2\_0.4\_0.6\_0.8 & 18.07743 & 18.54829 & 18.32186 & 24.28196 & 20.16116 & 23.50584 & 18.76463 & 20.26756 & 19.67425 \\ 
                    MMspline 0.2\_0.4\_0.6\_0.8 & 18.05857 & 18.53070 & 18.31701 & 24.23568 & 20.14488 & 23.46522 & 18.76060 & 20.24146 & 19.64336 \\ 
                    \hline
                    qrcmspline 0.1\_0.3\_0.5\_0.7\_0.9 & 18.02124 & 18.10853 & 21.09472 & 21.77022 & 21.85668 & 21.16040 & 21.50255 & 19.56462 & 19.57965 \\ 
                    MMspline 0.1\_0.3\_0.5\_0.7\_0.9 & 17.98476 & 18.09605 & 21.06962 & 21.74787 & 21.85921 & 21.15177 & 21.49179 & 19.56715 & 19.53355 \\   
				\hline
				\hline
				\multicolumn{10}{|c|}{\scalebox{1}{\bm{$e=5(B-0.5), \textbf{ where }B\sim \Beta(0.5,0.5)$}}}\\[2mm]
				\hline
				\textbf{Quantile Level}& $\bm{0.1}$ & $\bm{0.2}$ & $\bm{0.3}$ & $\bm{0.4}$ & $\bm{0.5}$ & $\bm{0.6}$ & $\bm{0.7}$ & $\bm{0.8}$ & $\bm{0.9}$ \\ 
				\hline
				rq & 4.42521 & 28.01759 & 73.96922 & 111.87383 & 131.72589 & 116.67583 & 69.64196 & 32.91205 & 5.38827 \\ 
                    qMM & 4.42837 & 28.01171 & 73.96328 & 111.87140 & 131.75659 & 116.53449 & 69.58970 & 32.90917 & 5.38638 \\ 
                    \hline
                    qrcmlogistic & 155.28794 & 312.24080 & 505.84052 & 268.19551 & 115.29391 & 302.30999 & 552.71613 & 338.06106 & 155.65009 \\ 
                    MMlogistic & 155.41237 & 312.13992 & 505.75909 & 268.14165 & 115.23746 & 302.23418 & 552.63602 & 338.01947 & 155.64352 \\ 
                    \hline
                    qrcmspline 0.1\_0.5\_0.9 & 21.09246 & 93.53950 & 180.12351 & 147.95729 & 120.02211 & 166.94315 & 203.29421 & 104.99504 & 20.22250 \\ 
                    MMspline 0.1\_0.5\_0.9 & 21.09814 & 93.49525 & 180.08445 & 147.95398 & 120.05107 & 166.97193 & 203.29181 & 104.96783 & 20.24161 \\ 
                    \hline
                    qrcmspline 0.2\_0.5\_0.8 & 21.28033 & 92.30125 & 178.07187 & 149.73724 & 125.40623 & 169.03744 & 200.98347 & 103.34772 & 20.42439 \\ 
                    MMspline 0.2\_0.5\_0.8 & 21.33202 & 92.23655 & 178.03494 & 149.74570 & 125.41385 & 169.00909 & 200.92911 & 103.30563 & 20.43941 \\ 
                    \hline
                    qrcmspline 0.1\_0.3\_0.7\_0.9 & 9.54570 & 32.62490 & 71.38348 & 89.49572 & 85.09412 & 86.33643 & 70.13905 & 36.30335 & 11.26571 \\ 
                    MMspline 0.1\_0.3\_0.7\_0.9 & 9.52102 & 32.59565 & 71.35463 & 89.47629 & 85.08785 & 86.32851 & 70.13407 & 36.31801 & 11.26977 \\ 
                    \hline
                    qrcmspline 0.2\_0.4\_0.6\_0.8 & 9.51700 & 43.04721 & 69.12940 & 127.14491 & 95.89035 & 117.20892 & 68.61221 & 49.21223 & 11.05473 \\ 
                    MMspline 0.2\_0.4\_0.6\_0.8 & 9.49450 & 43.05495 & 69.16367 & 127.14738 & 95.90020 & 117.18842 & 68.60905 & 49.21780 & 11.05151 \\ 
                    \hline
                    qrcmspline 0.1\_0.3\_0.5\_0.7\_0.9 & 9.67939 & 31.29348 & 65.70141 & 102.16686 & 124.76119 & 98.86606 & 64.62588 & 33.90553 & 11.48802 \\ 
                    MMspline 0.1\_0.3\_0.5\_0.7\_0.9 & 9.67236 & 31.27117 & 65.67984 & 102.16692 & 124.76465 & 98.87546 & 64.62972 & 33.90832 & 11.48771 \\  
				\hline
				\hline 
				\multicolumn{10}{|c|}{\scalebox{1}{\bm{$e=5(B-0.5), \textbf{ where }B\sim \Beta(2,5)$}}}\\[2mm]
				\hline
				\textbf{Quantile Level}& $\bm{0.1}$ & $\bm{0.2}$ & $\bm{0.3}$ & $\bm{0.4}$ & $\bm{0.5}$ & $\bm{0.6}$ & $\bm{0.7}$ & $\bm{0.8}$ & $\bm{0.9}$ \\ 
				\hline
				rq & 4.89481 & 6.08135 & 7.43628 & 7.73045 & 8.35502 & 9.98481 & 12.81717 & 15.32302 & 23.71795 \\ 
                    qMM & 4.88892 & 6.07928 & 7.43077 & 7.72922 & 8.35379 & 9.96288 & 12.81654 & 15.32294 & 23.71491 \\ 
                    \hline
                    qrcmlogistic & 3.91850 & 7.92200 & 8.30724 & 7.11437 & 7.15009 & 9.26427 & 12.52836 & 14.15726 & 18.65347 \\ 
                    MMlogistic & 3.92119 & 7.91587 & 8.30505 & 7.11371 & 7.14783 & 9.25865 & 12.52018 & 14.15267 & 18.67581 \\ 
                    \hline
                    qrcmspline 0.1\_0.5\_0.9 & 4.82828 & 6.26162 & 9.50787 & 9.25683 & 7.49254 & 10.96097 & 22.38225 & 26.32377 & 21.98446 \\ 
                    MMspline 0.1\_0.5\_0.9 & 4.85129 & 6.23945 & 9.47468 & 9.23587 & 7.49029 & 10.94572 & 22.29197 & 26.16270 & 22.10895 \\ 
                    \hline
                    qrcmspline 0.2\_0.5\_0.8 & 4.71063 & 5.68066 & 8.79059 & 9.45400 & 7.78548 & 10.82614 & 23.38320 & 27.67563 & 22.40786 \\ 
                    MMspline 0.2\_0.5\_0.8 & 4.72889 & 5.66403 & 8.76460 & 9.43690 & 7.78608 & 10.81187 & 23.28210 & 27.49450 & 22.53404 \\
                    \hline
                    qrcmspline 0.1\_0.3\_0.7\_0.9 & 4.41630 & 4.37253 & 7.39742 & 7.43065 & 8.10016 & 10.42985 & 9.97129 & 19.49951 & 24.01414 \\ 
                    MMspline 0.1\_0.3\_0.7\_0.9 & 4.40914 & 4.37456 & 7.38480 & 7.42593 & 8.09336 & 10.41179 & 9.97965 & 19.44002 & 23.88087 \\ 
                    \hline
                    qrcmspline 0.2\_0.4\_0.6\_0.8 & 4.35696 & 4.31579 & 6.58935 & 8.48875 & 8.59001 & 12.27445 & 10.28687 & 26.15131 & 21.26553 \\ 
                    MMspline 0.2\_0.4\_0.6\_0.8 & 4.34848 & 4.31617 & 6.57867 & 8.47396 & 8.57736 & 12.23050 & 10.28206 & 26.01887 & 21.18164 \\ 
                    \hline
                    qrcmspline 0.1\_0.3\_0.5\_0.7\_0.9 & 3.98354 & 5.43904 & 6.86927 & 7.90253 & 8.47386 & 10.17773 & 12.27454 & 16.68506 & 26.43939 \\ 
                    MMspline 0.1\_0.3\_0.5\_0.7\_0.9 & 3.97511 & 5.43345 & 6.86336 & 7.88910 & 8.47126 & 10.15742 & 12.24414 & 16.65205 & 26.25692 \\ 
				\hline
			\end{tabular}
		}
		\caption{IMSE (as in \Cref{imse}) calculated with the number of samples $N=350$ and sample size $n=500$ when the errors are distributed as in \Cref{3} (rescaled shifted Beta distributions with different representative parameters). The methods \texttt{qrcmlogistic} and \texttt{MMlogistic} denote respectively \texttt{qrcm} and \texttt{qrcmMM} with the logistic basis \eqref{logistic basis}; the methods \texttt{qrcmspline} and \texttt{MMspline} denote \texttt{qrcm} and \texttt{qrcmMM} with natural spline bases \eqref{natural spline basis} containing the specified knots. We used test quantiles $q=0.001,0.002,\dots,0.999$ to fit all MM models concerned.}
		\label{Beta imse}
	\end{table} 

\section{Discussion}\label{sect4}
This article explores MM algorithms in various linear quantile regression settings. Simulation studies comparing MM with existing tested methods and applications of MM to several real data sets showcase the effectiveness of our MM algorithms. MM's computation speed edge over some other prominent algorithms when estimating coefficients at different quantiles separately \parencite{Pietrosanu2017} has also been supported by our simulation study in \Cref{sect3}.
\par One major flaw of our MM is that it gets relatively slow compared to \texttt{qrcm} when estimating different quantile coefficients simultaneously, which should result from having to invert an $hp\times hp$ matrix $\sum_{a=1}^k \bm{D}^T(q_a)\bm{W}_a^{(t)}\bm{D}(q_a)$ (in \Cref{g' k}) at each iteration. It is, however, quite promising to address this current weakness, as some authors have successfully invented fast MM algorithms that bypass large matrix inversion problems (\cite{Langed2000}; \cite{Erdogan1999}). In situations where super fast algorithms are needed, a finite smoothing algorithm raised by \textcite{Chen2007}, which operates by directly minimizing a nice smooth approximate function of $L_q(\bm{\theta})=\sum_{i=1}^n\rho_q(r_i(\bm{\theta}))$, may also come to the rescue.


\appendix
\section{Additional Proofs and Figures}\label{appenA}
\textbf{Proposition A.1.}    
\begin{align*}
	&Q\left(\bm{\theta}|\bm{\theta}^{(t)}\right)=\sum_{i=1}^n \zeta_q\left(r_i(\bm{\theta})|r_i\big(\bm{\theta}^{(t)}\big)\right)=\sum_{i=1}^n\frac{1}{4}\left\lbrace \frac{r_i^2(\bm{\theta})}{\big|r_i\big(\bm{\theta}^{(t)}\big)\big|}+(4q-2)r_i(\bm{\theta})+\big|r_i\big(\bm{\theta}^{(t)}\big)\big|\right\rbrace \text{ majorizes }\\
	&L(\bm{\theta})=\sum_{i=1}^n \rho_q (r_i(\bm{\theta}))=\sum_{i=1}^n
	\left[qr_i(\bm{\theta})-r_i(\bm{\theta})\mathbbm{1}_{\{r_i(\bm{\theta})<0\}}\right] \text{ at } \bm{\theta}=\bm{\theta}^{(t)}.
\end{align*}
\begin{proof}
	\begin{align*}
		&Q\left( \bm{\theta}^{(t)}\big|\bm{\theta}^{(t)}\right)-L\big(\bm{\theta}^{(t)}\big)=\sum_{i=1}^n \left[\zeta_q\left(r_i\big(\bm{\theta}^{(t)}\big)\big|r_i\big(\bm{\theta}^{(t)}\big)\right)-\rho_q \left(r_i\big(\bm{\theta}^{(t)}\big)\right) \right]\\
		=&\sum_{i=1}^n\left( \frac{1}{4}\left[\big|r_i\big(\bm{\theta}^{(t)}\big)\big|-2r_i\big(\bm{\theta}^{(t)}\big)+\big|r_i\big(\bm{\theta}^{(t)}\big)\big|\right]+r_i\big(\bm{\theta}^{(t)}\big)\mathbbm{1}_{\left\lbrace r_i\big(\bm{\theta}^{(t)}\big)<0\right\rbrace } \right)\\
		=&\sum_{i=1}^n \left\lbrace r_i\big(\bm{\theta}^{(t)}\big)\left(- \frac{1}{2}+\mathbbm{1}_{\left\lbrace r_i\big(\bm{\theta}^{(t)}\big)<0\right\rbrace }\right)+\frac{1}{2}\big|r_i\big(\bm{\theta}^{(t)}\big)\big| \right\rbrace =0.
	\end{align*} 
	It remains to show $Q\big(\bm{\theta}|\bm{\theta}^{(t)}\big)$ satisfies \Cref{tangent ppty1}. $\forall \text{ residual } r,$ letting
	\begin{align}\label{A1h}
		h(r)=\,&\zeta_q\big(r|r^{(t)}\big)-\rho_q(r)=\frac{1}{4}\left( \frac{r^2}{\big|r^{(t)}\big|}-2r+\big|r^{(t)}\big|\right)+r\mathbbm{1}_{\{r<0\}}, \text{ we have }\nonumber \\
		h'(r)=\,&\frac{1}{2}\left(\frac{r}{\big|r^{(t)}\big|}-1\right)+\mathbbm{1}_{\{r<0\}}
		=\begin{cases}
			\frac{1}{2}\cdot \frac{r-\big|r^{(t)}\big|}{\big|r^{(t)}\big|}\;  \text{ if } r\geq 0 \\
			\frac{1}{2}\cdot \frac{r+\big|r^{(t)}\big|}{\big|r^{(t)}\big|}\;  \text{ if } r< 0 
		\end{cases}. 
	\end{align}
	Observe from \Cref{A1h} that $h\big(\pm r^{(t)}\big)=h'\big(\pm r^{(t)}\big)=0$ and 
	\begin{align*}
		& h'(r)\leq 0\: \text{  if } r \in \big(-\infty, -\big|r^{(t)}\big|\big]\cup \big[0,\big|r^{(t)}\big|\big] \text{ and }
		\geq 0\: \text{  if } r \in \big[-\big|r^{(t)}\big|,0\big]\cup \big[\big|r^{(t)}\big|,\infty\big) \\
		\iff & h(r)\: \text{ is non-increasing for } r \in \big(-\infty, -\big|r^{(t)}\big|\big]\cup \big[0,\big|r^{(t)}\big|\big] \\
		&\text{ and is non-decreasing for } r \in \big[-\big|r^{(t)}\big|,0\big]\cup \big[\big|r^{(t)}\big|,\infty\big)\\
		\Rightarrow\: & h(r)\geq \min\left\lbrace h\big(r^{(t)}\big),h\big(-r^{(t)}\big)\right\rbrace =0\; \forall \: r \quad
		\iff \quad \zeta_q\big(r|r^{(t)}\big)\geq \rho_q(r)\; \forall \: r. 
	\end{align*}
	Hence $Q\big(\bm{\theta | \theta}^{(t)}\big)\geq L(\bm{\theta}) \; \forall\; \bm{\theta}$ and \Cref{tangent ppty1} is satisfied. 
\end{proof}

\begin{figure}[ph!]
	\centering
	\includegraphics[width=0.9\textwidth]{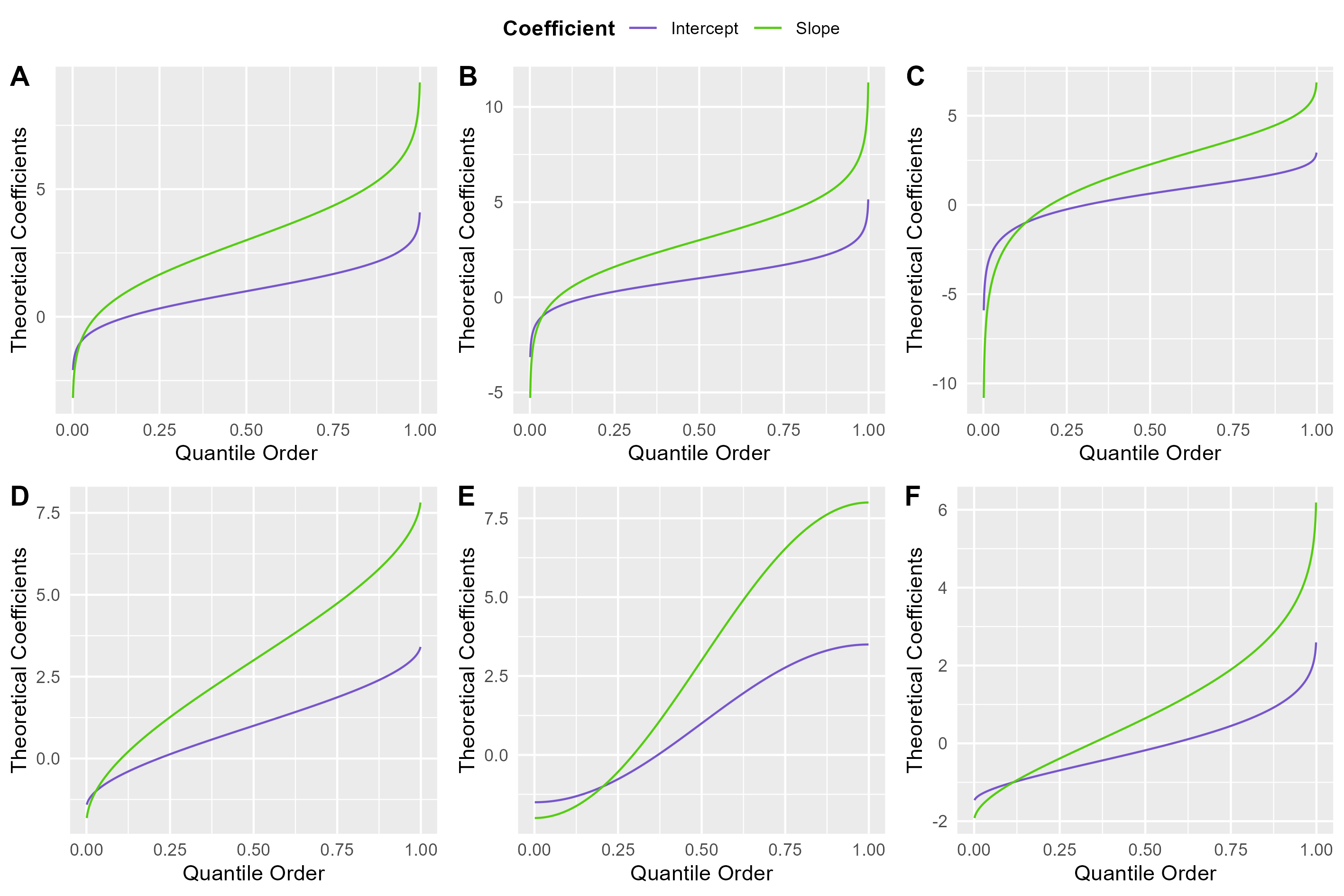}%
	\caption{Theoretical quantile coefficient functions for the six error distributions in \Cref{sect3} -- standard normal (\textbf{A}), $t(10)$ (\textbf{B}), standard log exponential (\textbf{C}), and short-tailed beta distributions $\Beta(2,2)$ (\textbf{D}), $\Beta(0.5,0.5)$ (\textbf{E}), $\Beta(2,5)$ (\textbf{F}).}
	\label{quantile theo coef function plots}
\end{figure}
\begin{figure}[ph!]
	\centering
	\includegraphics[width=1\textwidth]{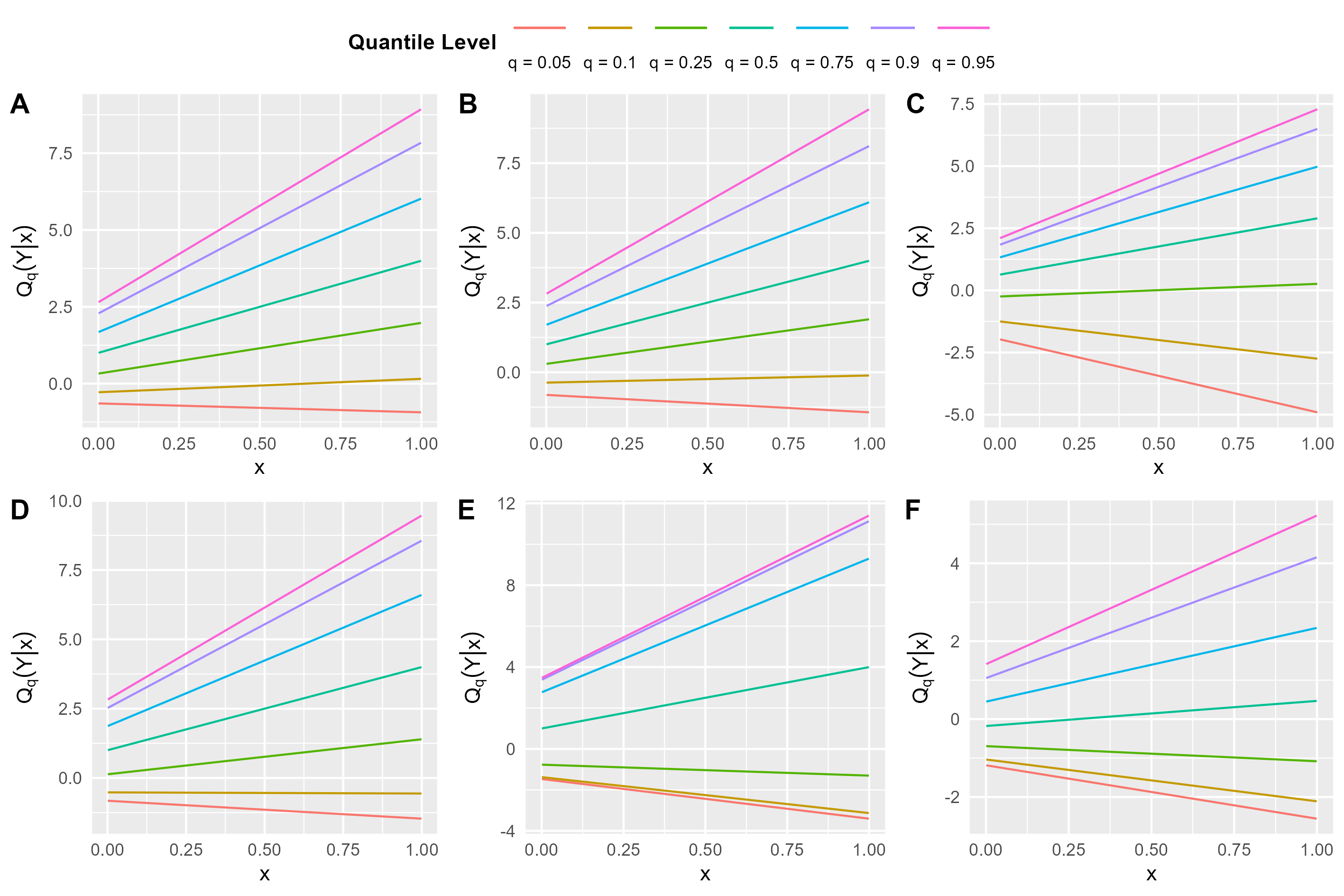}%
	\caption{Theoretical conditional quantile curves at $q=0.05,0.1,0.25,0.5,0.75,0.9,0.95$ for the six error distributions in \Cref{sect3} -- standard normal (\textbf{A}), $t(10)$ (\textbf{B}), standard log exponential (\textbf{C}), and short-tailed beta distributions $\Beta(2,2)$ (\textbf{D}), $\Beta(0.5,0.5)$ (\textbf{E}), $\Beta(2,5)$ (\textbf{F}).}
	\label{conditional quantile curves theo sect3c}
\end{figure}
\clearpage

\section{Adaptive Lasso for Heterogeneous Variable Selection in Regularized Quantile Regression}\label{appenB}
In diverse sparse modeling settings requiring efficient predictor selection, the adaptive lasso has been proven highly competent in consistent variable selection by imposing an adequately weighted $\mathit{l}_1$ penalty $\lambda\sum_{j=1}^{p} w_j|\beta_j|,\lambda\geq0$ \parencite{Zou2006}. Several authors (e.g., \cite{Wu2009}, \cite{Peng2014}) have extended this powerful technique to account for possible quantile-varying covariate effects by defining a penalized quantile regression problem at any quantile level $q\in(0,1)$ as follows. We write $\bm{\beta}=(\beta_1,\beta_2,\dots,\beta_p)^T$, where $\beta_1$ denotes the coefficient for the intercept term, and let the parameter estimate
\begin{align}\label{lasso L}
	\hat{\bm{\beta}}_{\lambda}(q)_{p\times 1}=&\underset{\bm{\beta}}{\text{ argmin}}\;L_q(\bm{\beta}),\text{  where}\nonumber\\
	L_q(\bm{\beta}) = & \sum_{i=1}^n \rho_q(r_i(\bm{\beta})) + \lambda \sum_{j=2}^p \frac{|\beta_j|}{\big|\tilde{\beta}_j(q)\big|}\text{  and}\\
	\tilde{\bm{\beta}}(q)_{p\times 1}=&\underset{\bm{\beta}}{\text{ argmin}}\;\sum_{i=1}^n \rho_q(r_i(\bm{\beta}))\text{ (the nonregularized quantile regression estimator).}\nonumber
\end{align}
\par Thanks to the AM-GM Inequality, 
\begin{equation}\label{lasso Q}	Q_{q,\lasso}\left(\bm{\beta}\bigg|\bm{\beta}^{(t)}\right)=\sum_{i=1}^n \zeta_q\left(r_i\bigg|r_i^{(t)}\right)+\lambda\sum_{j=2}^p \frac{1}{\left|\tilde{\beta}_j(q)\right|}\left[ \frac{1}{2}\left|\beta_j^{(t)}\right|+\frac{\beta_j^2}{2\left|\beta_j^{(t)}\right|}\right] 
\end{equation}
is a majorizer of \Cref{lasso L} at $\bm{\beta}^{(t)}$. To mend the flaw that \Cref{lasso Q} is undefined when $r_i^{(t)}=0$ for any $i\in\{1,2,\ldots,n\}$ or $\beta_j^{(t)}=0$ for any $j\in\{2,3,\dots,p\}$, a realistic concern as $r_i\big(\bm{\beta}^{(t)}\big)$ for some $f_i\big(\bm{\beta}^{(t)}\big)$, $i\in\{1,2,\ldots,n\}$ and $\beta_j^{(t)}$, $j\in\{2,3,\dots,p\}$ for penalized quantile regression may well converge to 0 as $t\rightarrow\infty$, we can adopt
\begin{align}\label{lasso Qepsilon}
	&Q^{\epsilon,\epsilon_l}_{q,\lasso}\left(\bm{\beta}\bigg|\bm{\beta}^{(t)}\right)=\sum_{i=1}^n \zeta_q^{\epsilon}\left(r_i\bigg|r_i^{(t)}\right)+\lambda\sum_{j=2}^p \frac{1}{\big|\tilde{\beta}_j(q)\big|}\left[ \frac{1}{2}\left(\big|\beta_j^{(t)}\big|+\epsilon_l\right)+\frac{\beta_j^2}{2\left(\big|\beta_j^{(t)}\big|+\epsilon_l\right)} \right],
\end{align}
an approximate majorizer at $\bm{\beta}^{(t)}$ of 
\begin{equation}\label{Lqepsilon}
	L_q^{\epsilon}(\bm{\beta})=\sum_{i=1}^n \rho_q^{\epsilon}(r_i(\bm{\beta})) + \lambda \sum_{j=2}^p \frac{|\beta_j|}{\big|\tilde{\beta}_j(q)\big|}=
	\sum_{i=1}^n \left\lbrace \rho_q(r_i(\bm{\beta}))-\frac{\epsilon}{2}\ln(\epsilon+|r_i(\bm{\beta})|) \right\rbrace + \lambda \sum_{j=2}^p \frac{|\beta_j|}{\big|\tilde{\beta}_j(q)\big|},
\end{equation}
which closely approximates $L_q(\bm{\beta})$ in \Cref{lasso L}. While it is still ensured that $L_q^{\epsilon}(\bm{\beta})\leq Q^{\epsilon,\epsilon_l}_{q,\lasso}\big(\bm{\beta}|\bm{\beta}^{(t)}\big)\;\forall\; \bm{\beta}$, introducing a perturbation $\epsilon_l>0$ makes $L_q^{\epsilon}\big(\bm{\beta}^{(t)}\big)\neq Q^{\epsilon,\epsilon_l}_{q,\lasso}\big(\bm{\beta}^{(t)}|\bm{\beta}^{(t)}\big)$. 
Let
\begin{align}\label{lassoMat}
	\bm{V}^{(t)}_{p\times p}&=
	\begin{pmatrix}
		0 & & & \\
		& \frac{1}{\big|\tilde{\beta}_2(q)\big|\cdot\big(\big|\beta_2^{(t)}\big|+\epsilon_l\big)} & & \\
		& & \ddots & \\
		& & &  \frac{1}{\big|\tilde{\beta}_p(q)\big|\cdot\big(\big|\beta_p^{(t)}\big|+\epsilon_l\big)}
	\end{pmatrix},\\
	\bm{W}^{(t)}_{n\times n}&=
	\begin{pmatrix}
		\frac{1}{\epsilon+\big|r_1\big(\bm{\beta}^{(t)}\big)\big|}& & & \\
		& \frac{1}{\epsilon+\big|r_2\big(\bm{\beta}^{(t)}\big)\big|} &  & \\
		& & \ddots & \\
		& & & \frac{1}{\epsilon+\big|r_n\big(\bm{\beta}^{(t)}\big)\big|}
	\end{pmatrix},\text{ and }\nonumber\\
	\bm{c}_{n\times 1}&=(4q-2,4q-2,\dots,4q-2)^T.\nonumber
\end{align} 
Then
\begin{equation}\label{simplifyLasso}
	Q^{\epsilon,\epsilon_l}_{q,\lasso}\big(\bm{\beta}|\bm{\beta}^{(t)}\big)=\frac{1}{4}\left( (\bm{y}-\bm{X\beta})^{T}\bm{W}^{(t)}(\bm{y}-\bm{X\beta})+\bm{c}^T(\bm{y}-\bm{X\beta})\right)+\frac{\lambda}{2}\bm{\beta}^{T}\bm{V}^{(t)}\bm{\beta}+\const.
\end{equation} 
Hence, minimizing $Q^{\epsilon,\epsilon_l}_{q,\lasso}\big(\bm{\beta}|\bm{\beta}^{(t)}\big)$ is equivalent to minimizing
\begin{align}\label{simplifyLassoh} 	 
	h\big(\bm{\beta}|\bm{\beta}^{(t)}\big)&=g\big(\bm{\beta}|\bm{\beta}^{(t)}\big)+2\lambda \bm{\beta}^{T}\bm{V}^{(t)}\bm{\beta},\text{ where  }\\
	g\big(\bm{\beta}|\bm{\beta}^{(t)}\big)&=(\bm{y}-\bm{X\beta})^{T}\bm{W}^{(t)}(\bm{y}-\bm{X\beta})+\bm{c}^T(\bm{y}-\bm{X\beta}).\nonumber
\end{align}
We have shown in \Cref{sect2b} that $\bm{X}^T\bm{W}^{(t)}\bm{X}$ is positive definite and thus invertible for a full rank $\bm{X}_{n\times p}$. It is also clear that $\bm{V}^{(t)}$ is positive semi-definite, as it is a diagonal matrix whose entries are all non-negative. Hence, for a full rank $\bm{X}_{n\times p}$, the Hessian
\begin{align}\label{hessianLasso}
	\frac{\partial^2}{\partial\bm{\beta}\partial\bm{\beta}^T}h\big(\bm{\beta}|\bm{\beta}^{(t)}\big)=2\bm{X}^T\bm{W}^{(t)}\bm{X}+4\lambda \bm{V}^{(t)}\;\text{ is positive definite since }\lambda\geq 0.
\end{align}
We can thus proceed to solve for $\bm{\beta}^{(t+1)}$ by directly minimizing \Cref{simplifyLassoh}:
\begin{align}
	\frac{\partial}{\partial\bm{\beta}}h\big(\bm{\beta}|\bm{\beta}^{(t)}\big)\bigg\rvert_{\bm{\beta}=\bm{\beta}^{(t+1)}} &=\left( 2\bm{X}^{T}\bm{W}^{(t)}\bm{X\beta}-2\bm{X}^T\bm{W}^{(t)}\bm{y}-\bm{X}^T\bm{c}+4\lambda \bm{V}^{(t)}\bm{\beta}\right)\bigg\rvert_{\bm{\beta}=\bm{\beta}^{(t+1)}}\nonumber\\
	&=2\left( \bm{X}^{T}\bm{W}^{(t)}\bm{X}+2\lambda\bm{V}^{(t)} \right) \bm{\beta}^{(t+1)}-2\bm{X}^T\bm{W}^{(t)}\bm{y}-\bm{X}^T\bm{c}=\bm{0}\nonumber\\
	\Rightarrow \bm{\beta}^{(t+1)}=\frac{1}{2}&\left( \bm{X}^{T}\bm{W}^{(t)}\bm{X}+2\lambda\bm{V}^{(t)} \right)^{-1}\bm{X}^T\left[2\bm{W}^{(t)}\bm{y}+\bm{c}\right].
\end{align}
$\left( \bm{X}^{T}\bm{W}^{(t)}\bm{X}+2\lambda\bm{V}^{(t)} \right)^{-1}$ indeed exists for a full rank $\bm{X}$ as \Cref{hessianLasso} $\Rightarrow 2\bm{X}^T\bm{W}^{(t)}\bm{X}+4\lambda \bm{V}^{(t)}=2\big(\bm{X}^T\bm{W}^{(t)}\bm{X}+2\lambda \bm{V}^{(t)}\big) \text{  is invertible  } \iff \bm{X}^T\bm{W}^{(t)}\bm{X}+2\lambda \bm{V}^{(t)}$ is invertible. 
\par It remains to find suitable choices of $\lambda>0$ for each $q\in(0,1)$ of interest. We followed \cite{Peng2014}'s idea and used a quantile regression BIC metric put forward by \textcite{Lee2014}:
\begin{equation}\label{BIC}
	{\BIC}_q(\lambda)={\BIC}_q\left(\hat{\bm{\beta}}_{\lambda}(q)\right)=\ln\left( \sum_{i=1}^n\rho_q\left(r_i\left(\hat{\bm{\beta}}_{\lambda}(q)\right)\right)\right) + \big|\mathit{S}_{q}^{\lambda}\big|\frac{\ln(n)}{2n},
\end{equation}
where $\big|\mathit{S}_q^{\lambda}\big|$ represents the size of $\mathit{S}_q^{\lambda}=\left\lbrace 2\leq j\leq p:\big|(\hat{\beta}_{\lambda})_j\big|>0\right\rbrace $. \\
\textbf{Proposition B.1.} 
Consider the linear quantile regression model
\begin{equation}\label{qmodel}
	Y_i=\bm{X}_i^T\bm{\beta}_q+U_i,\:i=1,2,\dots,n,\:q\in(0,1)
\end{equation}
with loss $\rho_q(u)=|u|\big[q\mathbbm{1}_{\{u\geq0\}}+(1-q)\mathbbm{1}_{\{u<0\}}\big]=qu-u\mathbbm{1}_{\{u<0\}}$ and error distribution 
\begin{equation}\label{errorlaplace}
	U_1,\dots,U_n\overset{\iid}{\sim} \text{ an asymmetric Laplace distribution with PDF }f_U(u)=\frac{q(1-q)}{\sigma}e^{-\frac{\rho_q(u)}{\sigma}}.
\end{equation}
Given a fixed $\lambda\geq 0$ and an obtained (penalized) coefficient estimator $\hat{\bm{\beta}}_q=\hat{\bm{\beta}}_{\lambda}(q)$ that attempts to minimize $L_q^{\epsilon}(\bm{\beta})$ in \Cref{Lqepsilon}, a BIC is given by 
\begin{equation}\label{BICunpenalized}
	{\BIC}_q\left(\hat{\bm{\beta}}_q\right)=\ln\left( \sum_{i=1}^n\rho_q\left(r_i\left(\hat{\bm{\beta}}_q\right)\right)\right) + |\mathit{S}_{q}|\frac{\ln(n)}{2n}
\end{equation}
if we approximate $\sigma$ by $\hat{\sigma}_{\mle}$.
\begin{proof}
	Given sample observations $\{(\bm{x}_i,y_i):1\leq i\leq n\}$, the log-likelihood is 
	\begin{equation}\label{loglikelihood}
		l(\bm{\beta}_q,\sigma)=\ln\left[\prod_{i=1}^n f_U(y_i-\bm{x}_i^T\bm{\beta}_q)\right]=-n\ln(\sigma)-\sum_{i=1}^n\frac{\rho_q(y_i-\bm{x}_i^T\bm{\beta}_q)}{\sigma}+\const.
	\end{equation}
	For any given $\hat{\bm{\beta}}_q$, we let $g(\sigma)=l\big(\hat{\bm{\beta}}_q,\sigma\big),\:\sigma\in(0,\infty)$. Since $g$ is differentiable everywhere on $(0,\infty)$ and 
	\begin{align}\label{sigma mle}
		g'(\sigma) &= \frac{\partial l\left(\bm{\beta}_q,\sigma\right)}{\partial \sigma}\bigg|_{\bm{\beta}_q=\hat{\bm{\beta}}_q}=-\frac{n}{\sigma}+\sum_{i=1}^n\frac{\rho_q\left(y_i-\bm{x}_i^T\hat{\bm{\beta}}_q\right)}{\sigma^2}=\frac{-n\sigma+\sum_{i=1}^n\rho_q\left(y_i-\bm{x}_i^T\hat{\bm{\beta}}_q\right)}{\sigma^2}\nonumber\\ 
		&\text{has a unique zero (and hence the unique critical point by Fermat's Theorem) at }\nonumber\\
		\sigma&=\hat{\sigma}\big(\hat{\bm{\beta}_q}\big)=\frac{\sum_{i=1}^n\rho_q\left(y_i-\bm{x}_i^T\hat{\bm{\beta}}_q\right)}{n} 
	\end{align}
	with $g'(\sigma)>0$ for $\sigma<\hat{\sigma}\big(\hat{\bm{\beta}}_q\big)$ and $g'(\sigma)<0$ for $\sigma>\hat{\sigma}\big(\hat{\bm{\beta}_q}\big)$, $\sigma=\hat{\sigma}\big(\hat{\bm{\beta}}_q\big)$ given in \Cref{sigma mle} is the unique local maxima. Since $g(\sigma)\rightarrow -\infty$ as $\sigma\rightarrow\infty$ or as $\sigma\rightarrow 0_+$, 
	\begin{equation*}
		\hat{\sigma}_{\mle}\left(\hat{\bm{\beta}}_q\right)=\frac{\sum_{i=1}^n\rho_q\left(y_i-\bm{x}_i^T\hat{\bm{\beta}}_q\right)}{n} \in (0,\infty)\text{ satisfies } g\left(\hat{\sigma}_{\mle}\left(\hat{\bm{\beta}}_q\right)\right) = \underset{\sigma\in(0,\infty)}{\max}\,g(\sigma) = \underset{\sigma\in(0,\infty)}{\max}\,l\left(\hat{\bm{\beta}}_q,\sigma\right)
	\end{equation*}
	and is the unique MLE of $\sigma$ given $\hat{\bm{\beta}}_q$. Setting $\sigma=\hat{\sigma}_{\mle}\left(\hat{\bm{\beta}}_q\right)=\frac{\sum_{i=1}^n\rho_q\left(y_i-\bm{x}_i^T\hat{\bm{\beta}}_q\right)}{n}$, we get
	\begin{align}\label{deriveBIC}
		{\BIC}_q\big(\hat{\bm{\beta}}_q\big)&=-2l\left(\hat{\bm{\beta}}_q,\hat{\sigma}_{\mle}\big(\hat{\bm{\beta}}_q\big)\right)+|\mathit{S}_{q}|\ln(n)=2n\ln\left[\hat{\sigma}_{\mle}\big(\hat{\bm{\beta}}_q\big)\right]+2n-2\const+|\mathit{S}_{q}|\ln(n)\nonumber\\
		&=2n\left(\ln\left[\frac{\sum_{i=1}^n\rho_q\big(y_i-\bm{x}_i^T\hat{\bm{\beta}}_q\big)}{n}\right]+|\mathit{S}_{q}|\frac{\ln(n)}{2n} \right)+2n-2\const\nonumber\\
		&=2n\left(\ln\left[\sum_{i=1}^n\rho_q\big(y_i-\bm{x}_i^T\hat{\bm{\beta}}_q\big)\right]+|\mathit{S}_{q}|\frac{\ln(n)}{2n} \right)+2n-2\const-2n\ln(n)\nonumber\\
		&={\const}_1\left(\ln\left[\sum_{i=1}^n\rho_q\big(y_i-\bm{x}_i^T\hat{\bm{\beta}}_q\big)\right]+|\mathit{S}_{q}|\frac{\ln(n)}{2n} \right)+{\const}_2,\text{ where }{\const}_1=2n>0.
	\end{align}
	Hence, \Cref{deriveBIC} is equivalent to \Cref{BICunpenalized} as a BIC metric. 
\end{proof}
Our algorithm thus proceeds according to the following steps:
\begin{enumerate}
	\item Choose a quantile level $q\in (0,1)$ of interest;
	\item Calculate a nonregularized estimator $\tilde{\bm{\beta}}(q)$ (using \texttt{rq} or MM);
	\item Propose a sequence of candidate $\lambda\in (0,1)$ choices;
	\item For each candidate $\lambda$, obtain a $\hat{\bm{\beta}}_{\lambda}(q)$ that attempts to minimize $L_q^{\epsilon}(\bm{\beta})$ in \Cref{Lqepsilon} via our MM algorithm detailed above;
	\item Evaluate ${\BIC}_q(\lambda)$ in \Cref{BIC} for this coefficient estimate $\hat{\bm{\beta}}_{\lambda}(q)$;
	\item Find $\lambda_{\opt}(q)$ among our picked sequence of candidate $\lambda$'s that yields the smallest ${\BIC}_q(\lambda)$; 
	\item Calculate our regularized estimator $\hat{\bm{\beta}}_{\lambda_{\opt}}(q)$ with this $\lambda_{\opt}$ at the desired quantile level $q$.
\end{enumerate}
\begin{table}[h]
	\centering
	\begin{tabular}{|cc|cc|cc|}
		\hline 
		\multicolumn{2}{|c|}{$\bm{q = 0.25}$} &\multicolumn{2}{|c|}{$\bm{q = 0.5}$} &\multicolumn{2}{|c|}{$\bm{q = 0.75}$}\\
		\hline
		$x_j$ & $\big(\hat{\beta}_{\lambda_{\opt}}\big)_j$ & $x_j$ & $\big(\hat{\beta}_{\lambda_{\opt}}\big)_j$ & $x_j$ & $\big(\hat{\beta}_{\lambda_{\opt}}\big)_j$ \\
		\hline
		$x_9$ & 0.05258 & $x_{12}$ & -0.03797 & $x_{12}$ &  -0.04276\\ 
		$x_{14}$ &  0.02368 & $x_9$ & 0.03291 &  $x_{13}$ & 0.03740 \\ 
		$x_3$ &  -0.02063 &  $x_{13}$ & 0.03278 &  $x_9$ & 0.03614 \\ 
		$x_{10}$ & -0.01457 &  $x_1$ & 0.01673 &   $x_6$ & -0.02074\\ 
		$x_7$ & -0.01440 & $x_{14}$ & 0.01394 & $x_2$ &  -0.01655 \\ 
		$x_5$ & -0.01329 & $x_8$ & 0.01008 &  $x_8$ & 0.00873  \\ 
		$x_1$ & 0.01167 &  $x_{10}$ & -0.00716 &  $x_4$ & 5.76206E-10 \\ 
		$x_{11}$  & -0.01087 & $x_2$ & -0.00308 &  $x_{14}$ &  4.89131E-10\\ 
		$x_2$ & -0.00938 & $x_3$ & -0.00300 &  $x_7$ &  -4.87400E-10  \\ 
		$x_{13}$ & -0.00933 & $x_5$ & -1.07803E-09  &  $x_1$ & 4.21575E-10 \\ 
		$x_{15}$ & -0.00780 & $x_4$ & 3.95779E-10 & $x_5$ & -4.98012E-11 \\ 
		$x_{12}$ & -1.09786E-10 &  $x_{11}$ & -3.90428E-11 &  $x_{15}$ & 4.23412E-11\\ 
		$x_8$ & -1.58700E-12 &  $x_{15}$ & 1.70300E-11 &  $x_{10}$ & 3.32322E-11 \\ 
		$x_4$ & 1.53641E-12 &  $x_6$ & -1.29810E-11 &  $x_3$ & -2.90341E-12 \\ 
		$x_6$ & 1.36344E-13 &  $x_7$ & -9.11257E-12 &  $x_{11}$ & -1.24519E-12 \\ 
		\hline
	\end{tabular}
	\caption{Coefficient estimates obtained via MM in \Cref{appenB} for regularized quantile regression with adaptive lasso penalty at $q=0.25,0.5,0.75$ on the ``pollution" data set. For each $q$, the covariates and their respective penalized coefficient estimates are presented in decreasing order of the magnitude of the absolute values of the estimators.}
	\label{lasso pollution}
\end{table}
\par We started off with the ``pollution" data set \parencite{McDonald1973} first raised in \Cref{sect2b}, as \textcite{Peng2014} have produced some ``guideline" results obtained by their adaptive lasso ``local" quantile regression methods $SS(q)$ at $q=0.25,0.5,0.75$. As presented in \Cref{lasso pollution}, our results obtained from MM do align quite closely with \cite{Peng2014}'s counterparts (\Cref{lasso PengXu}). We gained more confidence in our MM algorithm by further comparing our MM penalized estimates with those from \texttt{quantreg}'s \texttt{rq.fit.lasso} (\Cref{rqfitlassoVSmm}) with the penalty vector specified as $\left(0,\frac{\hat{\lambda}_{\opt}(q)}{\big|\tilde{\beta}_2(q)\big|},\dots,\frac{\hat{\lambda}_{\opt}(q)}{\big|\tilde{\beta}_p(q)\big|}\right)$, where $\hat{\lambda}_{\opt}(q)$ is the $\lambda\in(0,1)$ we found that yields the smallest $\BIC_q$.
\begin{table}[h]
	\centering
	\begin{tabular}{|*{2}{c|}}
		\hline
		\bf{Method} & \bf{Selected Variables}\\
		\hline
		$SS(0.25)$ & $x_1,x_3,x_9,x_{10},x_{13},x_{14}$ \\
		\hline
		$SS(0.5)$ & $x_1,x_8,x_9,x_{10},x_{12},x_{14}$ \\
		\hline
		$SS(0.75)$ & $x_6,x_8,x_9,x_{12},x_{14}$ \\
		\hline
	\end{tabular}
	\caption{Variables (of the ``pollution" data set) selected by penalized quantile regression methods with adaptive lasso penalty \parencite{Peng2014}.}
	\label{lasso PengXu}
\end{table}
\begin{table}[h]
	\centering
	\scalebox{0.9}{
		\begin{tabular}{|c|cc|cc|cc|}
			\hline
			\textbf{Effect} & \textbf{MM} & \textbf{\texttt{rq.fit.lasso}} & \textbf{MM} & \textbf{\texttt{rq.fit.lasso}} & \textbf{MM} & \textbf{\texttt{rq.fit.lasso}} \\ 
			\hline
			(intercept) & 6.82237 & 6.82237 & 6.84400 & 6.84499 & 6.87363 & 6.86302 \\ 
			prec & 0.01167 & 0.01167 & 0.01673 & 0.01735 & 4.22E-10 & 0.01056 \\ 
			jant & -0.00938 & -0.00938 & -0.00308 & -0.00557 & -0.01655 & -0.01672 \\ 
			jult & -0.02063 & -0.02063 & -0.00300 & -0.00237 & -2.90E-12 & -8.74E-11 \\ 
			ovr95 & 1.54E-12 & 7.20E-15 & 3.96E-10 & 1.23E-12 & 5.76E-10 & 3.32E-09 \\ 
			popn & -0.01329 & -0.01329 & -1.08E-09 & -0.00153 & -4.98E-11 & -0.00389 \\ 
			educ & 1.36E-13 & 1.34E-15 & -1.30E-11 & -3.51E-14 & -0.02074 & -0.01715 \\ 
			hous & -0.01440 & -0.01440 & -9.11E-12 & -2.50E-14 & -4.87E-10 & -0.00362 \\ 
			dens & -1.59E-12 & -1.03E-14 & 0.01008 & 0.00948 & 0.00873 & 0.00883 \\ 
			nonw & 0.05258 & 0.05258 & 0.03291 & 0.03455 & 0.03614 & 0.04168 \\ 
			wwdrk & -0.01457 & -0.01457 & -0.00716 & -0.00814 & 3.32E-11 & 4.11E-10 \\ 
			poor & -0.01087 & -0.01087 & -3.90E-11 & 7.51E-13 & -1.25E-12 & -0.00177 \\ 
			hc & -1.10E-10 & -1.63E-12 & -0.03797 & -0.03740 & -0.04276 & -0.05761 \\ 
			nox & -0.00933 & -0.00933 & 0.03278 & 0.03298 & 0.03740 & 0.05906 \\ 
			so & 0.02368 & 0.02368 & 0.01394 & 0.01366 & 4.89E-10 & 7.30E-10 \\ 
			humid & -0.00780 & -0.00780 & 1.70E-11 & -1.09E-14 & 4.23E-11 & 0.00272 \\ 
			\hline
		\end{tabular}
	}
	\caption{Lasso penalized quantile regression coefficient estimates obtained from MM and \texttt{rq.fit.lasso} for the ``pollution" data set at $q=0.25,0.5,0.75$.}
	\label{rqfitlassoVSmm}
\end{table}
\par We then implemented our MM algorithm on three other high-dimensional data sets at $q=0.1,0.25,0.5,0.75,0.9$ to uncover potential non-constant predictor effects in diverse fields.
{\par\centering\textbf{Economics Data}\par}
We constructed our first data set by compiling US monthly time series data for 9 economic variables (unemployment rate $y$, personal consumption expenditures (PCE) $x_1$, personal savings rate $x_2$, population $x_3$, Industrial Production Total Index $x_4$, Consumer Price Index (CPI) $x_5$, Policy Uncertainty Index based on US News $x_6$, Producer Price Index by Commodity $x_9$, and KC Fed Labor Market Conditions Index $x_{10}$) downloaded from \textcite{econdata} and US monthly Trade Policy Uncertainty (TPU), Geopolitical Risk (GPR) Indices $x_7,x_8$ constructed by \textcite{Caldara2020} and \textcite{Caldara2018}. We considered $n=347$ continuous months from January 1992 to November 2020 and all $p+1=11$ variables depicted above. We have ignored temporal dependence.
\begin{table}[h]
	\centering
	\begin{tabular}{|cc|cc|cc|cc|cc|}
		\hline 
		\multicolumn{2}{|c|}{$\bm{q = 0.1}$} & \multicolumn{2}{|c|}{$\bm{q = 0.25}$} &\multicolumn{2}{|c|}{$\bm{q = 0.5}$} &\multicolumn{2}{|c|}{$\bm{q = 0.75}$} & \multicolumn{2}{|c|}{$\bm{q = 0.9}$}\\
		\hline
		$x_j$ & $\big(\hat{\beta}_{\lambda_{\opt}}\big)_j$ & $x_j$ & $\big(\hat{\beta}_{\lambda_{\opt}}\big)_j$ & $x_j$ & $\big(\hat{\beta}_{\lambda_{\opt}}\big)_j$ & $x_j$ & $\big(\hat{\beta}_{\lambda_{\opt}}\big)_j$ & $x_j$ & $\big(\hat{\beta}_{\lambda_{\opt}}\big)_j$\\
		\hline
		$x_{1}$ & -6.55495 & $x_{3}$ & 9.06328 & $x_{3}$ & 11.07810 & $x_{3}$ & 10.27440 & $x_{3}$ & 8.19060 \\ 
		$x_{3}$ & 6.47031 & $x_{1}$ & -5.26884 & $x_{5}$ & -10.24142 & $x_{5}$ & -6.53847 & $x_{5}$ & -3.20889 \\ 
		$x_{4}$ & -1.64676 & $x_{5}$ & -4.93119 & $x_{9}$ & 3.14862 & $x_{4}$ & -3.07802 & $x_{4}$ & -3.15706 \\ 
		$x_{9}$ & 1.59533 & $x_{9}$ & 2.83706 & $x_{4}$ & -2.62308 & $x_{1}$ & -2.40785 & $x_{1}$ & -2.73452 \\ 
		$x_{8}$ & 0.04716 & $x_{4}$ & -1.89304 & $x_{1}$ & -1.43221 & $x_{9}$ & 1.77553 & $x_{9}$ & 0.97606 \\ 
		$x_{7}$ & 4.19E-07 & $x_{6}$ & 0.17831 & $x_{6}$ & 0.36607 & $x_{7}$ & -0.36456 & $x_{7}$ & -0.44530 \\ 
		$x_{10}$ & -4.15E-08 & $x_{8}$ & 0.09129 & $x_{10}$ & -0.24760 & $x_{6}$ & 0.16250 & $x_{6}$ & 0.21889 \\ 
		$x_{2}$ & 7.37E-09  & $x_{7}$ & 7.52E-07 & $x_{7}$ & -0.19279 & $x_{2}$ & 0.15769 & $x_{2}$ & 0.14194 \\ 
		$x_{5}$ & 4.78E-09 & $x_{10}$ & -6.26E-09 & $x_{8}$ & 1.02E-06 & $x_{10}$ & -0.14640 & $x_{10}$ & -2.99E-07 \\
		$x_{6}$ & 2.18E-09 & $x_{2}$ & -1.70E-11 & $x_{2}$ & 1.48E-08 & $x_{8}$ & 3.72E-10 & $x_{8}$ & -8.40E-08  \\ 
		\hline
	\end{tabular}
	\caption{Coefficient estimates obtained via MM in \Cref{appenB} for regularized quantile regression with adaptive lasso penalty at $q=0.1,0.25,0.5,0.75,0.9$ on the ``econ" data set. For each $q$, the covariates and their respective penalized coefficient estimates are presented in decreasing order of the magnitude of the absolute values of the estimators.}
	\label{lasso econ}
\end{table}
\par Our results, as shown in \Cref{lasso econ}, lend credence to non-constant predictor effects. Although the highest correlated predictor population $x_3$'s effects (positively influencing $y$) are relatively constant, 2 other critical predictors showcase manifest non-constant effects across various quantile orders: CPI $x_5$ is insignificant at $q=0.1$ yet counts a lot at higher quantiles; PCE $x_1$ appears clearly more significant (in negatively affecting $y$) at lower quantiles. Our finding that the inflation rate (measured by CPI $x_5$) negatively correlates to unemployment rate $y$ significantly when $y$ is larger aligns closely to the current widely accepted expectations-augmented Philips curve theory that states the inverse correlation between cyclical unemployment and unanticipated inflation. In addition, $x_4$ (negative correlation with $y$) seems to matter more at higher $q$ and $x_9$ (positive correlation with $y$) at lower $q$. These insightful findings, if backed by more evidence, may aid governments considerably in coming up with more effective and adequate macroeconomic policies to combat high unemployment rates in diverse situations.

{\par \centering\textbf{Social Data}\par}
Our second data set is a ``social" one obtained by combining certain columns of a public data set investigating worldwide country-level life expectancy \parencite{lifeExp} and other potential predictors' per-country data gathered from \textcite{otherWHO}, \textcite{fertility}, \textcite{unemprate}, and \textcite{cpi}. We removed variables with insufficient non-missing data and considered $n=111$ countries with no missing data on all remaining 21 (1 explanatory variable $y$, Life Expectancy, and 20 explanatory variables\footnote{Hence, $p=20+1=21$ (including the intercept). The 20 coviariates are crude suicide rate $x_1$, Corruption Perceptions Index (CPI) $x_2$, fertility rate $x_3$, mobilization resources $x_4$, unemployment rate $x_5$, adult mortality rate $x_6$, infant death frequency $x_7$, Hepatitis B immunization coverage $x_8$, measles frequency $x_9$, average Body Mass Index (BMI) $x_{10}$, under-five deaths $x_{11}$, polio and DTP3 immunization coverages $x_{12}$ and $x_{13}$, HIV/AIDS deaths $x_{14}$, GDP per capita $x_{15}$, population $x_{16}$, thinness frequency among 10-19-year-olds and among 5-9-year-olds $x_{17}$ and $x_{18}$, income composition of resources $x_{19}$, and number of school years $x_{20}$.}) variables. We have ignored spatial dependence.
\par Should there be drastic non-constant predictor effects, in which case regularized quantile regression may greatly improve fit, a country's leaders can then devise more potent policies to influence its life expectancy in its desired way given the quantile of life expectancy the country stands worldwide, by controlling the corresponding influential variables around that $q$.
\begin{table}[h]
	\centering
	\begin{tabular}{|cc|cc|cc|cc|cc|}
		\hline 
		\multicolumn{2}{|c|}{$\bm{q = 0.1}$} & \multicolumn{2}{|c|}{$\bm{q = 0.25}$} &\multicolumn{2}{|c|}{$\bm{q = 0.5}$} &\multicolumn{2}{|c|}{$\bm{q = 0.75}$} & \multicolumn{2}{|c|}{$\bm{q = 0.9}$}\\
		\hline
		$x_j$ & $\big(\hat{\beta}_{\lambda_{\opt}}\big)_j$ & $x_j$ & $\big(\hat{\beta}_{\lambda_{\opt}}\big)_j$ & $x_j$ & $\big(\hat{\beta}_{\lambda_{\opt}}\big)_j$ & $x_j$ & $\big(\hat{\beta}_{\lambda_{\opt}}\big)_j$ & $x_j$ & $\big(\hat{\beta}_{\lambda_{\opt}}\big)_j$\\
		\hline
		$x_{19}$ & 3.72627 & $x_{3}$ & -3.11098 & $x_{19}$ & 3.09372 & $x_{19}$ & 3.83252 & $x_{6}$ & -3.43217 \\ 
		$x_{6}$ & -2.44059 & $x_{19}$ & 2.23083 & $x_{6}$ & -2.54676 & $x_{6}$ & -2.02683 & $x_{19}$ & 3.20339 \\ 
		$x_{3}$ & -2.38420 & $x_{6}$ & -1.71024 & $x_{3}$ & -1.96873 & $x_{3}$ & -1.48831 & $x_{3}$ & -1.03290 \\ 
		$x_{8}$ & 1.36639 & $x_{1}$ & -0.81730 & $x_{1}$ & -0.87852 & $x_{1}$ & -0.98145 & $x_{14}$ & 0.98214 \\ 
		$x_{17}$ & -1.00701 & $x_{2}$ & 0.79936 & $x_{4}$ & 0.53576 & $x_{16}$ & -0.61017 & $x_{1}$ & -0.76843 \\ 
		$x_{11}$ & 0.85944 & $x_{17}$ & -0.77920 & $x_{16}$ & -0.43476 & $x_{8}$ & 0.58383 & $x_{8}$ & 0.57012 \\ 
		$x_{2}$ & 0.69624 & $x_{4}$ & 0.72327 & $x_{5}$ & -0.41536 & $x_{4}$ & 0.36708 & $x_{4}$ & 0.40426 \\ 
		$x_{13}$ & -0.63655 & $x_{8}$ & 0.67387 & $x_{2}$ & 0.31067 & $x_{18}$ & -0.25079 & $x_{18}$ & -0.34016 \\ 
		$x_{1}$ & -0.57268 & $x_{7}$ & 0.45569 & $x_{14}$ & -0.30373 & $x_{2}$ & 0.24719 & $x_{11}$ & -0.01845 \\ 
		$x_{14}$ & -0.50570 & $x_{16}$ & -0.31904 & $x_{8}$ & 0.15149 & $x_{5}$ & -0.20071 & $x_{16}$ & -1.90E-06\\ 
		$x_{4}$ & 0.27249 & $x_{14}$ & -0.29716 & $x_{17}$ & -0.14251 & $x_{12}$ & -0.07383 & $x_{5}$ & 6.12E-07 \\ 
		$x_{16}$ & -0.17622 & $x_{5}$ & -0.29464 & $x_{12}$ & -1.05E-06 & $x_{7}$ & 1.89E-08 & $x_{10}$ & 2.26E-08 \\ 
		$x_{10}$ & -7.54E-07 & $x_{9}$ & -4.42E-07 & $x_{7}$ & 3.98E-08 & $x_{15}$ & -8.70E-09 & $x_{20}$ & -7.03E-09 \\ 
		$x_{18}$ & 6.10E-08 & $x_{11}$ & 1.87E-08 & $x_{9}$ & 2.33E-08 & $x_{11}$ & -8.61E-09 & $x_{7}$ & -4.48E-09 \\ 
		$x_{5}$ & -1.43E-08 & $x_{15}$ & 1.21E-08 & $x_{10}$ & 1.52E-08 & $x_{17}$ & 4.08E-09 & $x_{2}$ & 3.94E-09 \\ 
		$x_{9}$ & -5.34E-09 & $x_{13}$ & -3.50E-09 & $x_{18}$ & -7.19E-09 & $x_{10}$ & 3.92E-09 & $x_{15}$ & 2.58E-09 \\ 
		$x_{7}$ & -2.98E-09 & $x_{12}$ & 1.50E-09 & $x_{20}$ & -6.09E-09 & $x_{14}$ & -8.14E-10 & $x_{12}$ & 1.55E-09 \\ 
		$x_{12}$ & -1.38E-09 & $x_{18}$ & 1.99E-10 & $x_{11}$ & 5.00E-09 & $x_{20}$ & 7.93E-10 & $x_{9}$ & 1.16E-09\\ 
		$x_{20}$ & 1.93E-10 & $x_{10}$ & 1.96E-10 & $x_{13}$ & -2.08E-09 & $x_{9}$ & 2.14E-10 & $x_{17}$ & -1.88E-10 \\ 
		$x_{15}$ & 2.80E-11 & $x_{20}$ & 3.68E-11 & $x_{15}$ & 6.90E-10 & $x_{13}$ & -3.07E-11 & $x_{13}$ & 1.48E-10 \\ 
		\hline
	\end{tabular}
	\caption{Coefficient estimates obtained via MM in \Cref{appenB} for regularized quantile regression with adaptive lasso penalty at $q=0.1,0.25,0.5,0.75,0.9$ on the ``social" data set. For each $q$, the covariates and their respective penalized coefficient estimates are presented in decreasing order of the magnitude of the absolute values of the estimators.}
	\label{lasso social}
\end{table}
\par It turns out that covariates for this ``social" data set do not seem to exhibit notably varying effects for different quantile orders (\Cref{lasso social}). The most attributable predictors ($x_{19}$ -- positively affects $y$; $x_6,x_3$ -- negatively correlate to $y$) seem to possess quite constant effects across quantiles. There are, however, covariates among second-tier influential predictors that do suggest possible varying effects. Effects of $x_{17}$ (which affects $y$ negatively) and $x_2$ (which influences $y$ positively) appear to decrease as $q$ increases, and $x_8$ (which positively correlates to $y$) seems less important in middle quantiles.

{\par \centering\textbf{Forest Fire Data}\par}
Our last data set is a real-world ``forest fire" data set collected from Northeast Portugal \parencite{Cortez2007}, which contains a response variable ``burned area" as well as potential spatial ($x_1,x_2$: x, y coordinates on the Montesinho park map), temporal ($x_3$: month of the year; $x_4$: day of the week), meteorological weather ($x_9$: temperature; $x_{10}$: relative humidity; $x_{11}$: wind speed; $x_{12}$: rainfall) explanatory variables and key Fire Weather Index (FWI) components including the Fine Fuel Moisture Code (FFMC) $x_5$, the Duff Moisture Code (DMC) $x_6$, the Drought Code (DC) $x_7$, and the Initial Spread Index (ISI) $x_8$. 
We used a subset (270 of the 517 instances when there is a forest fire) of the original data set for analysis and transformed the response variable using $y=\ln(x+0.1)$, as the original response variable ``area" is highly concentrated at $0_+$.
\par As small fires are less damaging but much more frequent and large fires are less frequent but much more destructive, it would significantly aid targeted firefighting and thus reduce fires' negative impacts if we can discern potential differentiated predictor effects across different fire scales and make more accurate predictions on a given initiated fire's possible damages. 

\begin{table}[h]
	\centering
	\begin{tabular}{|cc|cc|cc|cc|cc|}
		\hline 
		\multicolumn{2}{|c|}{$\bm{q = 0.1}$} & \multicolumn{2}{|c|}{$\bm{q = 0.25}$} &\multicolumn{2}{|c|}{$\bm{q = 0.5}$} &\multicolumn{2}{|c|}{$\bm{q = 0.75}$} & \multicolumn{2}{|c|}{$\bm{q = 0.9}$}\\
		\hline
		$x_j$ & $\big(\hat{\beta}_{\lambda_{\opt}}\big)_j$ & $x_j$ & $\big(\hat{\beta}_{\lambda_{\opt}}\big)_j$ & $x_j$ & $\big(\hat{\beta}_{\lambda_{\opt}}\big)_j$ & $x_j$ & $\big(\hat{\beta}_{\lambda_{\opt}}\big)_j$ & $x_j$ & $\big(\hat{\beta}_{\lambda_{\opt}}\big)_j$\\
		\hline
		$x_9$ & -0.20676 & $x_9$ & -0.18817  & $x_7$ & -0.15219 & $x_7$ &-0.58952 &$x_9$ & 0.26410 \\ 
		$x_8$ & -0.12203  & $x_6$ & 0.15870 & $x_3$ & 0.10327 & $x_9$ & 0.43747 & $x_6$ & 0.12177 \\ 
		$x_4$ & 0.05510 & $x_5$ & -0.10424 & $x_{10}$ & -0.03382 &$x_3$ & 0.37755 & $x_1$ & 0.08917 \\ 
		$x_6$ & 3.72E-06 & $x_7$ & -0.07071 & $x_9$ & -0.00555 & $x_6$ & 0.34871 & $x_8$ & -0.05010 \\ 
		$x_{12}$ & 8.44E-09 &$x_1$ & -8.61E-07 & $x_6$ & 2.86E-06 & $x_2$ & -0.32360 &$x_{11}$ & 1.04E-06\\ 
		$x_{10}$ & -2.12E-09 & $x_3$ & 4.06E-08 & $x_2$ & -1.96E-08 & $x_1$ & 0.32214 &$x_{7}$ & -8.77E-08 \\ 
		$x_{1}$ & 1.65E-09 & $x_{10}$ & -8.34E-09 &$x_1$ & 1.63E-08 & $x_8$ & -0.26278 &$x_3$ & -5.60E-09 \\ 
		$x_3$ & 1.09E-09 & $x_{4}$ & 3.69E-09 & $x_{4}$ & 1.34E-08 & $x_{4}$ & 0.18095 &$x_{12}$ & -4.32E-09\\ 
		$x_{7}$ & -8.94E-10 & $x_{12}$ & 2.83E-09 & $x_{12}$ & 7.85E-09 & $x_{11}$ & 0.14860 &$x_2$ & -3.21E-09 \\ 
		$x_{11}$ & 5.55E-10 & $x_{11}$ & 2.77E-09 & $x_8$ & -5.69E-09 &$x_{5}$ & 2.31E-07 &$x_{5}$ & -1.18E-09 \\ 
		$x_5$ & -1.67E-10 & $x_8$ & -1.14E-09 & $x_5$ & -1.75E-09 & $x_{12}$ & -2.74E-08 &$x_{10}$ & -9.12E-10\\ 
		$x_2$ & 6.91E-11 & $x_2$ & 8.35E-10 & $x_{11}$ & 6.04E-10 &$x_{10}$ & -2.76E-09 &$x_4$ & 4.13E-10  \\ 
		\hline
	\end{tabular}
	\caption{Coefficient estimates obtained via MM in \Cref{appenB} for regularized quantile regression with adaptive lasso penalty at $q=0.1,0.25,0.5,0.75,0.9$ on the ``forest fire" data set. For each $q$, the covariates and their respective penalized coefficient estimates are presented in decreasing order of the magnitude of the absolute values of the estimators.}
	\label{lasso forestfire}
\end{table}
\par As presented in \Cref{lasso forestfire}, it is likely that there are indeed noteworthy varying covariate effects. Temperature $x_9$ appears significant at all quantile levels but its effects switch sign. The Drought Code $x_7$ showcases prominent negative impacts on $y$ from middle to upper-middle $q$, and the Duff Moisture Code $x_6$ appears more influential in positively affecting $y$ at higher quantile levels. This finding, that higher DMC and lower DC result in more severe destructions for larger fires, match well to what \textcite{Amiro2004} found out. We also observed that smaller ($q=0.1$) fires' impacts seem to have much fewer key predictors whereas impairments from fires of an upper-middle scale ($q=0.75$) appear correlated to much more covariates.

\section{Nonparametric Quantile Regression}\label{appenC}
\par Despite its popularity and general competence, the typical linear quantile regression is inappropriate for data sets exhibiting manifest complex nonlinear patterns. Quantile regression for universal nonlinear data, however, has rarely been discussed. The handful of existing approaches in current literature (e.g., \cite{Yu2016}) rely on broader nonparametric methods like Kernel Estimation adapted to quantile regression instead of parametric methods convenient for interpretation and prediction. We attempt to bridge this gap by extending our MM algorithm constructed in \Cref{sect2c} to parametrically model this type of quantile regression problem. 
\begin{figure}[h]
	\centering
	\includegraphics[width=0.75\textwidth]{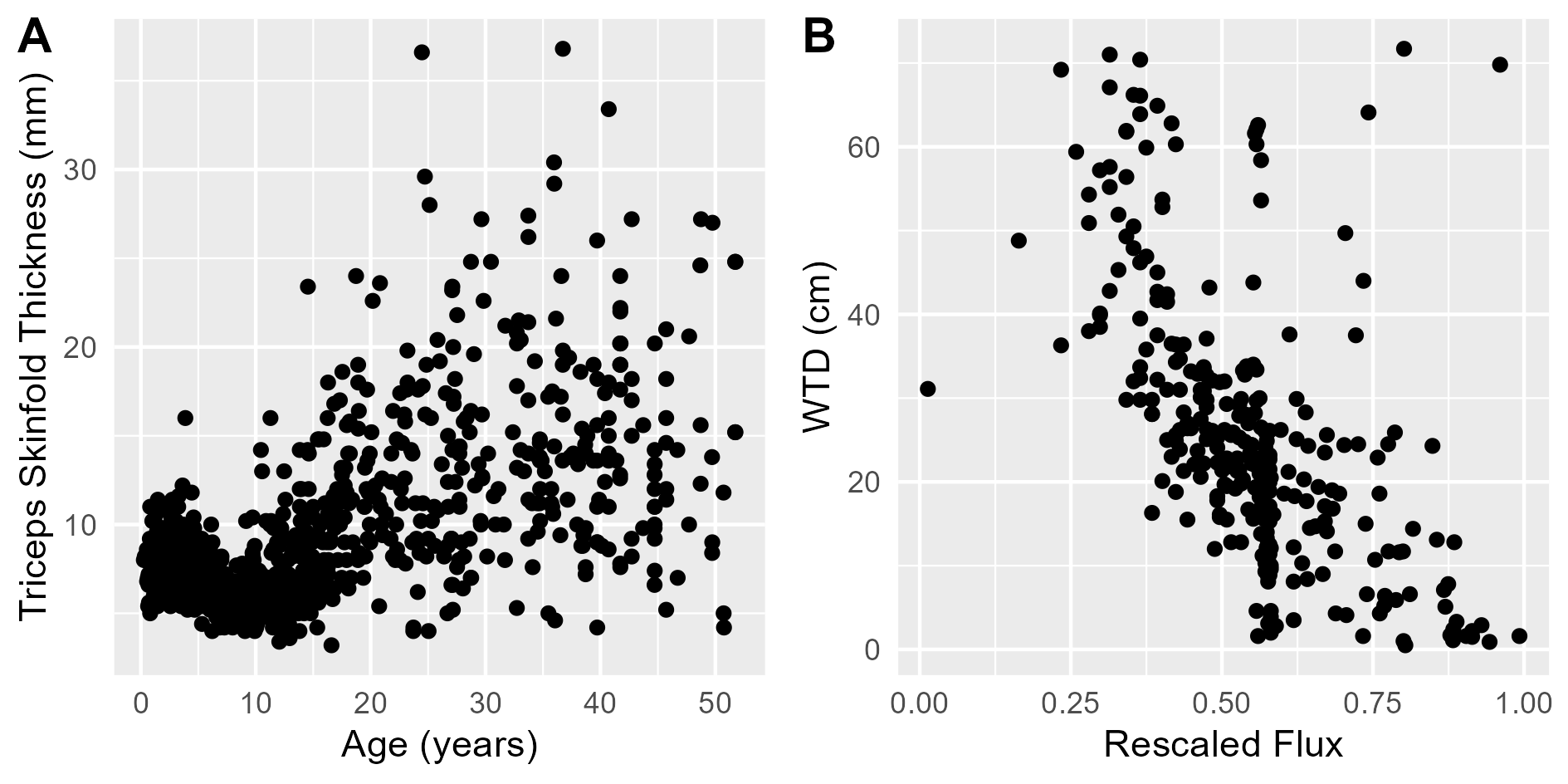}
	\caption{\footnotesize Scatter plots for two bivariate data sets. The ``triceps" data set (\textbf{A}) documents age and triceps skinfold thickness of 892 females under the age of 50 in West Africa's three villages. We used $\ln(\text{triceps})$ as the response variable and took exponential of the estimated values to get back triceps thickness's original scale. The ``WTD" data set (\textbf{B}) records 307 areas' Water Table Depth (WTD) and rainfall flux at one site in South Africa's Cape Town. We transformed the variables as described by \textcite{Noufaily2013}, i.e., take the absolute value of WTD and let Rescaled Flux = $\frac{1}{7.3}(\ln(\text{Flux }+6.9)+2.4)$.}
	\label{scatter plots} 
\end{figure}

\subsection{MM and Cross Validation}\label{appenC1}
\par We start with two real bivariate data sets displaying apparent complicated nonlinear correlations (\Cref{scatter plots}). It is in fact pretty straightforward to achieve the desired generalization via MM. The first step is fairly customary in ordinary regression settings -- replace the original sample observation matrix $\bm{X}_{n\times p}$ ($p=1+1=2$ here)  by some transformations of the original variables:
\begin{equation}\label{transform X}
	\bm{X}=(\bm{1}_n,\bm{x})=
	\begin{pmatrix}
		1 & x_1\\
		1 & x_2\\
		\vdots & \vdots \\
		1 & x_n
	\end{pmatrix}
	\overset{f_1,f_2,\dots,f_{p'-1}}{\longrightarrow}
	\bm{X}_{\text{new}}=
	\begin{pmatrix}
		1 & f_1(x_1) & f_2(x_1) & \dots & f_{p'-1}(x_1)\\
		1 & f_1(x_2) & f_2(x_2) & \dots & f_{p'-1}(x_2)\\
		\vdots & \vdots & \vdots & \ddots & \vdots\\
		1 & f_1(x_n) & f_2(x_n) & \dots & f_{p'-1}(x_n)
	\end{pmatrix}_{n\times p'},
\end{equation}
where $f_1,f_2,\dots,f_{p'-1}$ are a set of chosen functions.
With this new design matrix $\bm{X}_{\text{new}}$, we can proceed to calculate the corresponding $\bm{D}_{\text{new}}(q)_{n\times hp'}$s in \Cref{D(q)} and carry out our MM algorithm as delineated in \Cref{sect2c}. 
\par As the correlation types are not specified and the two scatter plots displayed in \Cref{scatter plots} do not suggest any particular simple nonlinear patterns, we would need highly flexible transformations applicable to universal nonlinear data sets. We hence considered the flexible, stable, and extensively utilized non-parametric natural spline transformations \parencite{Perperoglou2019} again, which were applied to covariates instead of quantile orders this time. 
\par
\begin{figure}[ph!]
	\centering
	\includegraphics[width=0.92\textwidth]{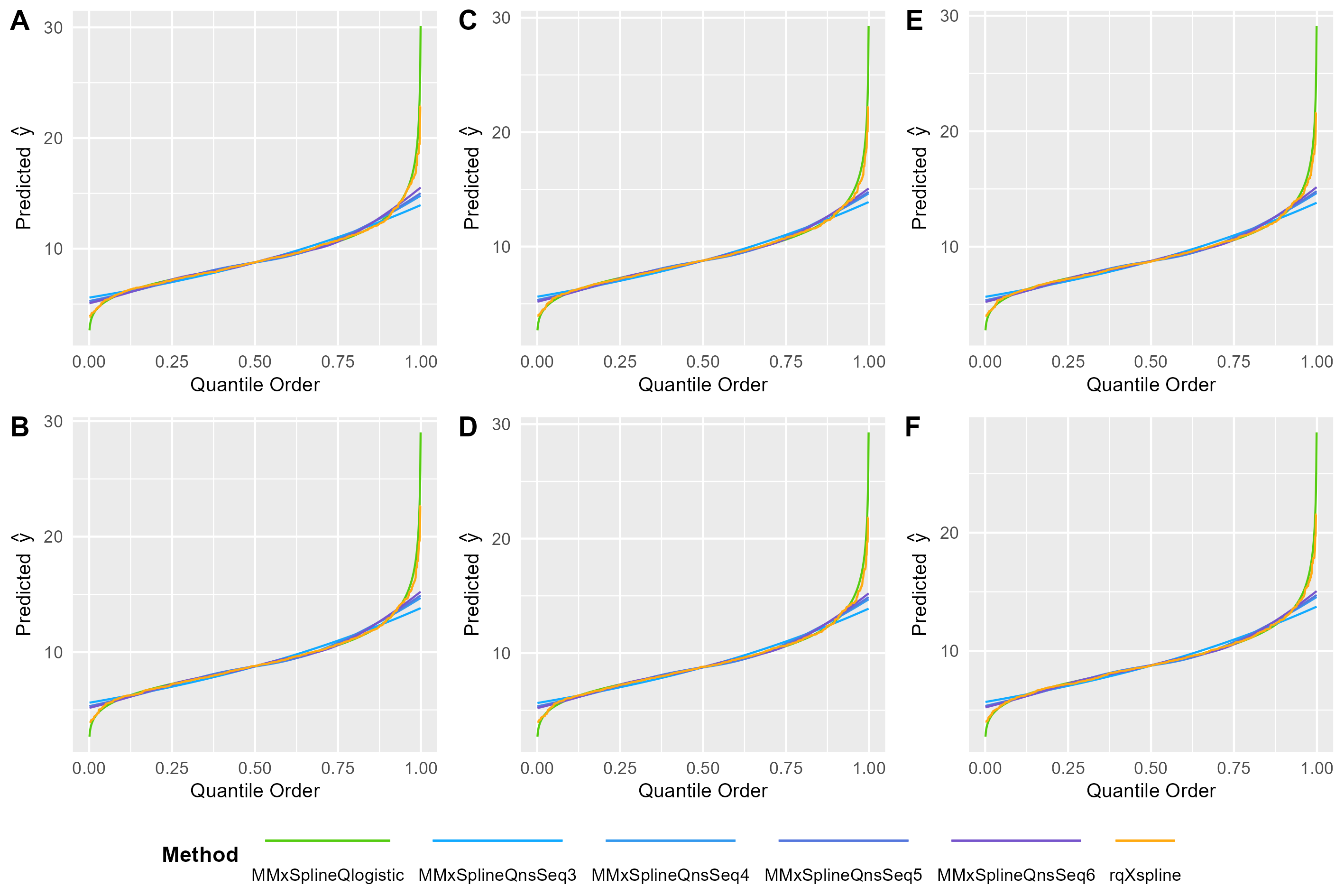}
	\caption{Quantile plots for the ``triceps" data set taking $x$ as the sample mean. We natural-spline-transformed $x$ using 3 (\textbf{A}), 4 (\textbf{B}), 5 (\textbf{C}), 6 (\textbf{D}), 7 (\textbf{E}), and 8 (\textbf{F}) equally spaced knots between 0 and 1. Corresponding to each covariate transformation, we adopted the logistic basis and natural spline bases with 3, 4, 5, and 6 equally spaced knots for quantiles on top of \texttt{rq}.}
	\label{triceps mean quantile plots}
\end{figure}

\begin{figure}[ph!]
	\centering
	\includegraphics[width=0.92\textwidth]{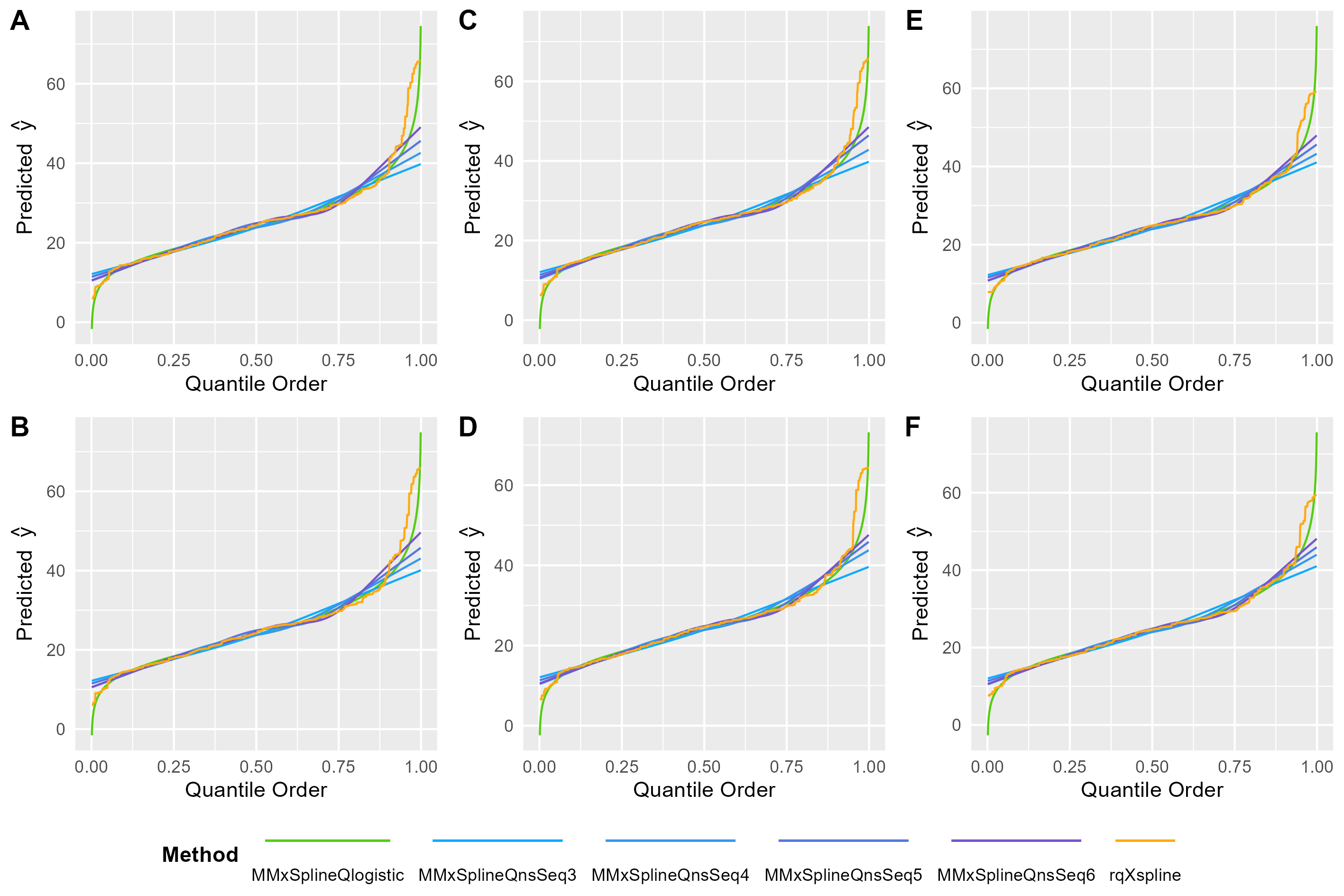}
	\caption{\footnotesize Quantile plots for the ``WTD" data set taking $x$ as the sample mean. We natural-spline-transformed $x$ using 3 (\textbf{A}), 4 (\textbf{B}), 5 (\textbf{C}), 6 (\textbf{D}), 8 (\textbf{F}) equally spaced knots between 0 and 1 and the asymmetric knots $(0.2, 0.4, 0.5, 0.55, 0.6, 0.7, 0.9)$ (\textbf{E}). Corresponding to each covariate transformation, we adopted the logistic basis and natural spline bases with 3, 4, 5, and 6 equally spaced knots for quantiles on top of \texttt{rq}.}
	\label{WTD mean quantile plots}
\end{figure}

\begin{figure}
	\centering
	\includegraphics[width=0.95\textwidth]{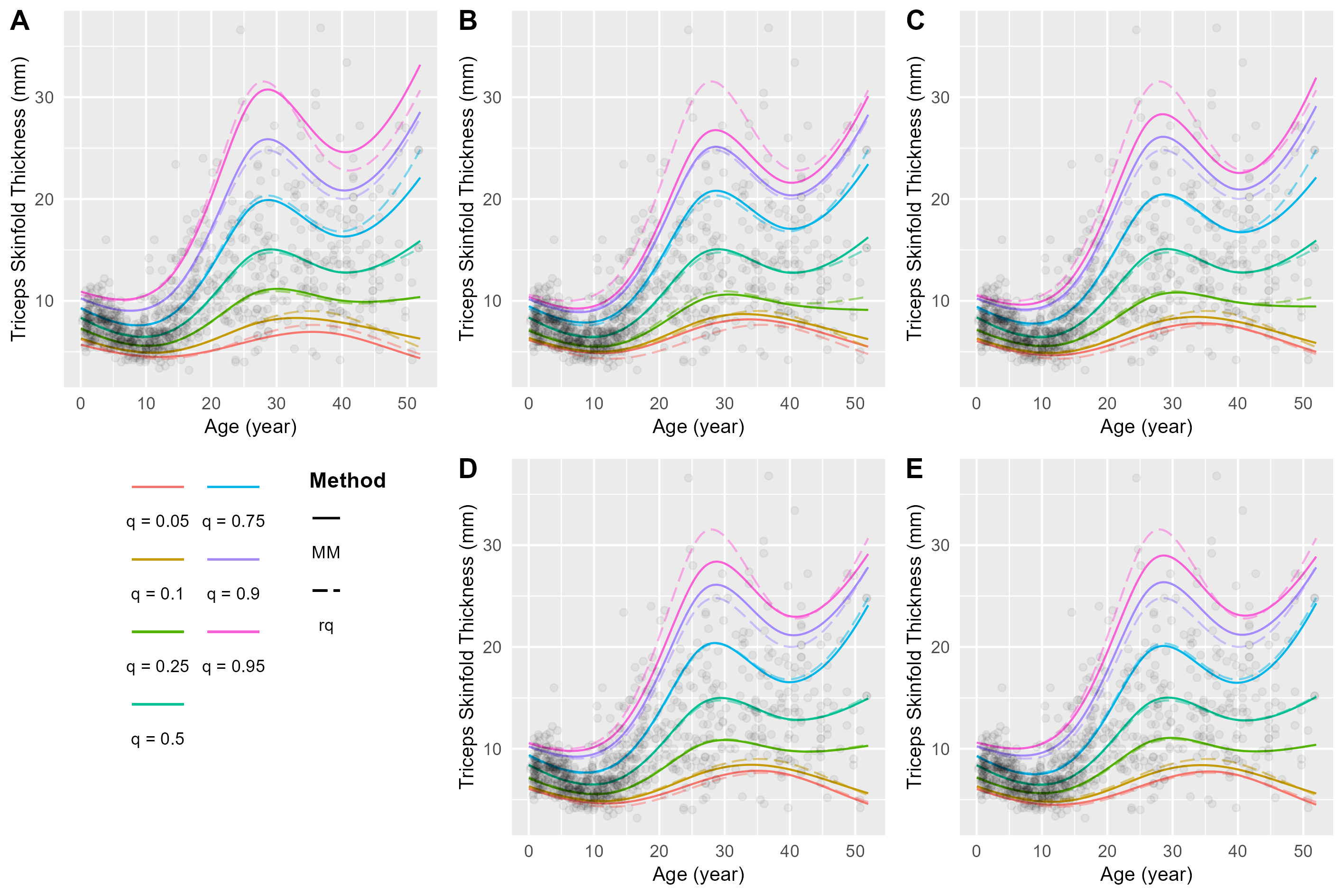}	
	\caption{Fitted quantile curves for the ``triceps" data set - natural spline transformation on $x$ with 3 equally spaced knots. We have adopted the logistic basis (\textbf{A}) and natural spline bases with 3 (\textbf{B}), 4 (\textbf{C}), 5 (\textbf{D}), and 6 (\textbf{E}) equally spaced knots for quantile levels.}
	\label{triceps quantile curves Xspline3}
\end{figure}

\begin{figure}
	\centering
	\includegraphics[width=0.95\textwidth]{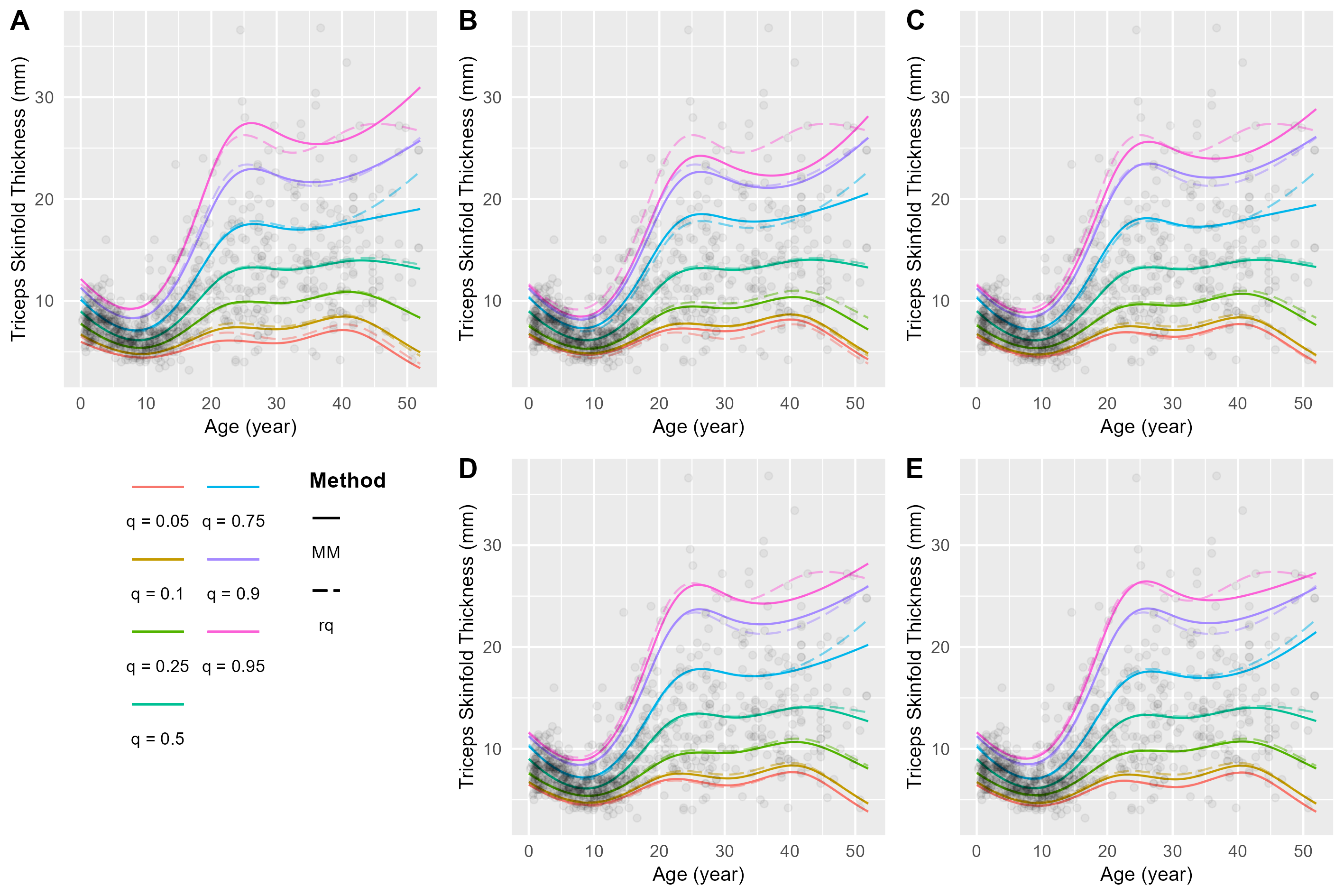}	
	\caption{Fitted quantile curves for the ``triceps" data set - natural spline transformation on $x$ with 4 equally spaced knots. We have adopted the logistic basis (\textbf{A}) and natural spline bases with 3 (\textbf{B}), 4 (\textbf{C}), 5 (\textbf{D}), and 6 (\textbf{E}) equally spaced knots for quantile levels.}
	\label{triceps quantile curves Xspline4}
\end{figure}

\begin{figure}
	\centering
	\includegraphics[width=0.95\textwidth]{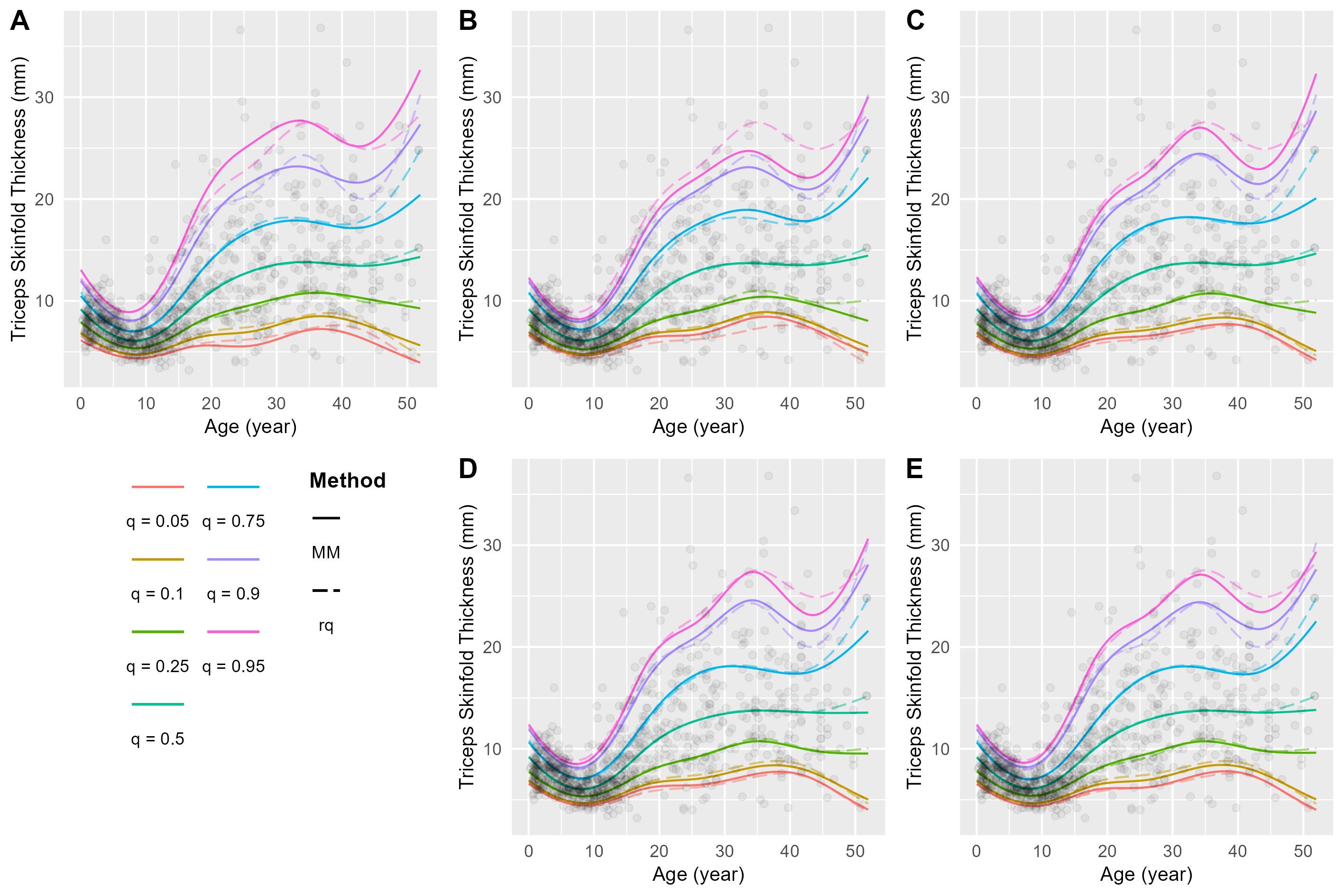}	
	\caption{Fitted quantile curves for the ``triceps" data set - natural spline transformation on $x$ with 5 equally spaced knots. We have adopted the logistic basis (\textbf{A}) and natural spline bases with 3 (\textbf{B}), 4 (\textbf{C}), 5 (\textbf{D}), and 6 (\textbf{E}) equally spaced knots for quantile levels.}
	\label{triceps quantile curves Xspline5}
\end{figure}

\begin{figure}[ph!]
	\centering
	\includegraphics[width=0.95\textwidth]{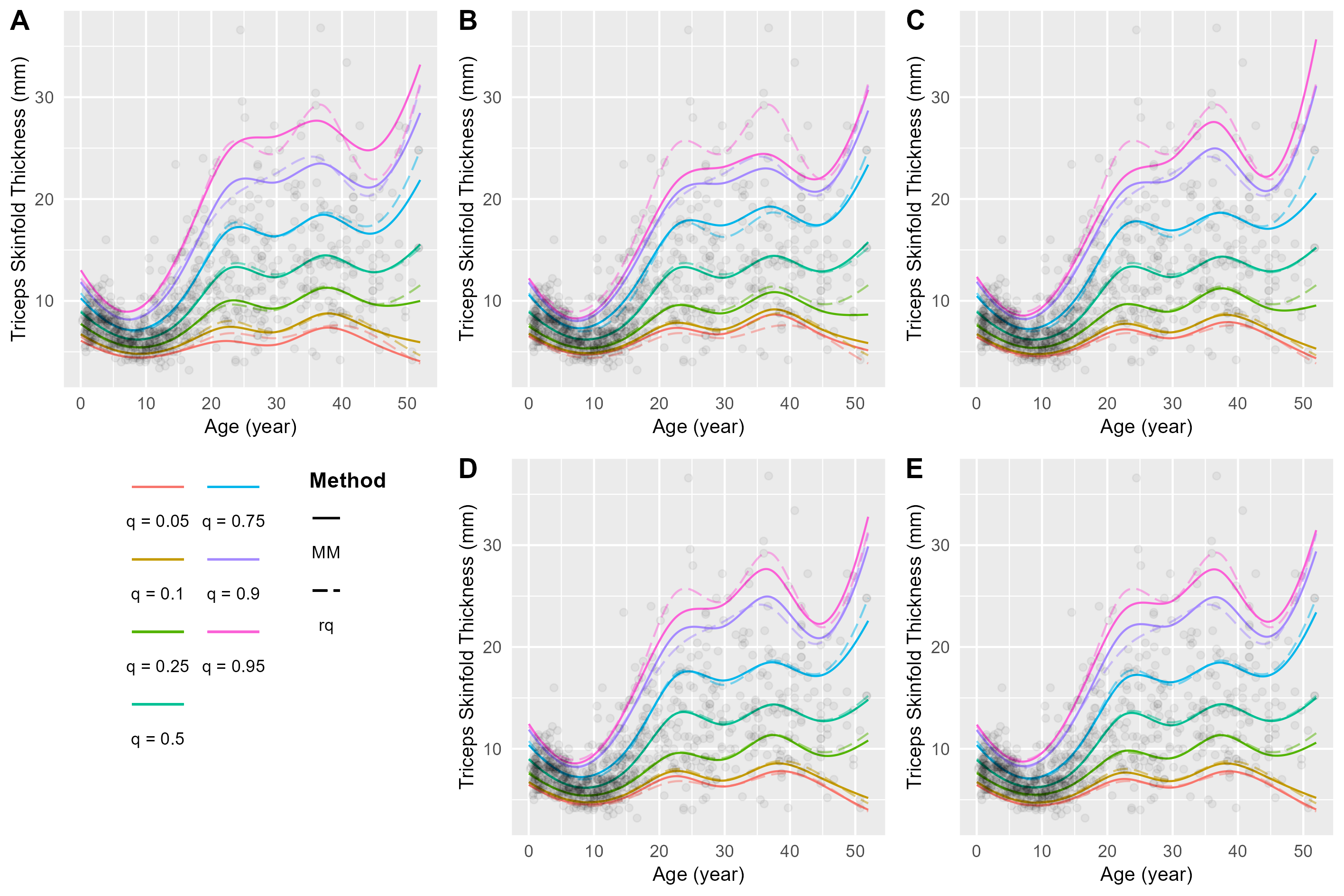}	
	\caption{Fitted quantile curves for the ``triceps" data set - natural spline transformation on $x$ with 6 equally spaced knots. We have adopted the logistic basis (\textbf{A}) and natural spline bases with 3 (\textbf{B}), 4 (\textbf{C}), 5 (\textbf{D}), and 6 (\textbf{E}) equally spaced knots for quantile levels.}
	\label{triceps quantile curves Xspline6}
\end{figure}

\begin{figure}
	\centering
	\includegraphics[width=0.95\textwidth]{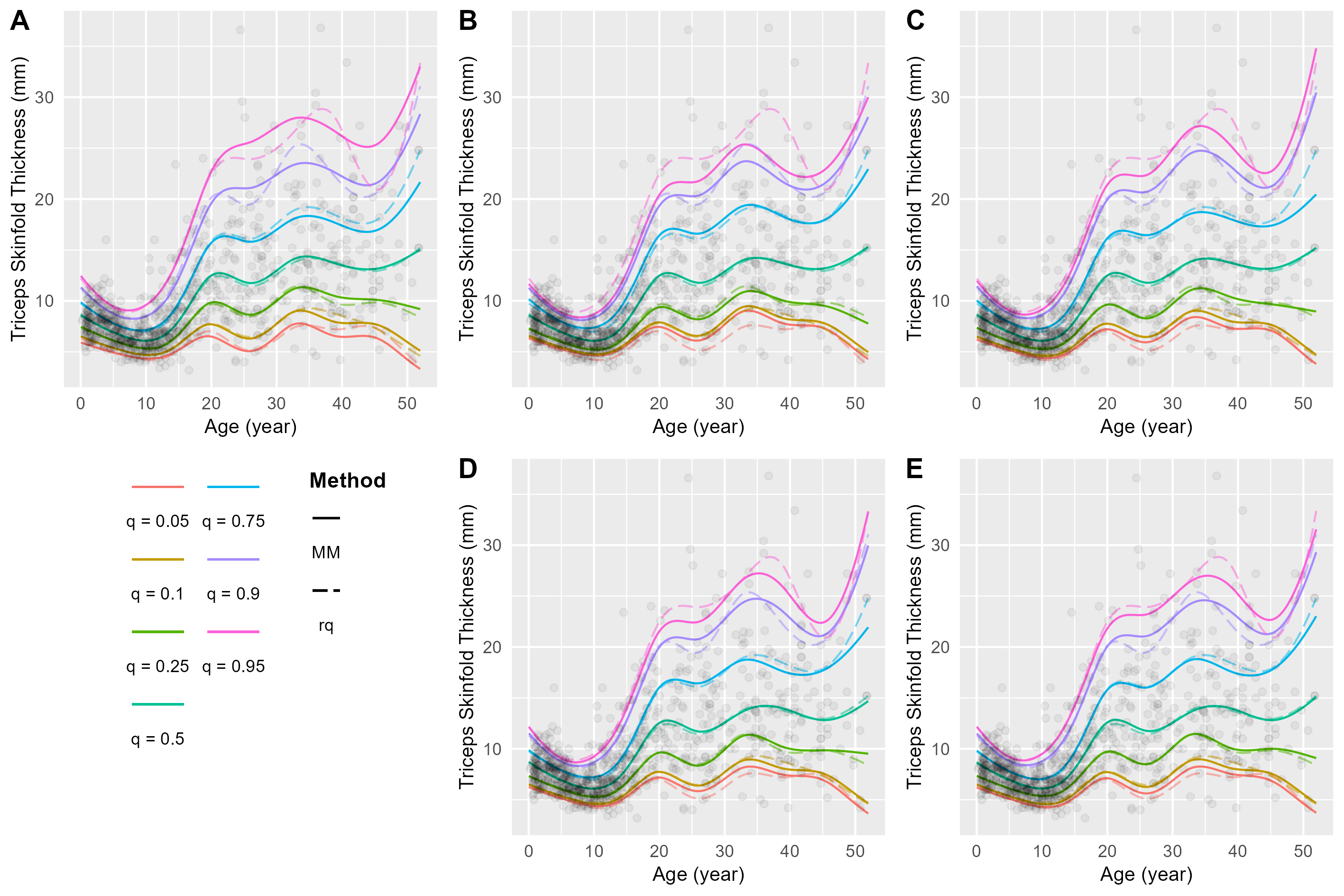}	
	\caption{Fitted quantile curves for the ``triceps" data set - natural spline transformation on $x$ with 7 equally spaced knots. We have adopted the logistic basis (\textbf{A}) and natural spline bases with 3 (\textbf{B}), 4 (\textbf{C}), 5 (\textbf{D}), and 6 (\textbf{E}) equally spaced knots for quantile levels.}
	\label{triceps quantile curves Xspline7}
\end{figure}

\begin{figure}
	\centering
	\includegraphics[width=0.95\textwidth]{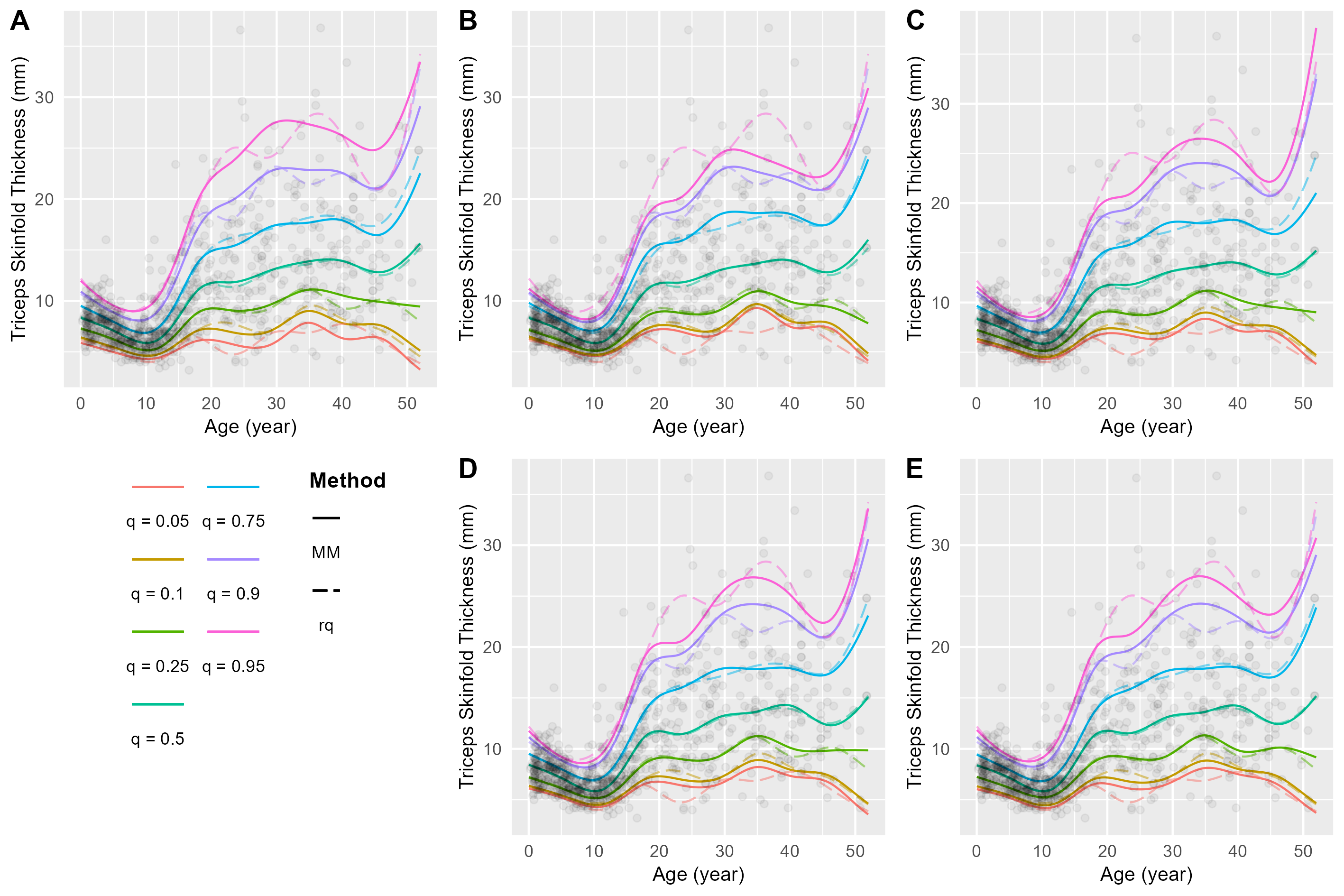}	
	\caption{Fitted quantile curves for the ``triceps" data set - natural spline transformation on $x$ with 8 equally spaced knots. We have adopted the logistic basis (\textbf{A}) and natural spline bases with 3 (\textbf{B}), 4 (\textbf{C}), 5 (\textbf{D}), and 6 (\textbf{E}) equally spaced knots for quantile levels.}
	\label{triceps quantile curves Xspline8}
\end{figure}

\begin{figure}
	\centering
	\includegraphics[width=0.95\textwidth]{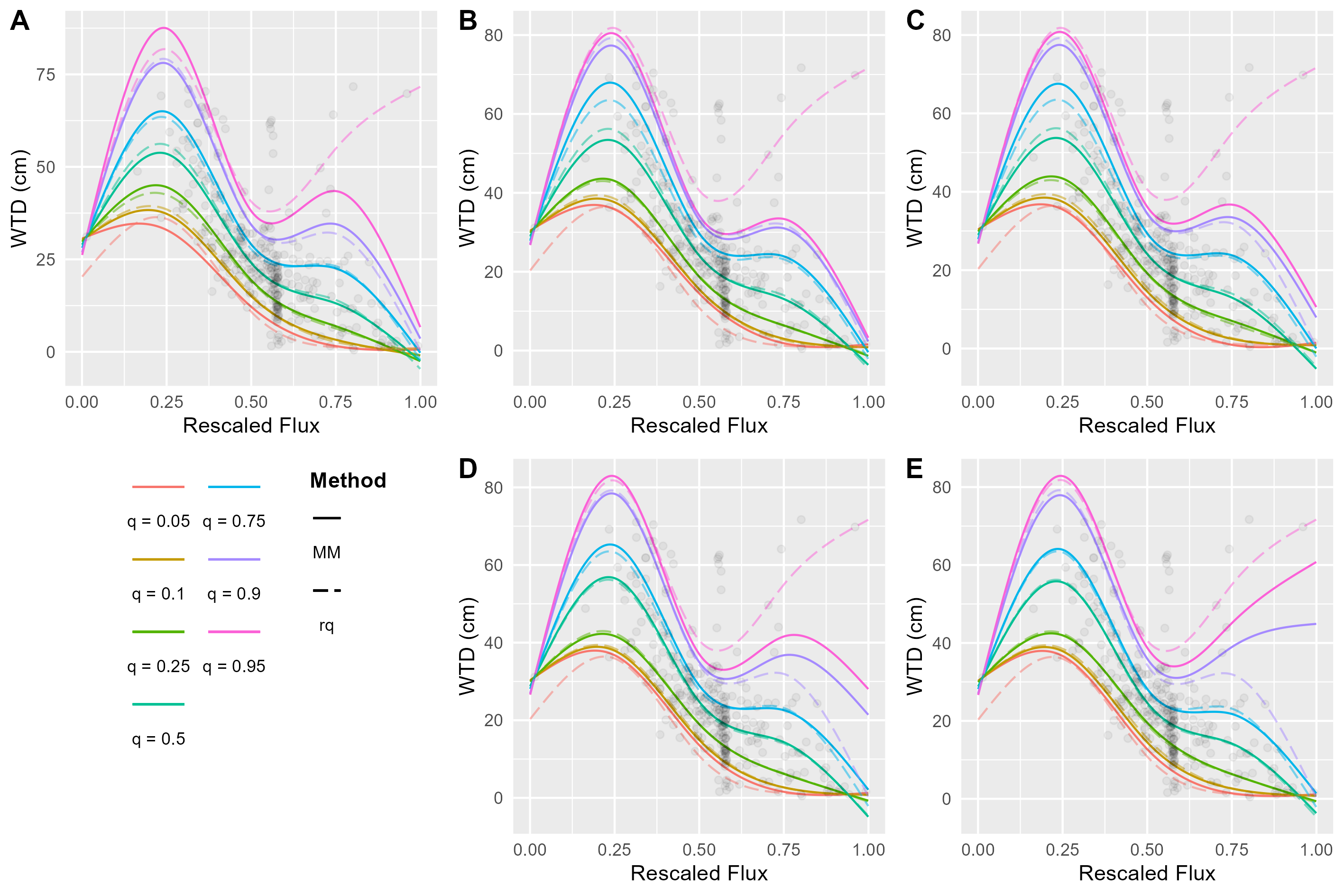}	
	\caption{Fitted quantile curves for the ``WTD" data set - natural spline transformation on $x$ with 3 equally spaced knots. We have adopted the logistic basis (\textbf{A}) and natural spline bases with 3 (\textbf{B}), 4 (\textbf{C}), 5 (\textbf{D}), and 6 (\textbf{E}) equally spaced knots for quantile levels.}
	\label{WTD quantile curves Xspline3}
\end{figure}

\begin{figure}
	\centering
	\includegraphics[width=0.95\textwidth]{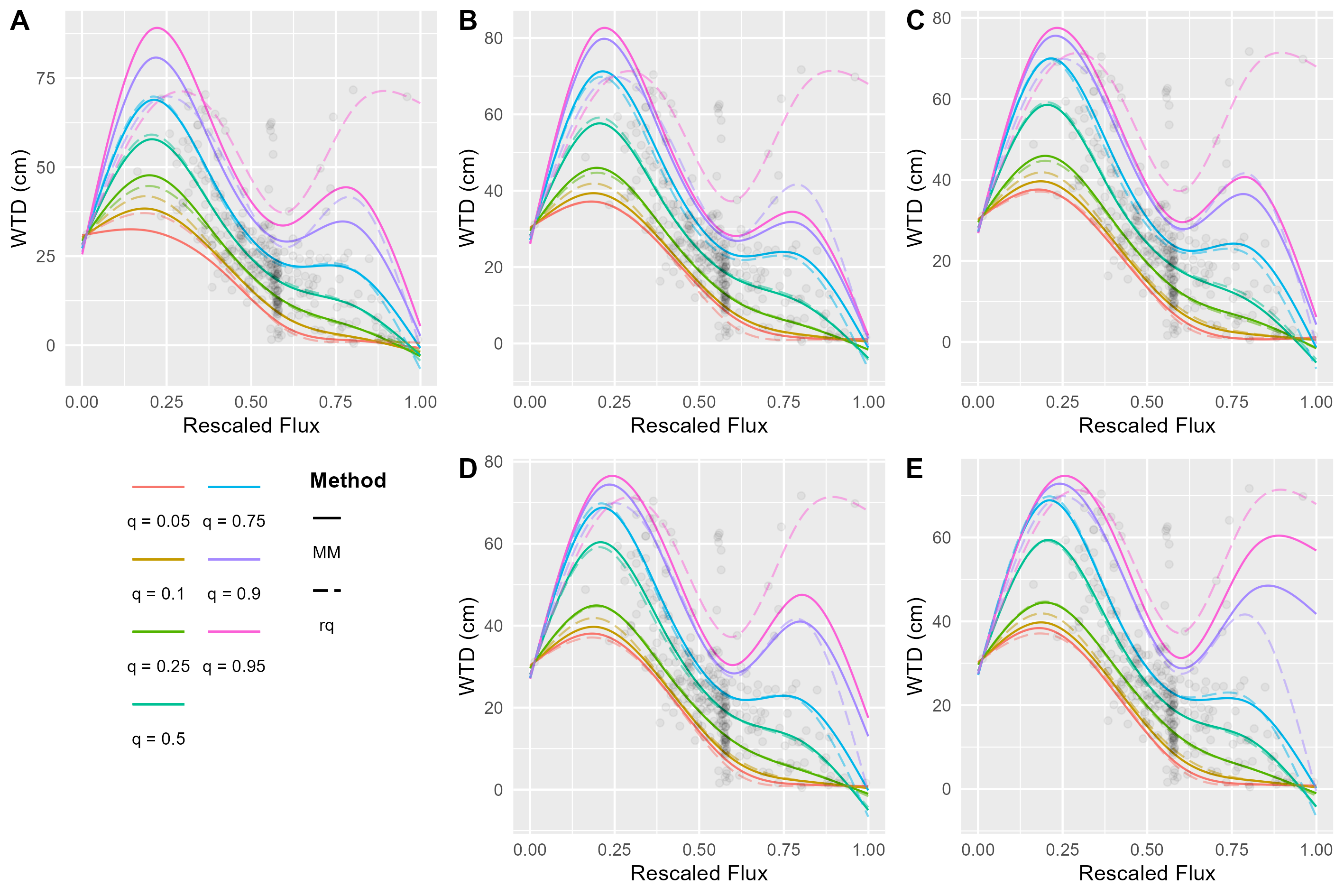}	
	\caption{Fitted quantile curves for the ``WTD" data set - natural spline transformation on $x$ with 4 equally spaced knots. We have adopted the logistic basis (\textbf{A}) and natural spline bases with 3 (\textbf{B}), 4 (\textbf{C}), 5 (\textbf{D}), and 6 (\textbf{E}) equally spaced knots for quantile levels.}
	\label{WTD quantile curves Xspline4}
\end{figure}

\begin{figure}
	\centering
	\includegraphics[width=0.95\textwidth]{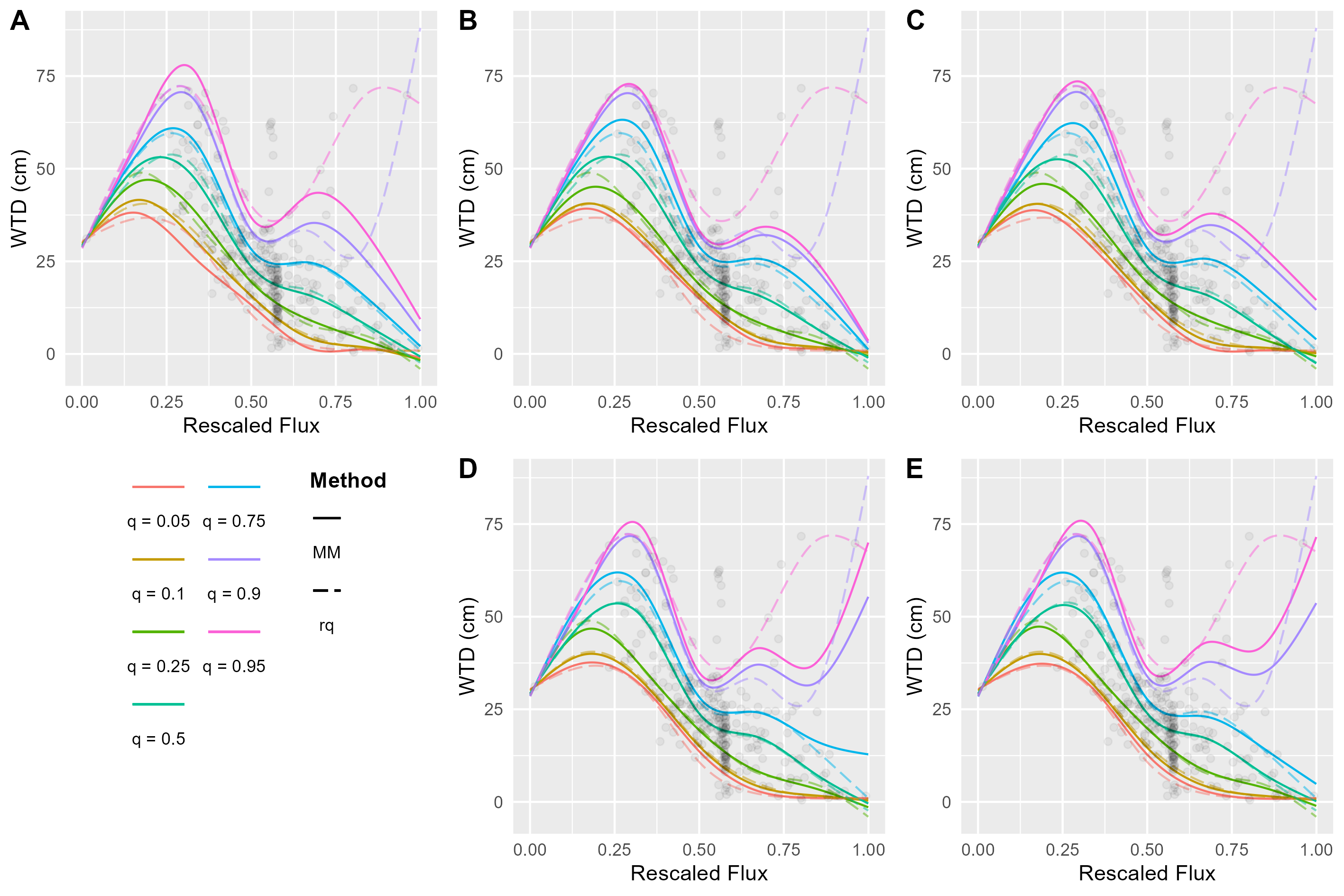}	
	\caption{Fitted quantile curves for the ``WTD" data set - natural spline transformation on $x$ with 5 equally spaced knots. We have adopted the logistic basis (\textbf{A}) and natural spline bases with 3 (\textbf{B}), 4 (\textbf{C}), 5 (\textbf{D}), and 6 (\textbf{E}) equally spaced knots for quantile levels.}
	\label{WTD quantile curves Xspline5}
\end{figure}

\begin{figure}
	\centering
	\includegraphics[width=0.95\textwidth]{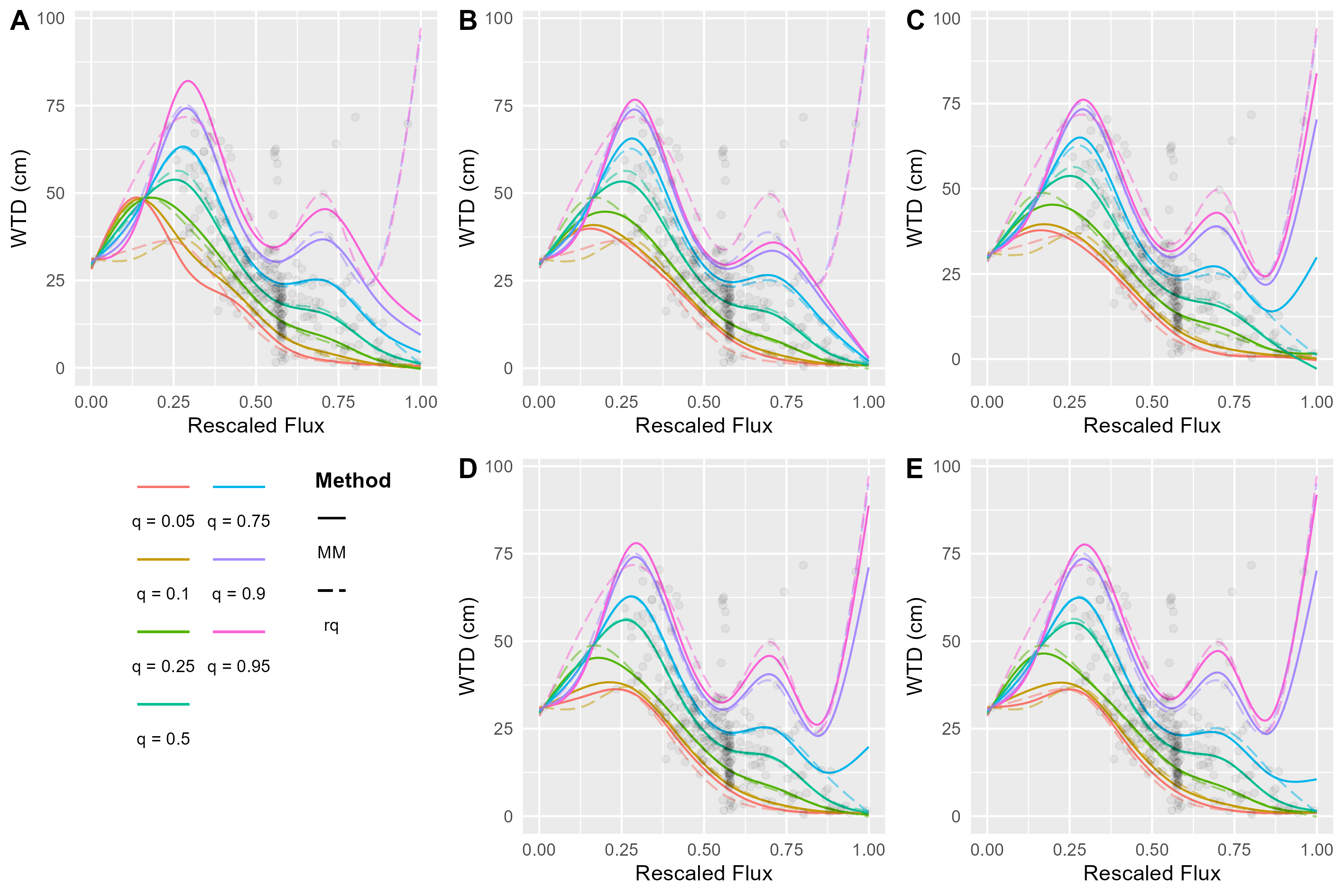}	
	\caption{Fitted quantile curves for the ``WTD" data set - natural spline transformation on $x$ with 6 equally spaced knots. We have adopted the logistic basis (\textbf{A}) and natural spline bases with 3 (\textbf{B}), 4 (\textbf{C}), 5 (\textbf{D}), and 6 (\textbf{E}) equally spaced knots for quantile levels.}
	\label{WTD quantile curves Xspline6}
\end{figure}

\begin{figure}[ph!]
	\centering
	\includegraphics[width=0.95\textwidth]{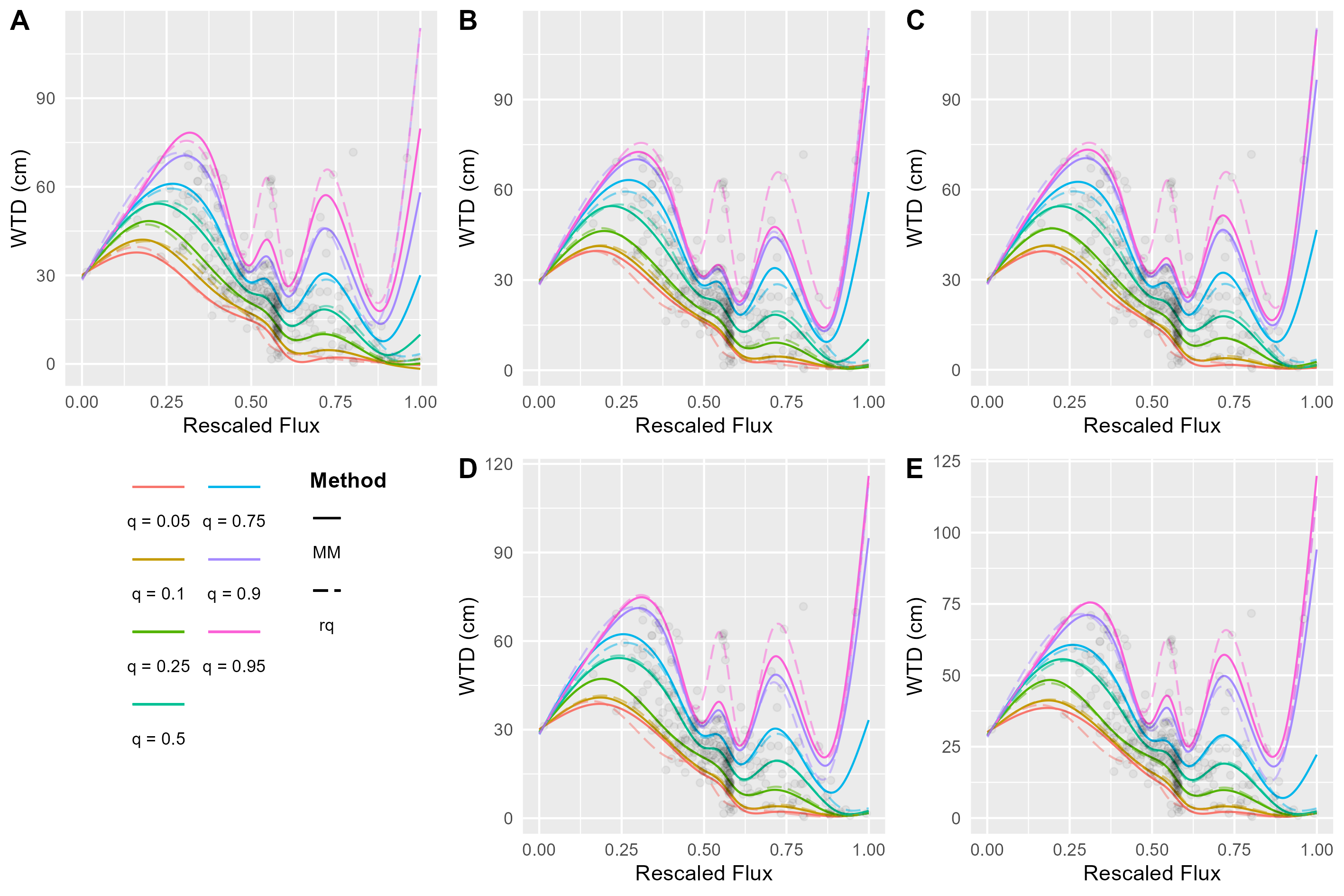}	
	\caption{Fitted quantile curves for the ``WTD" data set - natural spline transformation on $x$ with 7 knots $0.2, 0.4, 0.5, 0.55, 0.6, 0.7, 0.9$. We have adopted the logistic basis (\textbf{A}) and natural spline bases with 3 (\textbf{B}), 4 (\textbf{C}), 5 (\textbf{D}), and 6 (\textbf{E}) equally spaced knots for quantile levels.}
	\label{WTD quantile curves Xspline7}
\end{figure}

\begin{figure}
	\centering
	\includegraphics[width=0.95\textwidth]{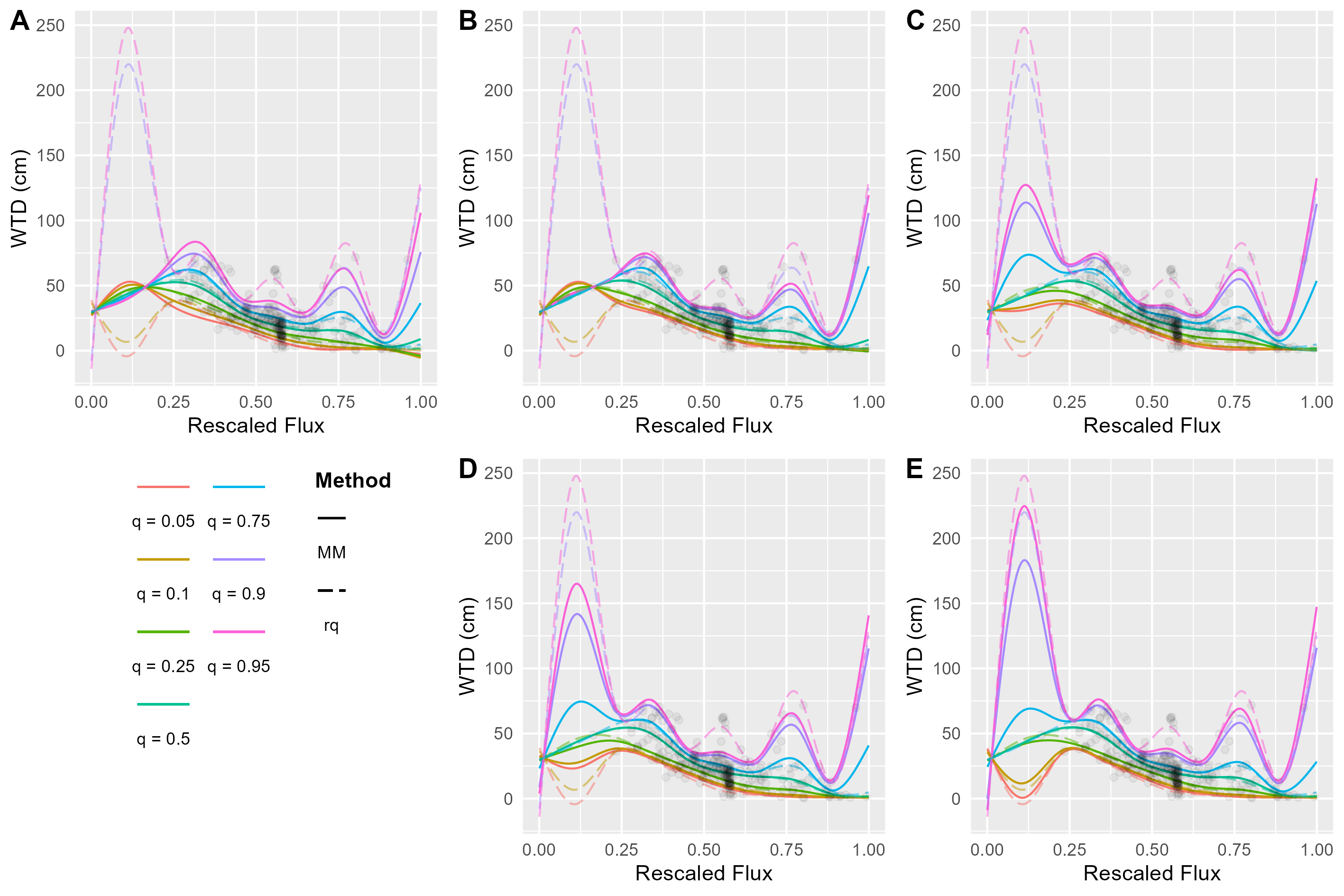}	
	\caption{Fitted quantile curves for the ``WTD" data set - natural spline transformation on $x$ with 8 equally spaced knots. We have adopted the logistic basis (\textbf{A}) and natural spline bases with 3 (\textbf{B}), 4 (\textbf{C}), 5 (\textbf{D}), and 6 (\textbf{E}) equally spaced knots for quantile levels.}
	\label{WTD quantile curves Xspline8}
\end{figure}

 We conducted our MM algorithms on the two data sets with diverse combinations of explanatory variable knots and quantile order bases. MM did appear to smooth out \texttt{rq} empirical estimates reasonably well in all quantile plots of $\hat{y}(q|x)=\hat Q_q(Y|x)$ versus $q$ that take $x$ as the sample mean (\Cref{triceps mean quantile plots,WTD mean quantile plots}). Since the two data sets are bivariate, we can also efficiently visualize MM's prediction accuracy and smoothing effects via estimated conditional quantile curves. As shown in \Cref{triceps quantile curves Xspline3,triceps quantile curves Xspline4,triceps quantile curves Xspline5,triceps quantile curves Xspline6,triceps quantile curves Xspline7,triceps quantile curves Xspline8} and \Cref{WTD quantile curves Xspline3,WTD quantile curves Xspline4,WTD quantile curves Xspline5,WTD quantile curves Xspline6,WTD quantile curves Xspline7,WTD quantile curves Xspline8} plotting $\hat y(x|q)$ against the concentrated $x$ values at selected representative quantile levels $q=0.05, 0.1, 0.25, 0.5, 0.75, 0.9, 0.95$, our MM algorithm did a good job in efficiently capturing the scatters' trends and smoothing the empirical estimates produced by \texttt{rq} for all combination choices.
\begin{table}[ph!]
	\centering
	\scalebox{0.9}{
		\centering
		\begin{tabular}{|c|ccccc|}
			\hline 
			\multicolumn{6}{|c|}{\textbf{``triceps"}}\\
			\hline
			\backslashbox{$x$}{$q$}& logistic & ns seq3 & ns seq4 &  ns seq5 &  ns seq6 \\ 
			\hline
			seq3 & 760.3107 & 761.9058 & 761.987 & 761.532 & 761.464 \\ 
			seq4 & 744.8911 & 746.7616 & 746.375 & 746.720 & 747.208 \\
			seq5 & 747.9345 & 749.9778 & 748.7668 & 749.5624 & 749.6911 \\  
			seq6 & 747.5314 & 749.8461 & 748.1148 & 748.9028 & 748.8585 \\ 
			seq7 & 741.6419 & 744.0620 & 742.6189 & 743.4991 & 743.1266 \\
			seq8 & 739.5723 & 741.6366 & 740.7900 & 741.5891 & 741.4301 \\   
			\hline
			\multicolumn{6}{|c|}{\textbf{``WTD"}}\\
			\hline
			\backslashbox{$x$}{$q$}& logistic & ns seq3 & ns seq4 &  ns seq5 &  ns seq6 \\ 
			\hline
			seq3 & 9318.931 & 9323.063 & 9367.188 & 9379.859 & 9387.506 \\ 
			seq4 & 9629.874 & 9638.909 & 9598.047 & 9620.840 & 9596.841 \\
			seq5 & 9427.951 & 9435.204 & 9558.181 & 9412.183 & 9489.295 \\  
			seq6 & 10461.212 & 10485.596 & 10531.684 & 10472.559 & 10460.843 \\ 
			asym7 & 9348.860 & 9307.616 & 9354.614 & 9378.879 & 9393.399 \\
			\hline
		\end{tabular}
	}
	\caption{\footnotesize CV loss (\Cref{CV loss}) of the ``triceps" and ``WTD" data sets for selected MM $x,q$ settings. For natural spline transformations on $x$, ``seq$h$" $(h\in \mathbb{N})$ denotes putting $h$ equally spaced knots between the minimum and maximum values of observed $x$'s and ``asym7" for ``WTD" represents the knots $(0.2, 0.4, 0.5, 0.55, 0.6, 0.7, 0.9)$. For quantile order $q$ bases, ``logistic" and ``ns seq$h$" $(h\in \mathbb{N})$ denote the logistic basis and natural spline bases with $h$ equally spaced knots. We used 999 quantile levels $0.001,0.002,\dots,0.999$ and $N=1000$ iterations to fit each \texttt{qrcmMMxspline} model in this chart.}
	\label{CV chart}
\end{table}
\par To gain a better picture of the relevant out-of-sample prediction powers for different covariate knots and quantile bases choices, we performed 10-fold cross-validation (CV) by randomly splitting the data sets into 10 approximately equal-sized subgroups $G_1,G_2,\dots,G_{10}\subset \{1,2,\dots,n\}$, picking $J=9$ validation quantiles $q_1=0.1,q_2=0.2,\ldots,q_J=0.9$, and calculating the loss
\begin{equation}\label{CV loss}
	l_{\text{CV}}=\sum_{g=1}^{10} \sum_{i\in G_g} \sum_{j=1}^{J} \rho_{q_j}\left(y_i-\bm{x}_i^T\hat{\bm{\beta}}^{(-G_g)}(q_j)\right),
\end{equation}
where each $\bm{x}_i$ denotes the $i^{\text{th}}$ row of $\bm{X}_{\text{new}}$ in \Cref{transform X} and each $\hat{\bm{\beta}}^{(-G_g)}(q_j)$ represents the estimated coefficient vector based on the other 9 subgroups for spline-transformed covariates at quantile level $q_j$. The smaller $l_{\text{CV}}$, the better the model. As \Cref{CV chart} suggests, basis choice for $q$ does not seem to matter as much as ns knots choice to transform $x$ does. 



\newpage 	 
\subsection{Double Kernel and Likelihood Cross Validation}\label{appenC2}
\par
\begin{figure}[ph!]
	\centering
	\includegraphics[width=\textwidth]{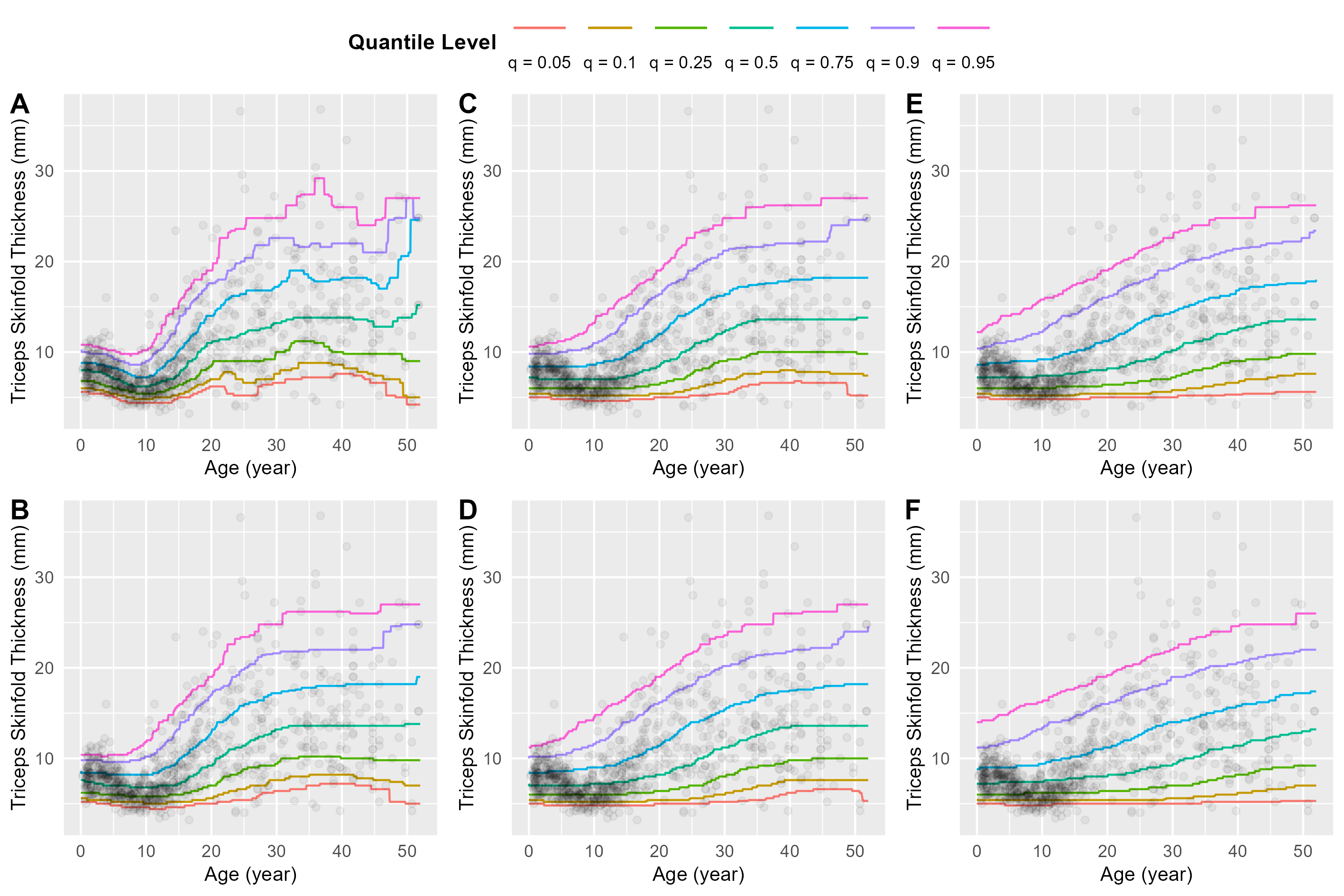}
	\caption{Quantile curves for the ``triceps" data set estimated by Double Kernel with $h_1=2.5$ (\textbf{A}), $5$ (\textbf{B}), $7.5$ (\textbf{C}), $10$ (\textbf{D}), $12.5$ (\textbf{E}), $15$ (\textbf{F}) and $h_2=10^{-4}$.}
	\label{triceps DK quantile curves}
\end{figure}

\begin{figure}[ph!]
	\centering
	\includegraphics[width=\textwidth]{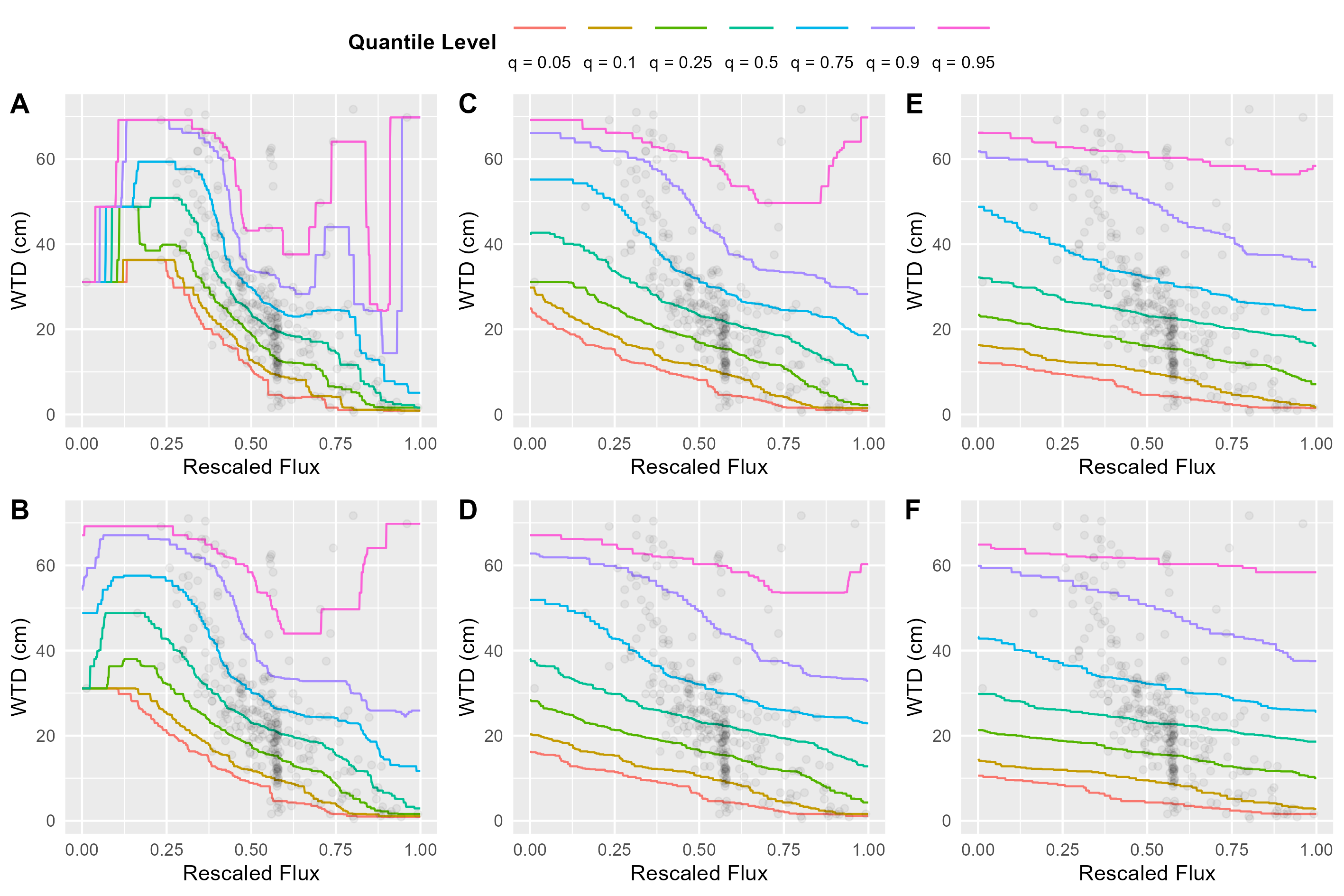}
	\caption{Quantile curves for the ``WTD" data set estimated by Double Kernel with $h_1=0.05$ (\textbf{A}), $0.1$ (\textbf{B}), $0.15$  (\textbf{C}), $0.2$ (\textbf{D}), $0.25$ (\textbf{E}), $0.3$ (\textbf{F}) and $h_2=10^{-4}.$}
	\label{WTD DK quantile curves}
\end{figure}

 To see how well our MM algorithms perform compared to existing nonparametric methods, we adopted widely-used Kernel Estimation techniques on the two bivariate real data sets to predict $\hat y(x|q)$ at $q=0.05,0.1,0.25,0.5,0.75,0.9,0.95.$ Given a finite sample with observed points $(x_i,y_i)$, $i=1,2,\dots,n$ and a new point with known $x$, the estimated conditional CDF of this point's $y$ coordinate is of the form
\begin{align}\label{SK F}
	\hat{F}(y|x) &= \sum_{i=1}^n w_i(x,h_1)\cdot I(y_i\leq y),\text{  where}\\
	w_i(x,h_1) &= \frac{K\left(\frac{x_i-x}{h_1} \right) }{\sum_{j=1}^n K\left(\frac{x_j-x}{h_1} \right)}.\nonumber
\end{align} 
To smooth our estimation in the $y$ direction, we replaced the deterministic form $I(y_i\leq y)$ in \Cref{SK F} by a standard normal CDF $N(y_i,h_2)=\Phi\left(\frac{y-y_i}{h_2}\right)$ and formed a double kernel conditional CDF estimate
\begin{equation}\label{DK F}
	\hat{\hat F}(y|x) = \sum_{i=1}^n w_i(x,h_1)\cdot\Phi\left(\frac{y-y_i}{h_2}\right) = \sum_{i=1}^n \left[ \frac{K\left(\frac{x_i-x}{h_1} \right) }{\sum_{j=1}^n K\left(\frac{x_j-x}{h_1} \right)}\cdot\Phi\left(\frac{y-y_i}{h_2}\right)\right] .
\end{equation}
The estimated conditional quantile function is thus
\begin{equation}\label{DK Q}
	\hat{\hat Q}_q(y|x)=\hat{\hat F}^{-1}(q|x),\:q\in(0,1).
\end{equation}
We adopted the normal kernel $\phi$ as the kernel function $K$ in \Cref{DK F} in light of the normal kernel's various favorable mathematical properties and proceeded to produce quantile curves of $\hat y(x|q)$ against $x$ at each $q\in \{0.05,0.1,0.25,0.5,0.75,0.9,0.95\}$ after specifying a set of sensible choices for the bandwidths $h_1,h_2$. We discovered and confirmed by Likelihood cross-validation (LCV) that different $h_2$ values generated estimates that are extremely alike, if not identical. As such, we present quantile curves with a set of $h_1$ choices and a fixed adequate $h_2=10^{-4}$ in \Cref{triceps DK quantile curves,WTD DK quantile curves}. The figures suggest that $h_1=10,12.5$ for ``triceps" and $h_1=0.15,0.2$ for ``WTD" most effectively and smoothly capture the scatters' trends. 
\par To decide optimal bandwidth choices for Double Kernel Estimation, we carried out leave-one-out likelihood cross-validation to quantitatively judge different bandwidths' out-of-sample prediction capabilities.
As the estimated conditional PDF is given by
\begin{equation}\label{DK f}
	\hat{\hat f}(y|x) = \sum_{i=1}^n w_i(x,h_1)\cdot\phi\left(\frac{y-y_i}{h_2}\right),
\end{equation}
a leave-one-out LCV likelihood can be expressed as 
\begin{equation}\label{LCV likelihood}
	\mathcal{L}_{\text{LCV}}=\prod_{j=1}^{n} \hat {\hat f}^{(-j)}(y_j|x_j),
\end{equation}
where $\hat {\hat f}^{(-j)}(y_j|x_j)$ denotes the estimated PDF for a new observation with $x=x_j$ derived using the entire sample except the point $(x_j,y_j)$. The larger $\mathcal{L}_{\text{LCV}}$, the better the method. 


As \Cref{LCV chart} suggests, $h_1\text{ approximately }\in (9,13)$ for ``triceps", whose sample $x\in(0,52)$, and $h_1\text{ approximately }\in (0.12,0.22)$ for ``WTD", whose sample $x\in(0,1)$, 
would be effective $x$ kernel bandwidth choices, corresponding well to what the quantile curves in \Cref{triceps DK quantile curves,WTD DK quantile curves} indicate. The values of $h_2$, on the other hand, do not seem to affect Double Kernel Estimation's prediction power much as long as $h_2$ is appropriately small. 
\begin{table}[h]
	\centering
	\scalebox{0.8}{
		\begin{tabular}{|c|ccc|c|ccc|}
			\hline 
			\multicolumn{4}{|c|}{\textbf{``triceps"}}&\multicolumn{4}{|c|}{\textbf{``WTD"}}\\[1.5mm]
			\hline
			\backslashbox{$h_1$}{$h_2$}& $10^{-4}$ & $10^{-5}$ & $10^{-6}$ & \backslashbox{$h_1$}{$h_2$}& $10^{-3}$ & $10^{-4}$ & $10^{-5}$\\ 
			\hline
			0.5 & 6.04978E-277 & 6.04978E-277 & 6.04978E-277 & 0.01 & 6.41081E-223 & 6.41081E-223 & 6.41081E-223\\ 
			2.5 & 3.44698E+53 & 3.44698E+53 & 3.44698E+53 & 0.05 & 2.19999E-102 & 2.19999E-102 & 2.19999E-102\\ 
			5 & 1.48980E+111 & 1.48980E+111 & 1.48980E+111 & 0.1 & 3.51314E-75 & 3.51314E-75 & 3.51314E-75 \\ 
			7.5 & 1.67443E+130 & 1.67443E+130 & 1.67443E+130 & 0.15 & 9.16157E-73 & 9.16157E-73 & 9.16157E-73 \\ 
			10 & 6.95286E+136 & 6.95286E+136 & 6.95286E+136 & 0.2 & 1.14691E-74 & 1.14691E-74 & 1.14691E-74\\ 
			12.5 & 8.20678E+135 & 8.20678E+135 & 8.20678E+135 & 0.25 & 1.20404E-76 & 1.20404E-76 & 1.20404E-76 \\ 
			15 & 3.26345E+130 & 3.26345E+130 & 3.26345E+130 & 0.3 & 3.63631E-78 & 3.63631E-78 & 3.63631E-78 \\ 
			\hline
		\end{tabular}
	}
	\caption{Scaled LCV likelihood (\Cref{LCV likelihood}) calculated using some suitable bandwidth choices for the ``triceps" and ``WTD" data sets. We have scaled the likelihood product in \Cref{LCV likelihood} to make sure that all resulting values can be expressed in R. Hence, it is the relative magnitude rather than the absolute magnitude that counts here.}
	\label{LCV chart}
\end{table}

\subsection{Simulation Studies}\label{appenC3}
\par The quantile curves produced by Double Kernel seem to appear ``rougher" than those obtained from MM. To more convincingly and objectively compare the two approaches, we resort to simulation studies again. We considered the three-parameter generalized gamma (GG) distribution, as it incorporates many common fewer-parameter distributions (e.g., Gamma, Weibull, Exponential distributions).
\par A generalized gamma random variable Y with parameters $\theta,\beta,k$, or $\mu,\sigma,k$, where 
\begin{equation}\label{theta beta mu sigma}
	\begin{cases}
		\mu=\ln(\theta)+\frac{\ln(k)}{\beta},\qquad \sigma=\frac{1}{\beta \sqrt{k}}\\
		\theta=\frac{e^{\mu}}{k^{\sigma \sqrt{k}}},\qquad \qquad \beta=\frac{1}{\sigma \sqrt{k}}
	\end{cases},
\end{equation}
has probability density function 
\begin{equation}\label{f GG}
	f(y;\theta,\beta,k)=\frac{\beta}{\theta^{k\beta}\Gamma(k)}y^{k\beta-1}\exp\left\lbrace-\left(\frac{y}{\theta}\right)^{\beta}\right\rbrace,\qquad y>0
\end{equation}
and quantile function 
\begin{equation}\label{Q GG}
	Q(q;\mu,\sigma,k)=e^{\mu}\left\lbrace \frac{r(q;k)}{k}\right\rbrace ^{\sigma \sqrt{k}},\qquad q\in(0,1),
\end{equation}
where $r(q;k)$ is the quantile function of the $\Ga(\theta=1,k)$, or $\GG(\theta=1,\beta=1,k)$ distribution. It is virtually the $\frac{1}{\beta}^{\text{th}}$ power of a $\Ga\left( \theta^{\beta},k\right)$ random variable and can be generated either by substituting a $\unif(0,1)$ random variable as $q$ into \Cref{Q GG} or by generating a $\Ga\left( \theta^{\beta},k\right)$ random variable and taking the $\frac{1}{\beta}^{\text{th}}$ power of it. For implementation in software like R, the first method is preferred, as the second may fail for some parameter choices causing extreme values.
\par We modeled the parameters $\mu,\sigma,k$ as suggested by \textcite{Noufaily2013} -- assume $\mu(x)=a+bx$, $\sigma(x)=e^{c+dx}$, $k(x)=e^{f+gx}$, where $\{a,b,c,d,f,g\}$ is the new set of parameters. 
For each chosen parameter set $\{a,b,c,d,f,g\}$, we formed $N$ bivariate samples of size $n$ by 
\begin{enumerate}
	\item Sample $x_1,x_2,\dots,x_n$ independently from a uniform distribution;
	\item for all $i\in\{1,2,\dots,n\},$ generate $y_i$ from $\GG(\mu(x_i),\sigma(x_i),k(x_i))$ using one of the two methods described in the previous paragraph.
\end{enumerate} 
We then applied to each sample MM/Double Kernel with a variety of covariate knots and quantile order bases/bandwidths choices. Integrated Mean Squared Error (IMSE), given in \Cref{imse}, was calculated using the theoretical quantile function \Cref{Q GG} at $q=0.1,0.25,0.5,0.75,0.9$ for each approach. Estimated quantile curves based on the mean conditional predictions $\hat{y}(x|q)=\hat Q_q(y|x)$ were also made to compare with the theoretical ones.

\begin{figure}[ph!]
	\centering
	\includegraphics[width=\textwidth]{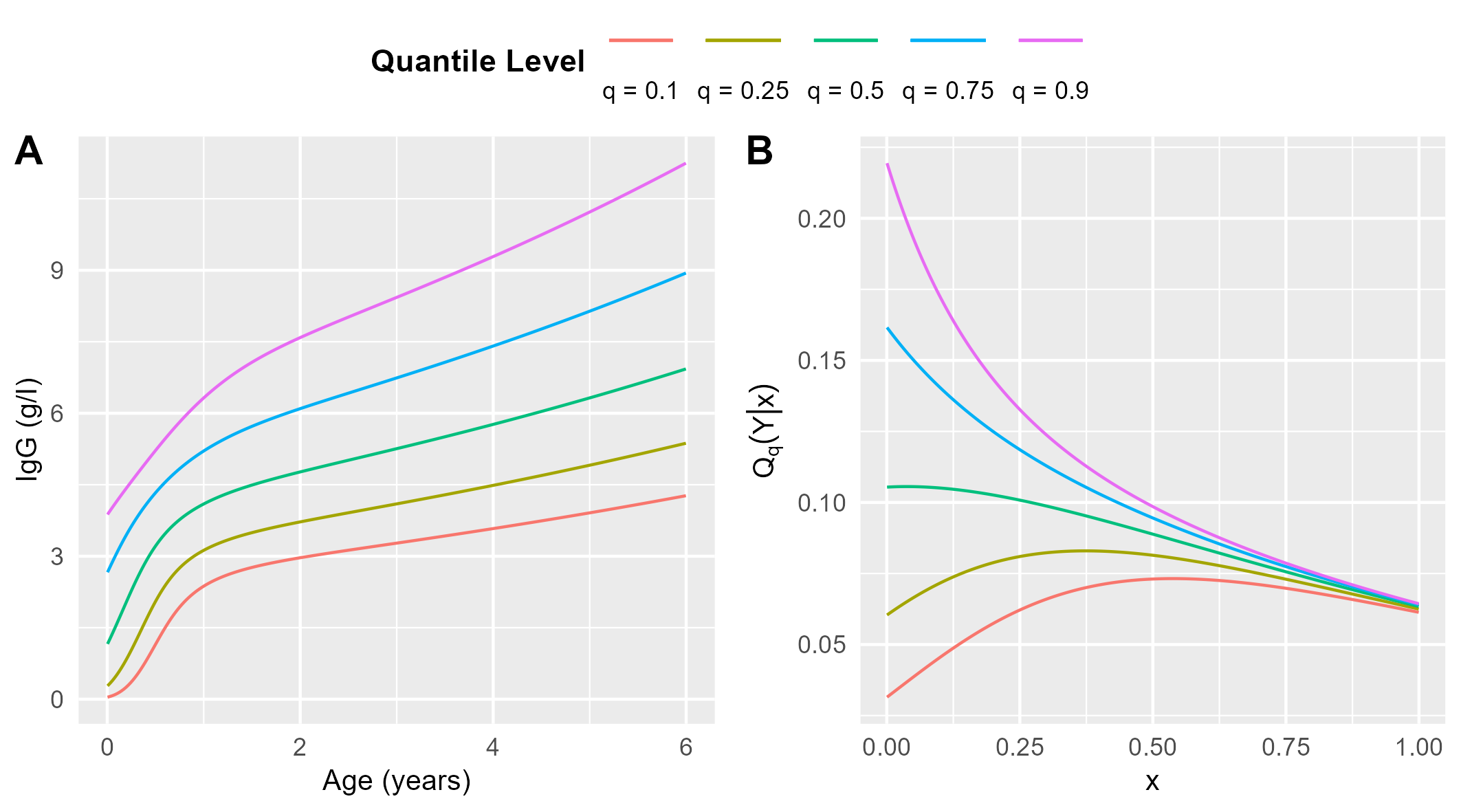}		
	\caption{Theoretical quantile curves at $q=0.1,0.25,0.5,0.75,0.9$ for generalized gamma distributions with parameters specified in \Cref{GGb} (\textbf{A}) and \Cref{GGd} (\textbf{B}).}
	\label{GG theoretical quantile curves}
\end{figure}

\par We tried out two sets of parameter choices for $\{a,b,c,d,f,g\}$
\begin{enumerate}\label{GG parameters}
	\item $a=1.384,b=0.092,c=-1.021,d=0.008,f=-3.493,g=4.766$;\label{GGb}
	\item $a = -2, b = -0.75, c = -0.5, d = -4, f = -0.2, g = -1$.\label{GGd}
\end{enumerate}
The former consists of approximate parameter estimates obtained by \textcite{Noufaily2013} when fitting generalized gamma distributions to an ``IgG" data set \parencite{Isaacs1983} measuring concentrations of serum immunoglobulin G for 298 kids under six, and the latter was argued by \textcite{Noufaily2013} to produce a particular interesting GG distribution scenario. Theoretical quantile curves at $q=0.1,0.25,0.5,0.75,0.9$ for the two generalized gamma distributions are displayed in \Cref{GG theoretical quantile curves}.

\par
\begin{figure}[ph!]
	\centering
	\includegraphics[width = 1\textwidth]{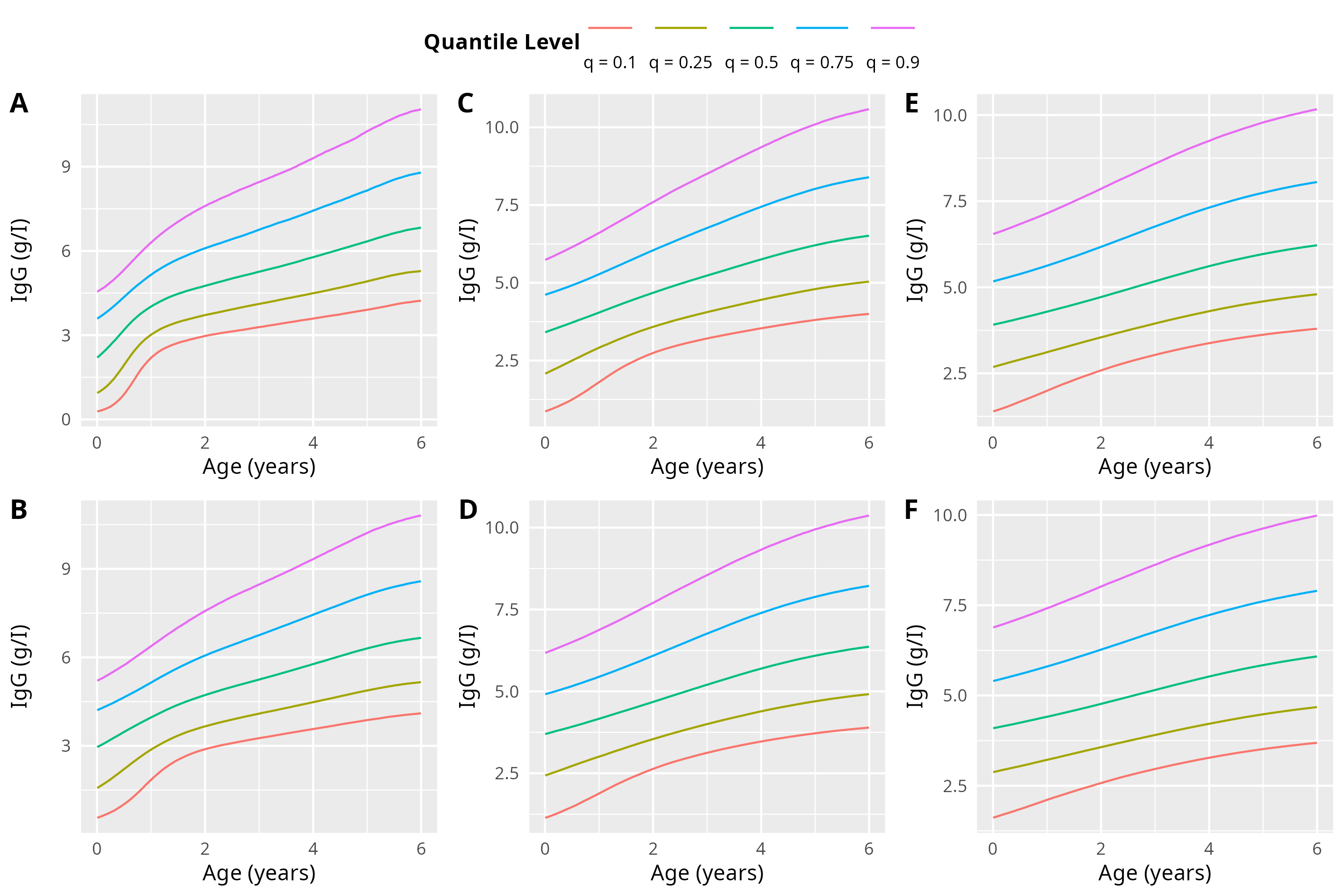}
	\caption{Mean quantile curves estimated by Double Kernel with $h_2=10^{-4}$ and $h_1=0.3$ (\textbf{A}), $0.6$ (\textbf{B}), $0.9$ (\textbf{C}), $1.2$ (\textbf{D}), $1.5$ (\textbf{E}), $1.8$ (\textbf{F}) for simulated GG data from \Cref{GGb} (\Cref{GG theoretical quantile curves}\textbf{A}). }
	\label{IgG DK estimated quantile curves}
\end{figure}

\begin{figure}[ph!]
	\centering
	\includegraphics[width = 0.96\textwidth]{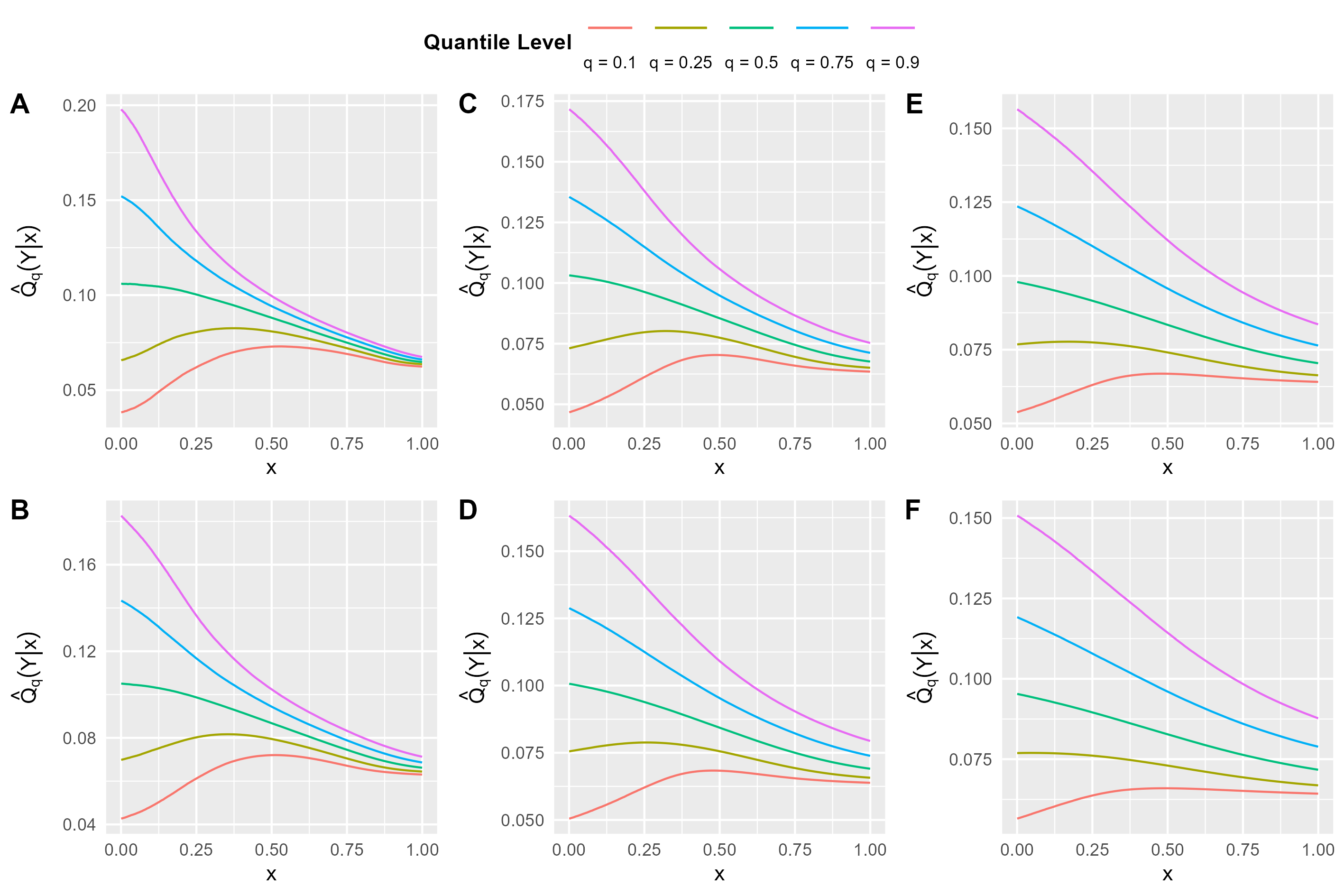}
	\caption{Mean quantile curves estimated by Double Kernel with $h_2=10^{-4}$ and $h_1=0.05$\\(\textbf{A}), $0.1$ (\textbf{B}), $0.15$ (\textbf{C}), $0.2$ (\textbf{D}), $0.25$ (\textbf{E}), $0.3$ (\textbf{F}) for simulated GG data from \Cref{GGd} (\Cref{GG theoretical quantile curves}\textbf{B}). }
	\label{D62 DK estimated quantile curves}
\end{figure}

\begin{figure}
	\centering
	\includegraphics[width = 0.96 \textwidth]{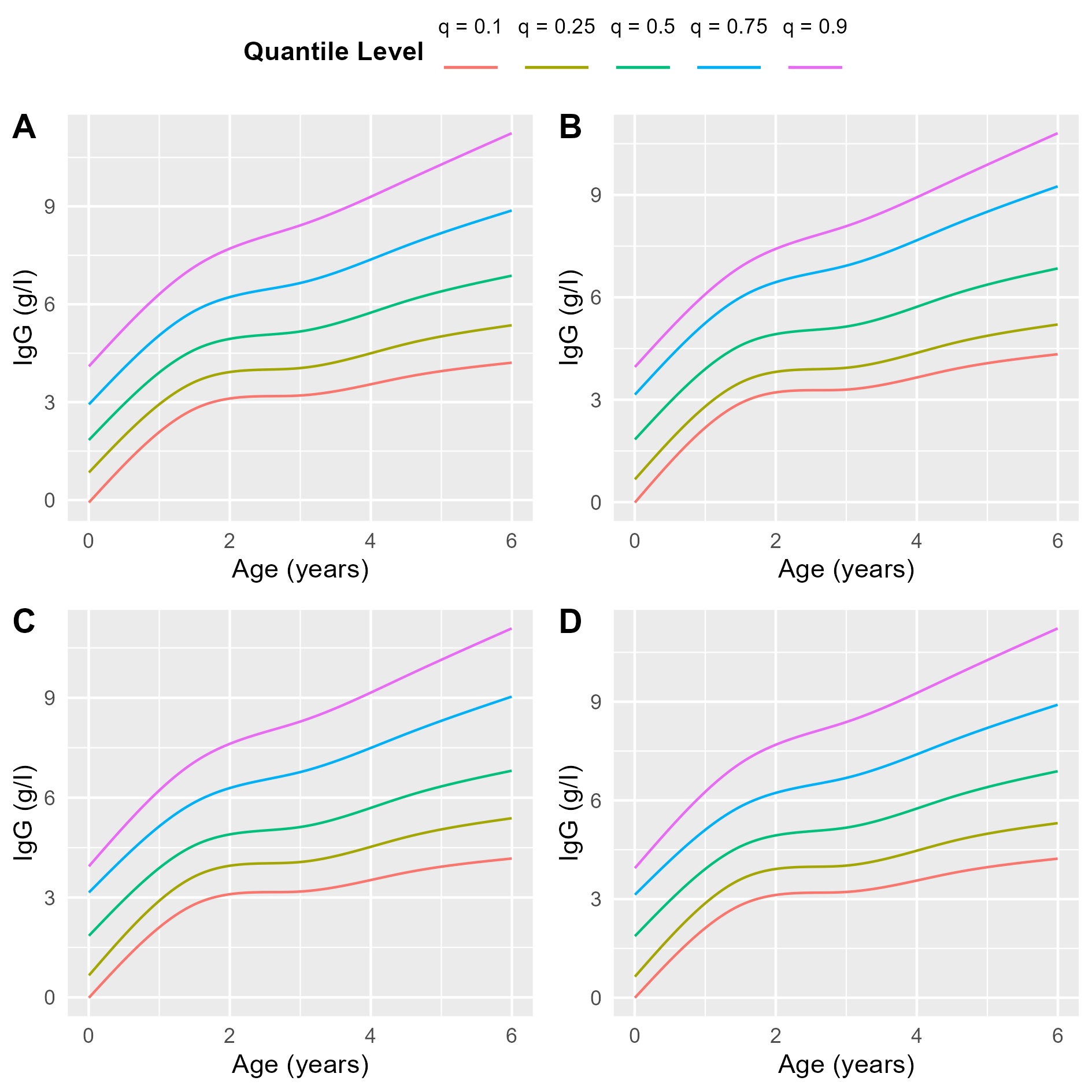}
	\caption{Mean quantile curves estimated by MM with logistic (\textbf{A}) and natural spline (ns) with 3 (\textbf{B}), 4 (\textbf{C}), 5 (\textbf{D}) equally spaced knots quantile level bases for simulated GG data from \Cref{GGb} (\Cref{GG theoretical quantile curves}\textbf{A}). We used 3 equally spaced knots to natural-spline-transform $x$.}
	\label{IgG MM xseq3 estimated quantile curves}
\end{figure}

\begin{figure}
	\centering
	\includegraphics[width = \textwidth]{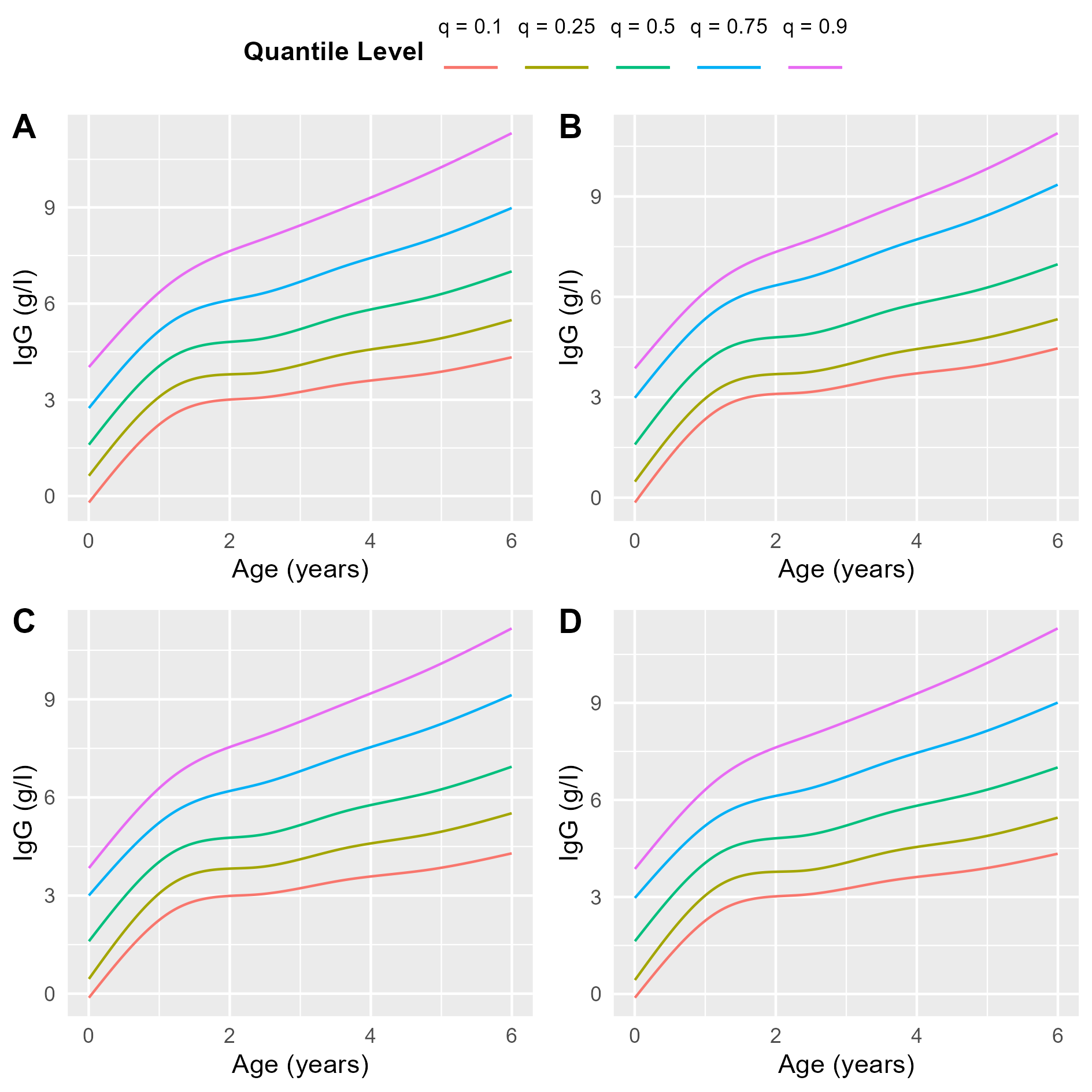}
	\caption{Mean quantile curves estimated by MM with logistic (\textbf{A}) and natural spline (ns) with 3 (\textbf{B}), 4 (\textbf{C}), 5 (\textbf{D}) equally spaced knots quantile level bases for simulated GG data from \Cref{GGb} (\Cref{GG theoretical quantile curves}\textbf{A}). We used 4 equally spaced knots to natural-spline-transform $x$.}
	\label{IgG MM xseq4 estimated quantile curves}
\end{figure}

\begin{figure}[ph!]
	\centering
	\includegraphics[width = \textwidth]{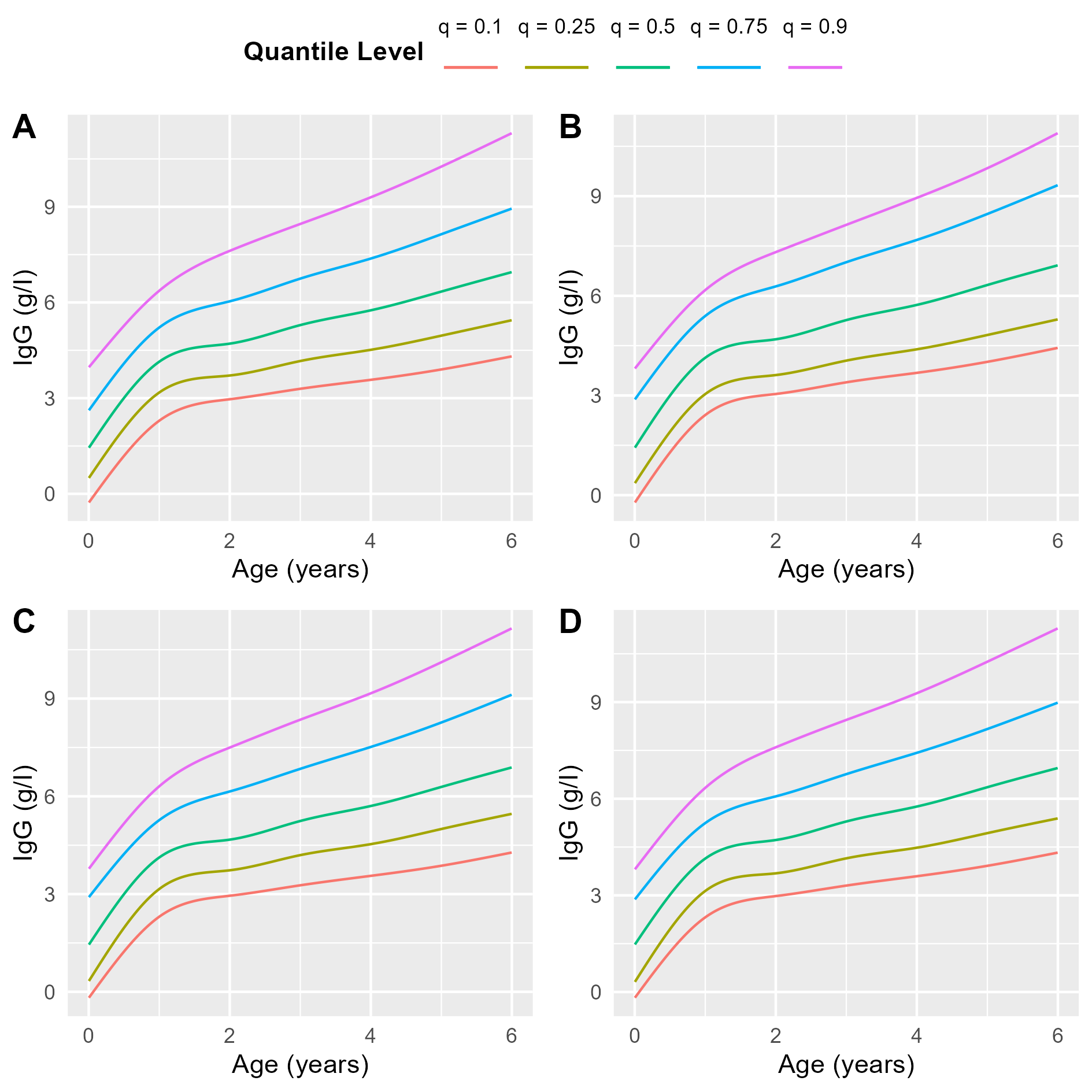}
	\caption{Mean quantile curves estimated by MM with logistic (\textbf{A}) and natural spline (ns) with 3 (\textbf{B}), 4 (\textbf{C}), 5 (\textbf{D}) equally spaced knots quantile level bases for simulated GG data from \Cref{GGb} (\Cref{GG theoretical quantile curves}\textbf{A}). We used 5 equally spaced knots to natural-spline-transform $x$.}
	\label{IgG MM xseq5 estimated quantile curves}
\end{figure}

\begin{figure}[ph!]
	\centering
	\includegraphics[width = \textwidth]{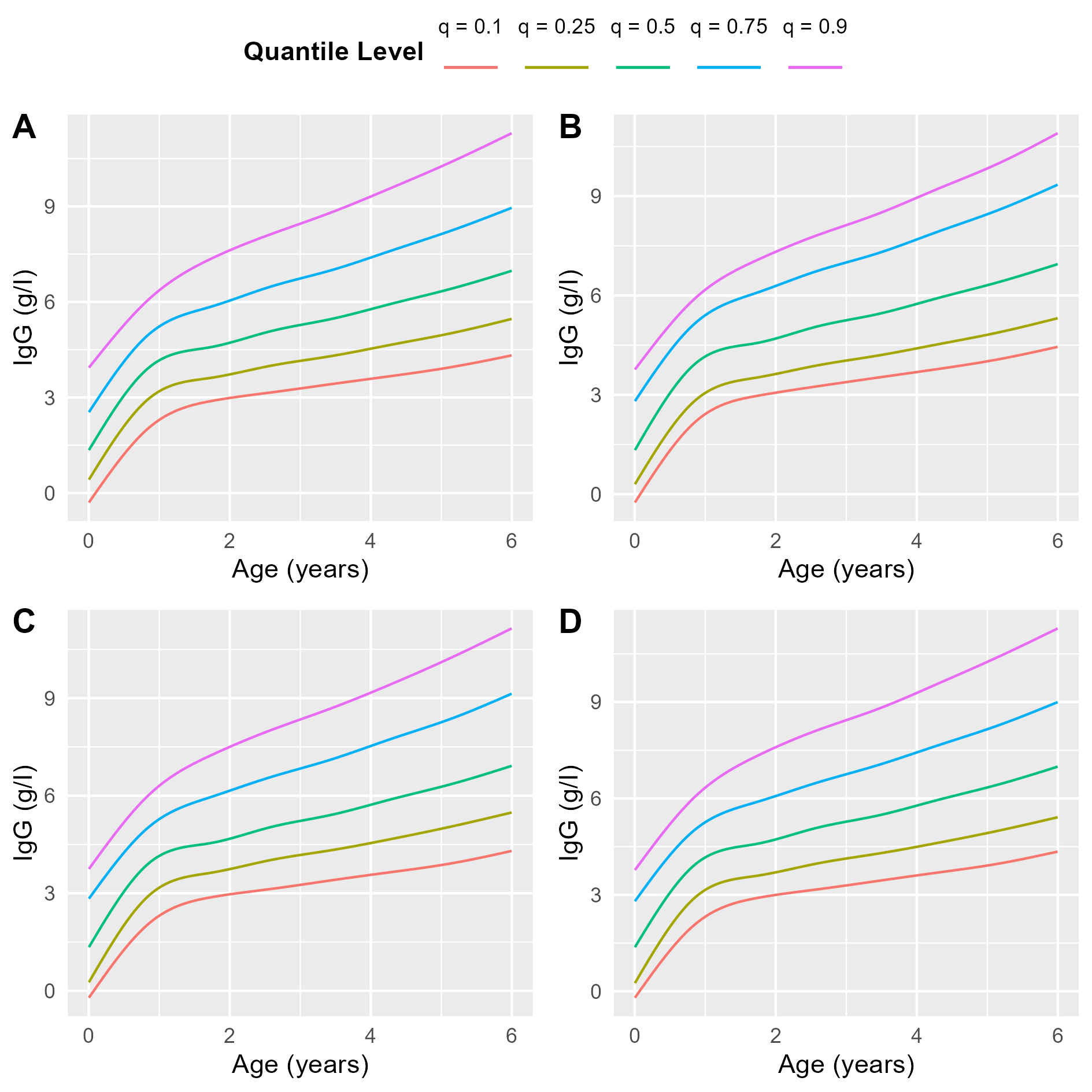}
	\caption{Mean quantile curves estimated by MM with logistic (\textbf{A}) and natural spline (ns) with 3 (\textbf{B}), 4 (\textbf{C}), 5 (\textbf{D}) equally spaced knots quantile level bases for simulated GG data from \Cref{GGb} (\Cref{GG theoretical quantile curves}\textbf{A}). We used 6 equally spaced knots to natural-spline-transform $x$.}
	\label{IgG MM xseq6 estimated quantile curves}
\end{figure}

\begin{figure}[ph!]
	\centering
	\includegraphics[width = \textwidth]{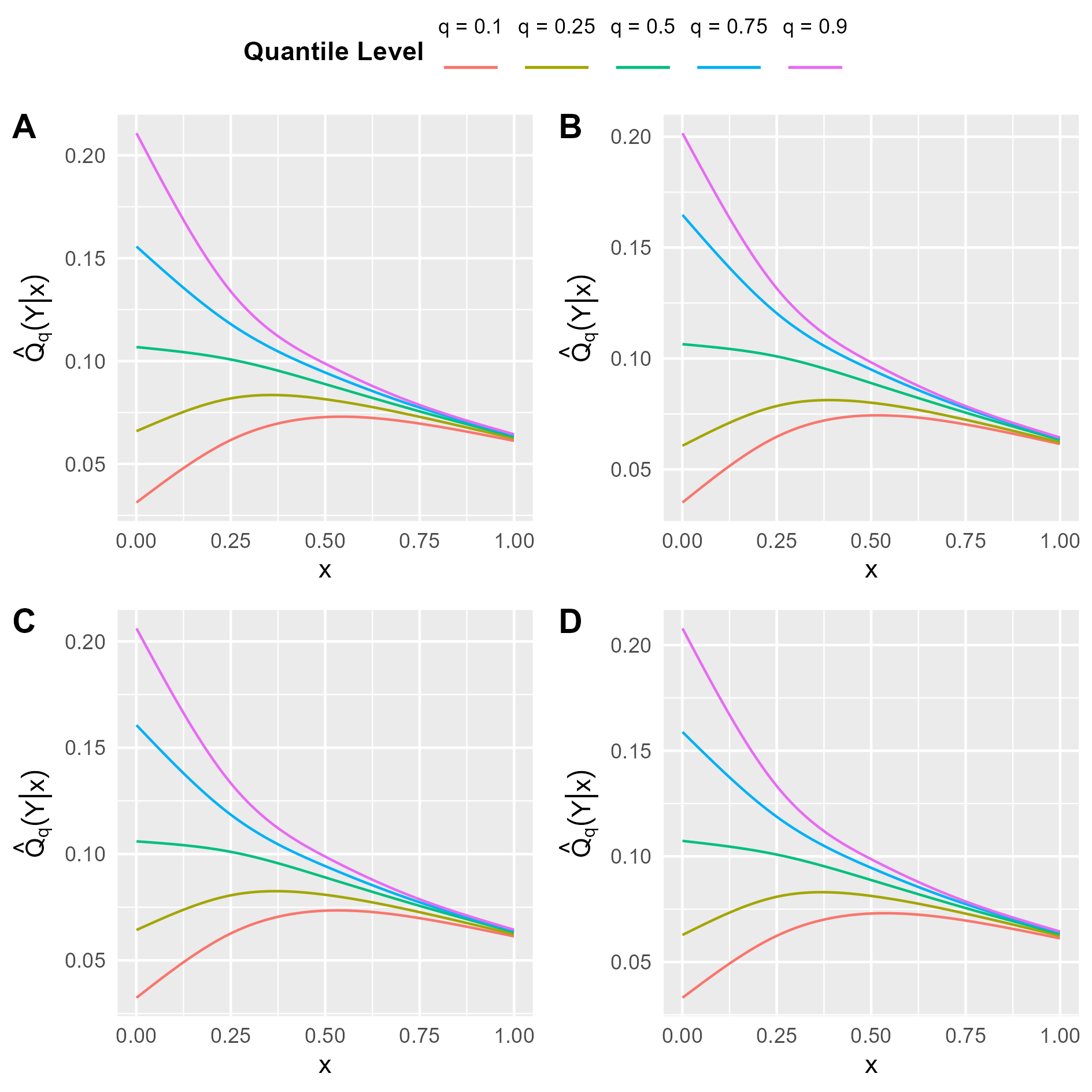}
	\caption{Mean quantile curves estimated by MM with logistic (\textbf{A}) and natural spline (ns) with 3 (\textbf{B}), 4 (\textbf{C}), 5 (\textbf{D}) equally spaced knots quantile level bases for simulated GG data from \Cref{GGd} (\Cref{GG theoretical quantile curves}\textbf{B}). We used 3 equally spaced knots to natural-spline-transform $x$.}
	\label{D62 MM xseq3 estimated quantile curves}
\end{figure}

\begin{figure}[ph!]
	\centering
	\includegraphics[width = \textwidth]{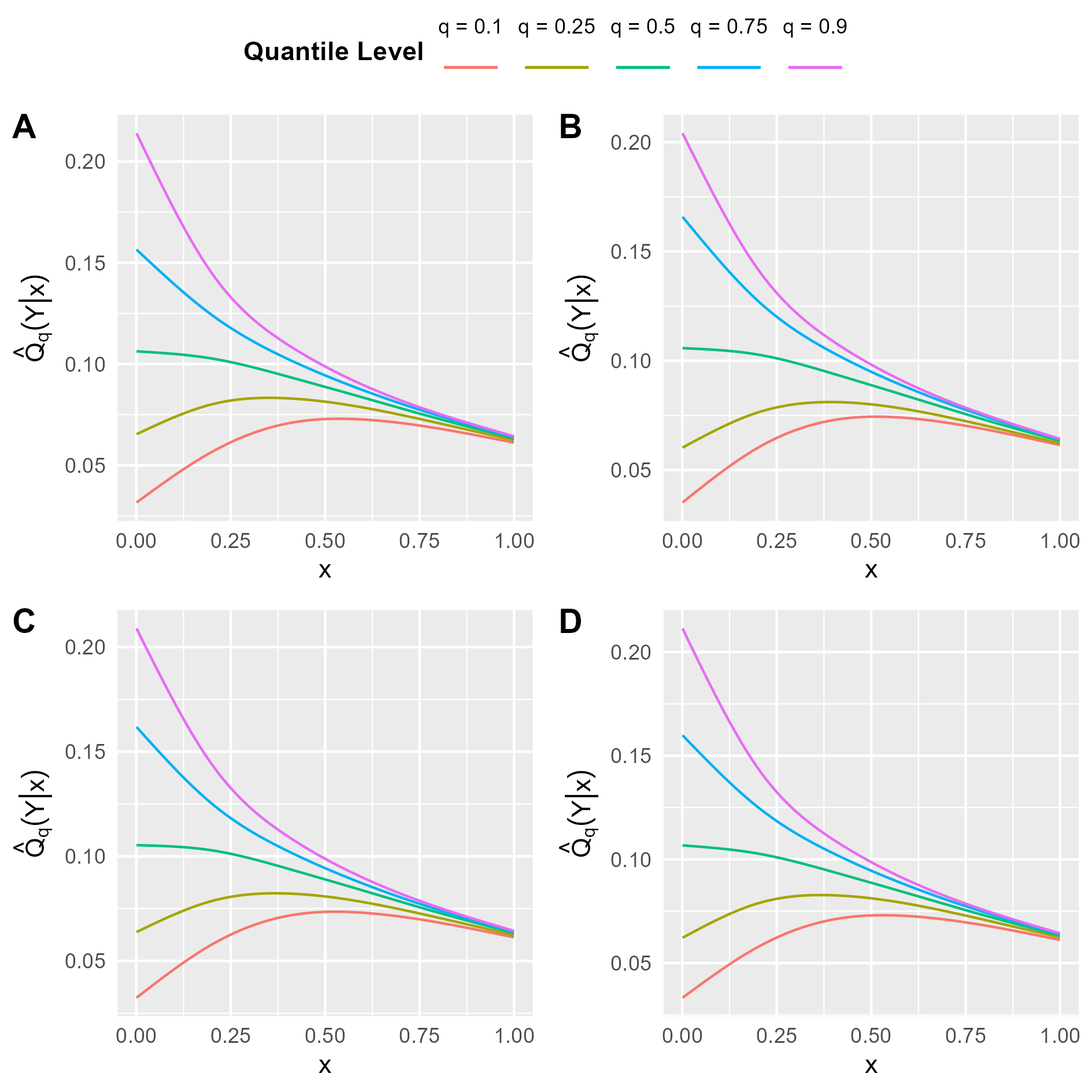}
	\caption{Mean quantile curves estimated by MM with logistic (\textbf{A}) and natural spline (ns) with 3 (\textbf{B}), 4 (\textbf{C}), 5 (\textbf{D}) equally spaced knots quantile level bases for simulated GG data from \Cref{GGd} (\Cref{GG theoretical quantile curves}\textbf{B}). We used 4 equally spaced knots to natural-spline-transform $x$.}
	\label{D62 MM xseq4 estimated quantile curves}
\end{figure}

\begin{figure}[ph!]
	\centering
	\includegraphics[width = \textwidth]{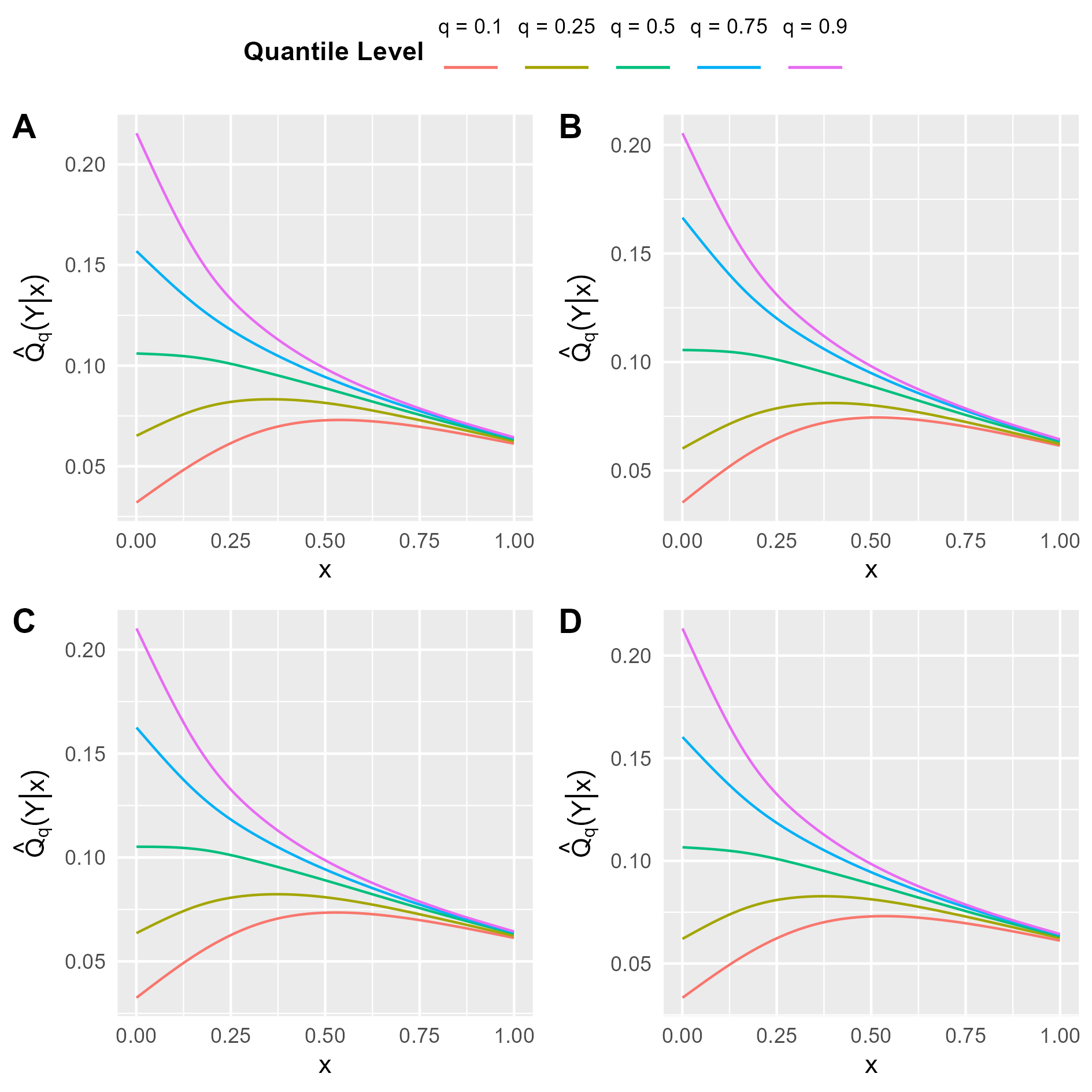}
	\caption{Mean quantile curves estimated by MM with logistic (\textbf{A}) and natural spline (ns) with 3 (\textbf{B}), 4 (\textbf{C}), 5 (\textbf{D}) equally spaced knots quantile level bases for simulated GG data from \Cref{GGd} (\Cref{GG theoretical quantile curves}\textbf{B}). We used 5 equally spaced knots to natural-spline-transform $x$.}
	\label{D62 MM xseq5 estimated quantile curves}
\end{figure}

\begin{figure}[ph!]
	\centering
	\includegraphics[width = \textwidth]{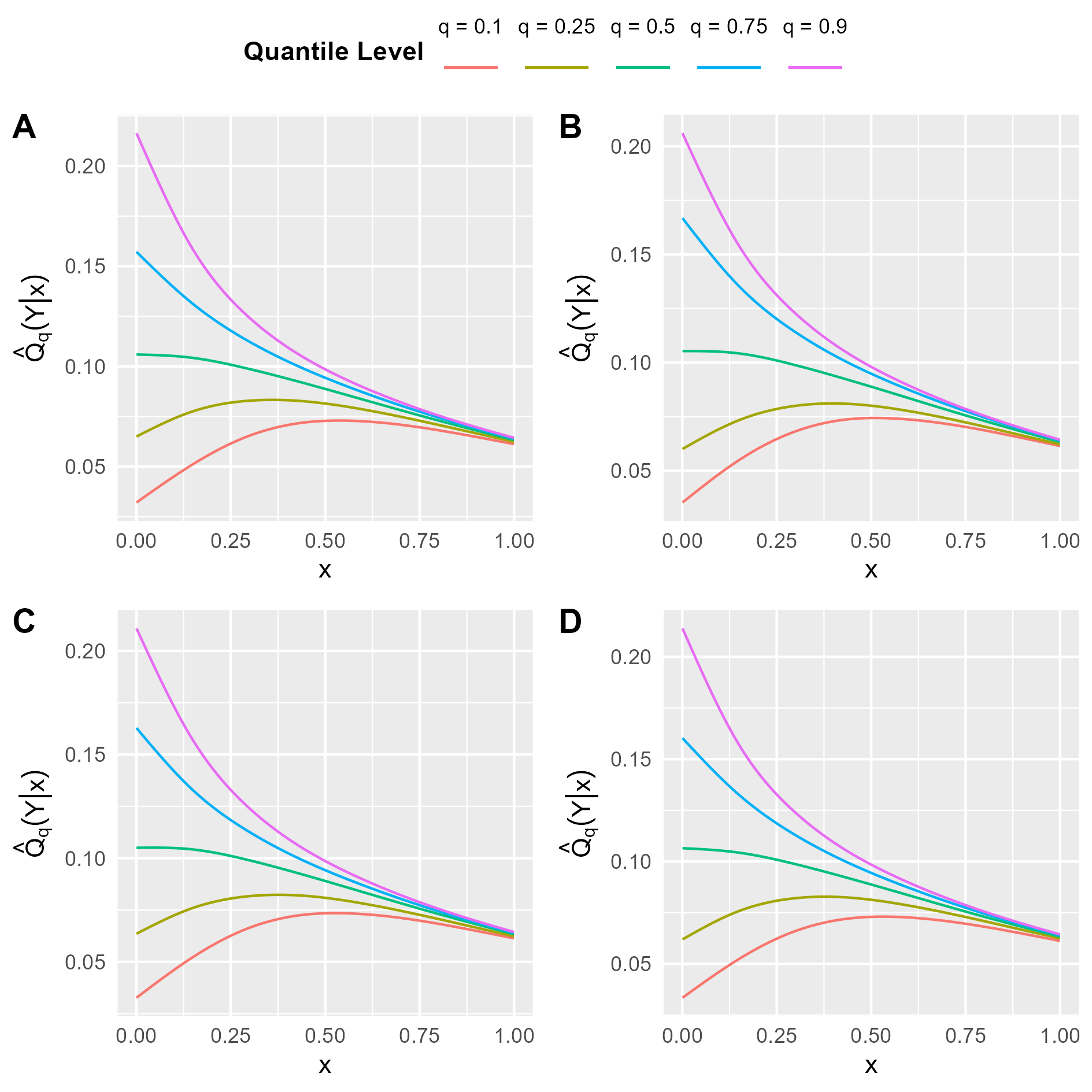}
	\caption{Mean quantile curves estimated by MM with logistic (\textbf{A}) and natural spline (ns) with 3 (\textbf{B}), 4 (\textbf{C}), 5 (\textbf{D}) equally spaced knots quantile level bases for simulated GG data from \Cref{GGd} (\Cref{GG theoretical quantile curves}\textbf{B}). We used 6 equally spaced knots to natural-spline-transform $x$.}
	\label{D62 MM xseq6 estimated quantile curves}
\end{figure}

\begin{figure}[ph!]
	\centering
	\includegraphics[width = \textwidth]{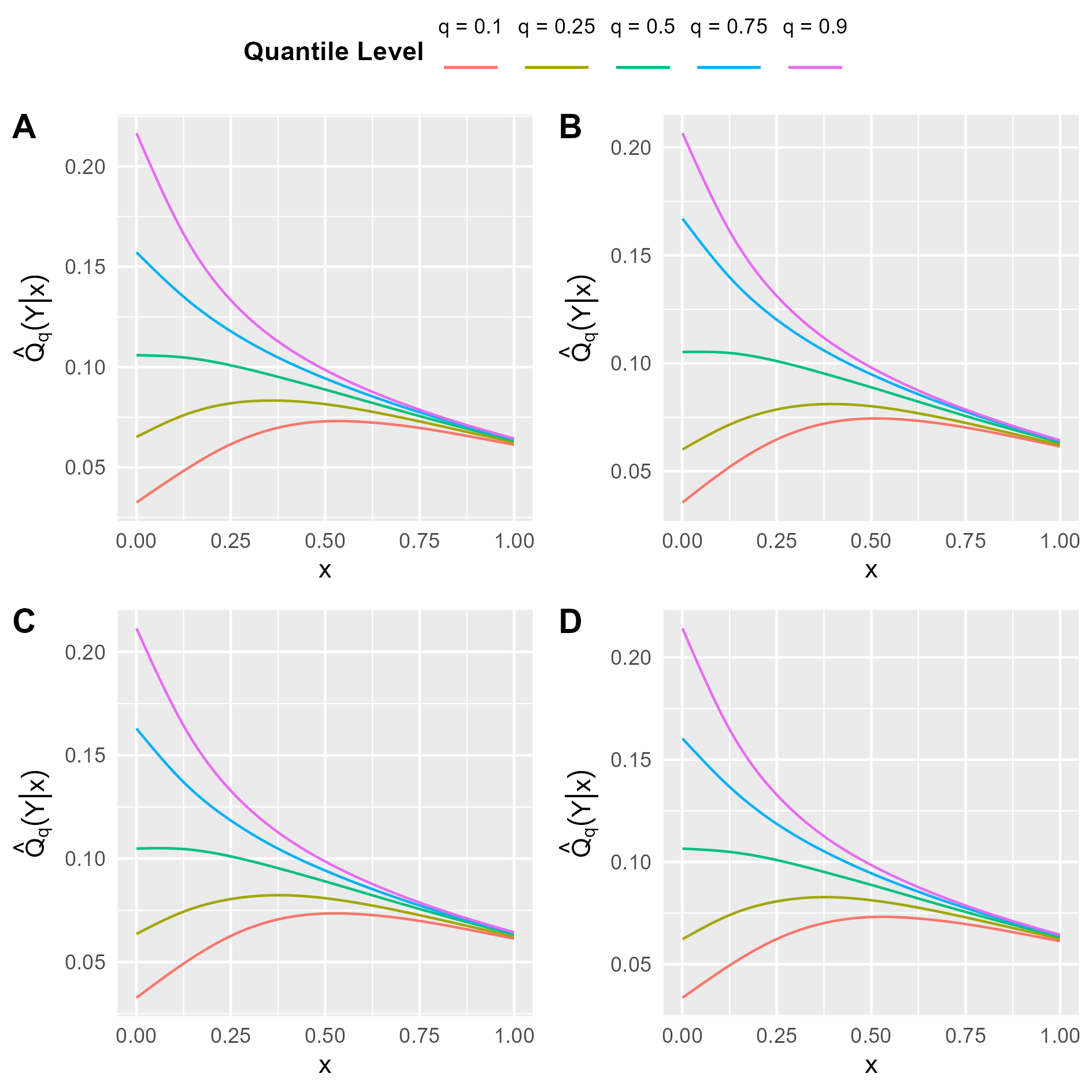}
	\caption{Mean quantile curves estimated by MM with logistic (\textbf{A}) and natural spline (ns) with 3 (\textbf{B}), 4 (\textbf{C}), 5 (\textbf{D}) equally spaced knots quantile level bases for simulated GG data from \Cref{GGd} (\Cref{GG theoretical quantile curves}\textbf{B}). We used 7 equally spaced knots to natural-spline-transform $x$.}
	\label{D62 MM xseq7 estimated quantile curves}
\end{figure}

\begin{figure}[ph!]
	\centering
	\includegraphics[width = \textwidth]{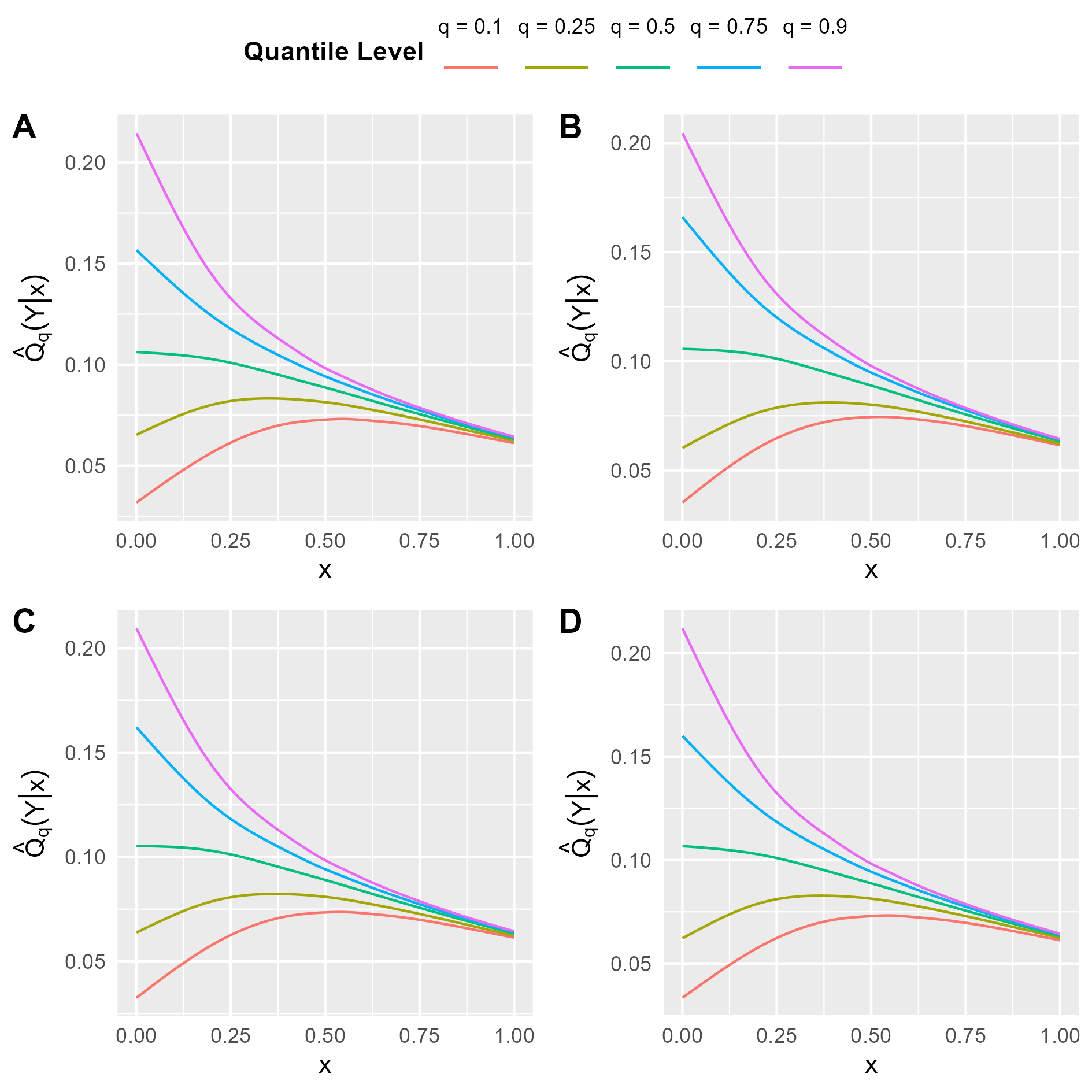}
	\caption{Mean quantile curves estimated by MM with logistic (\textbf{A}) and natural spline (ns) with 3 (\textbf{B}), 4 (\textbf{C}), 5 (\textbf{D}) equally spaced knots quantile level bases for simulated GG data from \Cref{GGd} (\Cref{GG theoretical quantile curves}\textbf{B}). We used knots $0.2, 0.4, 0.5, 0.55, 0.6, 0.7, 0.9$ to natural-spline-transform $x$.}
	\label{D62 MM xasym7 estimated quantile curves}
\end{figure}

 IMSE charts produced using both \Cref{GGb,GGd} (\Cref{IgG IMSE,D62 IMSE}) largely favor MM over Double Kernel (DK). Performances of DK with only optimal bandwidth choices are comparable to those of some MM scenarios we have carried out. MM's superiority is also quite evidently showcased in the estimated quantile curves at $q=0.1,0.25,0.5,0.75,0.9$ (\Cref{IgG DK estimated quantile curves,IgG MM xseq3 estimated quantile curves,IgG MM xseq4 estimated quantile curves,IgG MM xseq5 estimated quantile curves,IgG MM xseq6 estimated quantile curves} for \Cref{GGb} and \Cref{D62 DK estimated quantile curves,D62 MM xseq3 estimated quantile curves,D62 MM xseq4 estimated quantile curves,D62 MM xseq5 estimated quantile curves,D62 MM xseq6 estimated quantile curves,D62 MM xseq7 estimated quantile curves,D62 MM xasym7 estimated quantile curves} for \Cref{GGd}), as MM's plots resemble the theoretical ones more.

\begin{table}[ph]
	\centering
	\scalebox{1}{
		\begin{tabular}{|c|ccccc|}
			\hline
			\textbf{Quantile Level} & \textbf{0.1} & \textbf{0.25} & \textbf{0.5} & \textbf{0.75} & \textbf{0.9} \\ 
			\hline
			MM xseq3qlogistic & 29.53798 & 33.15703 & 40.64669 & 51.95018 & 130.46971 \\ 
			MM xseq3qseq3 & 32.22690 & 32.55102 & 43.72007 & 90.98207 & 152.24396 \\ 
			MM xseq3qseq4 & 27.94898 & 37.05314 & 44.98011 & 63.38843 & 128.00275 \\ 
			MM xseq3qseq5 & 27.77062 & 36.51548 & 46.18318 & 64.00380 & 141.33597 \\ 
			\hline
			MM xseq4qlogistic & 30.26293 & 30.45600 & 40.51418 & 59.50063 & 156.27963 \\ 
			MM xseq4qseq3 & 33.31403 & 28.99899 & 44.41032 & 100.23370 & 177.29709 \\ 
			MM xseq4qseq4 & 28.80074 & 34.55267 & 46.04952 & 73.07086 & 152.24233 \\ 
			MM xseq4qseq5 & 29.07826 & 33.64704 & 47.13464 & 74.15304 & 171.70107 \\ 
			\hline
			MM xseq5qlogistic & 34.21044 & 31.65504 & 42.58306 & 67.03314 & 182.34407 \\ 
			MM xseq5qseq3 & 37.80912 & 29.47901 & 46.90506 & 109.76913 & 197.75112 \\ 
			MM xseq5qseq4 & 32.42703 & 35.81286 & 48.62030 & 82.87317 & 178.59810 \\ 
			MM xseq5qseq5 & 32.36871 & 35.04934 & 50.42919 & 84.64958 & 199.15493 \\ 
			\hline
			MM xseq6qlogistic & 39.33538 & 34.85782 & 46.57569 & 75.94048 & 210.34923 \\ 
			MM xseq6qseq3 & 42.88264 & 32.13177 & 51.64003 & 121.31662 & 224.35882 \\ 
			MM xseq6qseq4 & 37.81620 & 39.15473 & 52.90108 & 93.84072 & 205.46926 \\ 
			MM xseq6qseq5 & 37.95851 & 38.25552 & 55.37623 & 96.60661 & 232.71490 \\ 
			\hline
			DK $h_1$=0.06 & 134.62546 & 138.60452 & 190.93208 & 349.77425 & 877.77262 \\ 
			DK $h_1$=0.3 & 34.52883 & 38.27516 & 53.44602 & 80.50243 & 185.68365 \\ 
			DK $h_1$=0.6 & 41.75013 & 53.99047 & 72.34377 & 77.52675 & 143.73672 \\ 
			DK $h_1$=0.9 & 59.20143 & 90.26966 & 108.56249 & 106.52192 & 185.49303 \\ 
			DK $h_1$=1.2 & 81.05519 & 130.80855 & 148.14726 & 150.72910 & 267.49941 \\ 
			DK $h_1$=1.5 & 109.49982 & 173.51105 & 193.46701 & 211.09502 & 383.41182 \\ 
			DK $h_1$=1.8 & 144.86249 & 218.13756 & 245.50189 & 285.77092 & 527.16545 \\ 
			\hline
		\end{tabular}
	}
	\caption{IMSE (\Cref{imse}) for MM and Double Kernel (DK) calculated using $N=1000$ Generalized Gamma samples of size $n=500$ with parameters specified in \Cref{GGb} and $X\sim \unif(0,6)$. For MM, we used equally spaced knots to natural-spline-transform $x$ throughout (numbers of knots for $x$ are specified after ``xseq"). For DK, $h_2=10^{-4}$ for all choices of $h_1$.}
	\label{IgG IMSE}
\end{table}
\begin{table}[ph]
	\centering
	\scalebox{1}{
		\begin{tabular}{|c|ccccc|}
			\hline
			\textbf{Quantile Level} & \textbf{0.1} & \textbf{0.25} & \textbf{0.5} & \textbf{0.75} & \textbf{0.9} \\ 
			\hline
			MM xseq3qlogistic & 0.00707 & 0.00538 & 0.00551 & 0.00674 & 0.01490 \\ 
			MM xseq3qseq3 & 0.00835 & 0.00586 & 0.00599 & 0.00875 & 0.01608 \\ 
			MM xseq3qseq4 & 0.00680 & 0.00586 & 0.00612 & 0.00743 & 0.01435 \\ 
			MM xseq3qseq5 & 0.00705 & 0.00623 & 0.00625 & 0.00820 & 0.01509 \\ 
			\hline
			MM xseq4qlogistic & 0.00853 & 0.00641 & 0.00675 & 0.00814 & 0.01739 \\ 
			MM xseq4qseq3 & 0.00982 & 0.00681 & 0.00728 & 0.01041 & 0.01811 \\ 
			MM xseq4qseq4 & 0.00807 & 0.00706 & 0.00744 & 0.00908 & 0.01678 \\ 
			MM xseq4qseq5 & 0.00852 & 0.00762 & 0.00768 & 0.01013 & 0.01755 \\ 
			\hline
			MM xseq5qlogistic & 0.00977 & 0.00736 & 0.00789 & 0.00937 & 0.01991 \\ 
			MM xseq5qseq3 & 0.01127 & 0.00765 & 0.00856 & 0.01184 & 0.02049 \\ 
			MM xseq5qseq4 & 0.00941 & 0.00809 & 0.00864 & 0.01061 & 0.01932 \\ 
			MM xseq5qseq5 & 0.00995 & 0.00888 & 0.00900 & 0.01180 & 0.02046 \\
			\hline
			MM xseq6qlogistic & 0.01116 & 0.00833 & 0.00910 & 0.01064 & 0.02256 \\ 
			MM xseq6qseq3 & 0.01279 & 0.00859 & 0.00983 & 0.01326 & 0.02319 \\ 
			MM xseq6qseq4 & 0.01078 & 0.00923 & 0.01005 & 0.01208 & 0.02220 \\ 
			MM xseq6qseq5 & 0.01129 & 0.00998 & 0.01020 & 0.01360 & 0.02302 \\ 
			\hline
			MM xseq7qlogistic & 0.01266 & 0.00927 & 0.01012 & 0.01179 & 0.02500 \\ 
			MM xseq7qseq3 & 0.01412 & 0.00945 & 0.01092 & 0.01452 & 0.02524 \\ 
			MM xseq7qseq4 & 0.01230 & 0.01037 & 0.01112 & 0.01349 & 0.02453 \\ 
			MM xseq7qseq5 & 0.01275 & 0.01137 & 0.01140 & 0.01529 & 0.02600 \\ 
			\hline
			MM xasym7qlogistic & 0.01038 & 0.00732 & 0.00762 & 0.00895 & 0.01883 \\ 
			MM xasym7qseq3 & 0.01169 & 0.00779 & 0.00829 & 0.01128 & 0.01972 \\ 
			MM xasym7qseq4 & 0.01000 & 0.00810 & 0.00845 & 0.01017 & 0.01832 \\ 
			MM xasym7qseq5 & 0.01052 & 0.00900 & 0.00870 & 0.01119 & 0.01912 \\ 
			\hline
			DK h1=0.01 & 0.04036 & 0.03357 & 0.03332 & 0.04248 & 0.06991 \\ 
			DK h1=0.05 & 0.00997 & 0.00782 & 0.00703 & 0.01029 & 0.02347 \\ 
			DK h1=0.1 & 0.00925 & 0.00666 & 0.00449 & 0.01384 & 0.04093 \\ 
			DK h1=0.15 & 0.01346 & 0.00971 & 0.00689 & 0.02932 & 0.07780 \\ 
			DK h1=0.2 & 0.02018 & 0.01410 & 0.01302 & 0.05181 & 0.12667 \\ 
			DK h1=0.25 & 0.02782 & 0.01807 & 0.02189 & 0.07828 & 0.18427 \\ 
			DK h1=0.3 & 0.03514 & 0.02095 & 0.03221 & 0.10712 & 0.24698 \\ 
			\hline
		\end{tabular}
	}
	\caption{IMSE (\Cref{imse}) for MM and Double Kernel (DK) calculated using $N=1000$ Generalized Gamma samples of size $n=500$ with parameters specified in \Cref{GGd} and $X\sim \unif(0,1)$. For MM, we used equally spaced knots (numbers of knots for $x$ specified after ``xseq") and knots $(0.2,0.4,0.5,0.55,0.6,0.7,0.9)$ (``MM xasym7") to natural-spline-transform $x$. For DK, $h_2=10^{-4}$ for all choices of $h_1$.}
	\label{D62 IMSE}
\end{table}




\printbibliography
\end{document}